\pdfoutput=1
\RequirePackage{rotating}
\documentclass{raa}           
\usepackage{graphicx,times}
\usepackage{natbib}
\usepackage{amssymb,amsmath}
\bibpunct{(}{)}{;}{a}{}{,}
\usepackage{rotating}
\usepackage{mathptmx}
\usepackage{ae,aecompl}
\usepackage{txfonts}
\usepackage{pifont}

\usepackage[colorlinks,linkcolor=red,anchorcolor=green,citecolor=blue]{hyperref}
\begin{document}

   \title{Evolution of High-energy Electron Distribution in Pulsar Wind Nebulae}

 \volnopage{ {\bf 20XX} Vol.\ {\bf X} No. {\bf XX}, 000--000}
   \setcounter{page}{1}

   \author{Yi-Ming Liu
   \inst{1,2}, Hou-Dun Zeng\inst{1}, Yu-Liang Xin\inst{3}, Si-Ming Liu
      \inst{3},  Yi Zhang\inst{1}
   }

   \institute{ Key Laboratory of Dark Matter and Space Astronomy, Purple Mountain Observatory, 
Chinese Academy of Sciences, Nanjing 210023,
Peoples Republic of China; {\it zhd@pmo.ac.cn, zhangyi@pmo.ac.cn}\\
        \and
             School of Astronomy and Space Science, University of Science and Technology of China, Hefei 230026, China\\
	\and
School of Physical Science and Technology, Southwest Jiaotong University, Chengdu 610031, Peoples Republic of China: {\it liusm@swjtu.edu.cn}\\
\vs \no
   {\small Received 20XX Month Day; accepted 20XX Month Day}
}

\abstract{In this paper, we analyze the spectral energy distributions (SEDs) of 17 powerful (with a spin-down luminosity greater than $10^{35}$ erg s$^{-1}$) young (with an age less than 15000 yrs) pulsar wind nebulae (PWNe) using a simple time-independent one-zone emission model. Our aim is to investigate correlations between model parameters and the ages of the corresponding PWNe, thereby revealing the evolution of high-energy electron distributions within PWNe. Our findings are as follows: (1) The electron distributions in PWNe can be characterized by a double power-law with a superexponential cutoff; (2) As PWNe evolve, the high-energy end of the electron distribution spectrum becomes harder with the index decreasing from approximately 3.5 to 2.5, while the low-energy end spectrum index remains constant near 1.5; (3) There is no apparent correlation between the break energy or cutoff energy and the age of PWNe. (4) The average magnetic field within PWNe decreases with age, leading to a positive correlation between the energy loss timescale of electrons at the break energy or the high-energy cutoff, and the age of the PWN. 
(5) The total electron energy within PWNe remains constant near  $2 \times 10^{48}$ erg, while the total magnetic energy decreases with age. 
}

   \authorrunning{Y.-M. Liu et al. }            
   \titlerunning{Evolution of High-energy Electron Distribution in PWNe}  
   \maketitle

%
\section{Introduction}           
\label{sect:intro}

Pulsar wind nebulae (PWNe) are dynamic and energetic structures in the cosmos driven by the rotational energy released from pulsars. These nebulae consist of bubbles filled with magnetized relativistic particles, primarily electrons and positrons, which interact with the surrounding medium to emit a wide range of electromagnetic radiation from radio waves to TeV gamma-rays \cite[e.g.][]{2017hsn..book.2159S,2024arXiv240210912A}. Particle acceleration occurs at the termination shock within the PWN \citep{1984ApJ...283..694K}, creating a complex system where particle transport and radiation processes shape their observed properties. PWNe are considered as potential sources of high-energy electrons and positrons of cosmic rays \citep[e.g.][]{2022MNRAS.511.1439F}. The observational results from the Large High Altitude Air Shower Observatory (LHAASO) provide strong evidence for the particle acceleration capability of PWNe. The high-energy gamma-ray photons above 1 PeV detected from Crab Nebula not only confirm that PWNe are efficient particle accelerators in the Galaxy, but also make it to be the first officially recognized PeV-scale leptonic sources \citep{doi:10.1126/science.abg5137}. Moreover, most of the very-high-energy gamma-ray sources observed by LHAASO are associated with pulsars \citep{2024ApJS..271...25C}.

Theoretical models, including pure leptonic scenarios, have been proposed to explain the observed multiband nonthermal photon spectra emitted by PWNe \citep[e.g.][]{1984ApJ...283..694K,2008ApJ...689L.125D,2008ApJ...676.1210Z,2009ApJ...703.2051G,2010A&A...515A..20F,2010ApJ...715.1248T,2011MNRAS.410..381B,2012MNRAS.427..415M,2014JHEAp...1...31T,2018A&A...609A.110Z,2021A&A...655A..41Z,Zhu_2023}. These models often involve two distinct particle populations that are accelerated within the PWN, such as the pulsar wind termination shock \citep[][]{1974MNRAS.167....1R,1984ApJ...283..694K} or the nebula's equatorial region \citep[e.g.][]{1996MNRAS.278..525A,2019MNRAS.489.2403L}. While synchrotron radiation is believed to generate radio to X-ray photons, and the GeV to TeV gamma-rays are typically produced through inverse Compton scattering processes involving ultrarelativistic electrons and positrons. Additionally, a possible hadronic origin for gamma-ray emission has been suggested \cite[][]{1996MNRAS.278..525A,1997PhRvL..79.2616B,2003A&A...402..827A,2003A&A...405..689B}, highlighting the complexity of particle interactions within PWNe.

Overall, PWNe serve as important astrophysical laboratories for studying the dynamics of relativistic particles, magnetic fields, and shock physics in extreme environments. By analyzing the broadband spectral energy distributions and evolution of PWNe, We can deepen the understanding of cosmic ray production and propagation mechanisms. 
Unlike the currently prevailing time-dependent models accounting of the nebula evolution and electron losses \citep[e.g.][]{2014JHEAp...1...31T,2018A&A...609A.110Z}, here we used the simplest phenomenological model (time-independent one-zone model) to fit the multiband spectra of 16 PWNe. Attempting to reveal the evolution of electrons in PWNe through the correlation between model parameters and age. In Section 2, we briefly describe our selected sample sources and models. Section 3 shows the results of the spectral fitting and their implications.  Summary and discussion are drawn in Section 4.



\section{Sample and Model}
\label{sect:Obs}
\subsection{Sample}
In the supernova remnant source catalog\footnote{SNRcat is the online high-energy catalogue of supernova remnants (SNRs), \url{http://snrcat.physics.umanitoba.ca}}  containing 383 sources \citep[][up-to-date database]{2012AdSpR..49.1313F}, there are 111 sources potentially housing PWNe or displaying PWNe characteristics. \cite{2022arXiv220911855E} suggests the detection of at least 125 PWNe from radio to TeV bands, primarily through radio or X-ray surveys \citep{2013uean.book..359K}. Notably, around 70 PWNe were discerned via observations with the Chandra telescope \citep{2008AIPC..983..171K}. In recent years, the number of pulsars discovered in the gamma-ray band has also increased. The fourth Fermi Large Area Telescope (LAT) catalog (4FGL-DR4) contains 19 sources, including 11 confirmed and 8 associated PWNe \citep{2022ApJS..260...53A}. A further 36 PWNe are documented in the TeV source catalog specialized in TeV Astronomy\footnote{\url{http://tevcat2.uchicago.edu/}}. \cite{2018A&A...612A...2H} reported 14 identified PWNe and 10 candidate PWNe. However, only a handful of PWNe possess comprehensive multi-band data, spanning radio, X-ray, GeV and TeV ranges. Our goal is to study the evolution of electron distributions in PWNe by using a time-independent one-zone model to fit multi-band data of PWNe. To well constrain the electron distributions and magnetic fields in PWNe, the data with at least three wavelengths are required for each source in the sample. 
Considering the sources in \cite{2014JHEAp...1...31T} and \cite{2018A&A...609A.110Z}, we obtain a supplementary sample comprising 17 PWNe. 
It is obvious that Crab nebula possesses a complete spectrum covering all wavebands from radio to PeV. We note that PWNe such as 3C 58, HESS J1813-178, MSH 15-52 and VER J2227+608 contain complete data across radio, X-ray, GeV and TeV bands. However, many PWNe only have flux upper limits in the radio band, for example, HESS J1640-465, HESS J1356-645, HESS J1418-609, HESS J1420-607, HESS J1427-608, and HESS J1303-637. Five PWNe have spectroscopic data in radio, X-ray and TeV bands but have not been obtained by the Fermi satellite, namely Kes75, G54.1+0.3, G25.1-0.9, G0.9+0.1, and N157B. Notably, G25.1-0.9 also has infrared band data. Furthermore, even though CTA 1 only has energy upper limit data in the radio band, its X-ray and TeV band data make it a strong candidate for parameter constraint studies.

\subsection{Model Description and Spectral Fitting Strategy}

The model we considered here is the simplest one-zone time-independent leptonic model, which considers the present distribution of leptons. With the results of fitting the SEDs of 17 PWNe, we can reveal the evolution of leptons in these nebulae by correlating the physical parameters of the model with PWNe's age. The distribution function of leptons is assumed to be a broken power law form:
\begin{eqnarray}
f(E) = A \times {e^{ - (E/E_{\rm c})}}^\beta 
 \left\{
\begin{array}{lcl}
 (E/{E_0})^{ -\alpha _1}&& {\rm if}~~~ E \le E_{\rm b} \\
(E_{\rm b}/E_0)^{\alpha _2 - \alpha _1} (E/{E_0})^{- \alpha _2} && {\rm if}~~~E> E_{\rm b}\;,
\end{array}
\right.
\label{eq:elec distribution}
\end{eqnarray}
where $E_0$ is the fiducial energy, and we choose it to be $E_0=1$ TeV; $E_b$ is the break energy, and $\alpha _1$ and $\alpha _2$ are low- and high-energy spectral indices, respectively. $E_c$ is the high-energy cutoff of the leptons, and the index of exponential cutoff $\beta$ is set at 2.0 in consideration of the electron energy losses\citep{2007A&A...465..695Z}. $A$ is the normalization of the particle distribution, which can be determined by the total energy content of leptons above 1 GeV, denoted as $W_e$. We defined $E_{\rm min}=100$ MeV and $E_{\rm max}=50$ PeV to ensure that the electrons remain relativistic and capable of efficiently producing synchrotron emission with the mean magnetic field $B$ as a variable, allowing for the fitting of nonthermal radio to X-ray data. Additionally, these electrons are able to generate gamma-ray emission through inverse Compton scattering (ICS) with ambient low-energy background photons such as the cosmic microwave background radiation (CMB), diffuse Galactic infrared background, or starlight. Note that whenever necessary, self-synchrotron Compton scattering is introduced, such as for Crab nebula. It should be noted that the background photon field varies for each PWN apart from CMB component, and such values are adopted from \cite{2018A&A...609A.110Z} for most PWNe, except for HESS J1640-465,  VER J2227+608 and HESS J1427-608. The information of the background photon field of these two PWNe comes from \cite{2018ApJ...867...55X},
\cite{2021Innov...200118G} 
and \cite{2017ApJ...835...42G}, respectively.
The distance, characteristic age and spin-down power of the pulsar of PWN, as well as the age of the PWN, can be obtained from the related literature. In this model, there are six free parameters: $\alpha_1$, $\alpha _2$, $E_b$, $E_c$, $W_e$ for the electron distribution, and the magnetic field $B$ for synchrotron radiation. The \textit{naima} package
by \cite{naima} is utilized for computing non-thermal radiation from relativistic electron populations and fitting the observed spectra with Markov Chain Monte Carlo (MCMC) algorithm.

\section{The Results}
Figure 1-17 illustrate the best-fit SEDs with the one-dimensional (1D) posterior probability distribution and two-dimensional (2D) confidence contours of parameters for 17 PWNe. The thick black line represents the total SED, while the green solid line depicts the synchrotron emission. The dashed, dash–dotted and dotted lines correspond to the ICS components with CMB and other background photon fields, respectively. References for the spectra and background photon field parameters can be found in the figure caption.

\begin{sidewaystable}
\begin{center}
\caption[]{Sample and Spectral Fitting Parameters.}\label{Tab:1}
\setlength{\tabcolsep}{3.0pt}
\begin{tabular}{lcccccccccccc}
  \hline 
Name & $D$ & $R^a$ & $\tau_c$ & $L(t)$& $T_{\rm age}$& $\alpha_1$& $\alpha_2$& $E_{br}$&  $E_{cut}$&  $B$&  $W_e$& $\frac{\chi2}{N-n}$      \\
& kpc & pc & year & erg/s & year &&& GeV & TeV & $\mu$ G & erg &\\
  \hline
G21.5$-$0.9 & 4.1 & $<4.0$ & 4850 & $3.38 \times 10^{37}$ & $900^{+1100}_{-30}$ & $1.21^{+0.16}_{-0.18}$ &$3.06^{+0.05}_{-0.06}$ & $84^{+36}_{-22}$ & $410^{+320}_{-130}$ & $59^{+8.7}_{-6.9}$ & $6.7^{+1.7}_{-1.6} \times 10^{47}$  & $\frac{20.12}{37-6}=0.65$\\
Crab nebula & 2.0 & $<3.0$ & 1296 & $4.53 \times 10^{38}$ & $940$ & $1.74^{+0.001}_{-0.001}$ &$3.44^{+0.003}_{-0.003}$ & $1034^{+11}_{-11}$ & $2423^{+22}_{-22}$ & $113^{+1}_{-1}$ & $6.25^{+0.08}_{-0.08} \times 10^{48}$  & $\frac{3144.0}{298-6}=10.77$\\
Kes 75 & 5.8 & $<3.0$ & 726 & $8.21 \times 10^{36}$ & $1000^{+770}_{-20}$ & $1.71^{+0.09}_{-0.09}$ &$2.99^{+0.10}_{-0.11}$ & $900^{+460}_{-340}$ & $>504$ & $15^{+2}_{-2}$ & $3.7^{+0.7}_{-0.7} \times 10^{47}$  & $\frac{6.62}{14-6}=0.83$\\
HESS J1640$-$465 & 10.0 & $25 \pm 8$ & 3100 & $4.40 \times 10^{36}$ & $3094^{+406}_{-1994}$ & $1.45^{+0.34}_{-0.49}$ &$3.43^{+0.06}_{-0.08}$ & $1530^{+280}_{-230}$ & $250^{+570}_{-140}$ & $3.7^{+0.8}_{-0.7}$ & $4.6^{+2.7}_{-1.4} \times 10^{48}$  & $\frac{26.62}{27-6}=1.27$\\
3C 58 & 2.0 & $<5.0$ & 5397 & $2.68 \times 10^{37}$ & $2400^{+500}_{-500}$ & $1.04^{+0.03}_{-0.04}$ &$3.17^{+0.01}_{-0.01}$ & $31^{+3}_{-2}$ & $155^{+14}_{-12}$ & $17.7^{+0.9}_{-0.8}$ & $1.9^{+0.1}_{-0.1} \times 10^{48}$  & $\frac{79.0}{56-6}=1.58$\\
HESS J1813$-$178 & 4.7 & $4.0 \pm 0.3$& 5624 & $5.56 \times 10^{37}$ & $\sim 2500$ & $1.92^{+0.03}_{-0.03}$ &$3.0^{+0.1}_{-0.1}$ & $4600^{+1900}_{-1400}$ & $>1800$ & $4.7^{+0.8}_{-0.7}$ & $1.6^{+0.3}_{-0.2} \times 10^{48}$  & $\frac{15.60}{14-6}=1.95$\\
G54.1$+$0.3 & 7.0 & $<9.0$ & 2889 & $1.16 \times 10^{37}$ & $2600^{+3400}_{-900}$ & $1.35^{+0.09}_{-0.09}$ &$3.0^{+0.2}_{-0.2}$ & $240^{+200}_{-120}$ & $>450$ & $9.4^{+2.6}_{-2.0}$ & $1.8^{+0.4}_{-0.4} \times 10^{48}$  & $\frac{6.82}{12-6}=1.14$\\
G0.9$+$0.1 & 13.3 & $<7.0$ & 5305 & $4.32 \times 10^{37}$ & $3000^{+0}_{-1000}$ & $1.24^{+0.34}_{-0.33}$ &$2.95^{+0.11}_{-0.11}$ & $54^{+43}_{-20}$ & $>400$ & $17.5^{+3.9}_{-3.0}$ & $5.8^{+1.1}_{-1.0} \times 10^{48}$  & $\frac{3.20}{13-6}=0.46$\\
MSH 15$-$52 & 4.4 & $11.1 \pm 2.0$ & 1585 & $1.75 \times 10^{37}$ & $4000^{+0}_{-2500}$ & $1.4^{+0.1}_{-0.2}$ &$2.90^{+0.06}_{-0.06}$ & $540^{+370}_{-190}$ & $620^{+430}_{-170}$ & $15^{+2}_{-1}$ & $1.9^{+0.4}_{-0.6} \times 10^{48}$  & $\frac{18.06}{35-6}=0.62$\\
N 157B & 53.7 & $<94$ & 4982 & $4.83 \times 10^{38}$ & $4600^{+400}_{-0}$ & $1.34^{+0.07}_{-0.08}$ &$3.34^{+0.09}_{-0.09}$ & $180^{+52}_{-38}$ & $170^{+240}_{-61}$ & $27^{+2}_{-2}$ & $1.5^{+0.2}_{-0.2} \times 10^{50}$  & $\frac{48.48}{38-6}=1.52$\\
HESS J1356$-$645 & 2.5 & $10.1+0.9$ & 7310 & $3.10 \times 10^{36}$ & $6500^{+1500}_{-500}$ & $0.40^{+0.27}_{-0.24}$ &$2.56^{+0.05}_{-0.05}$ & $32^{+9}_{-7}$ & $570^{+2200}_{-380}$ & $3.3^{+0.6}_{-0.4}$ & $6.6^{+0.9}_{-1.0} \times 10^{47}$  & $\frac{15.86}{16-6}=1.59$\\
VER J2227$+$608 & 0.8 & 3.5 & 10500 & $2.25 \times 10^{37}$ & $\sim 7000$ & $2.53^{+0.01}_{-0.01}$ &$4.16^{+0.48}_{-0.42}$ & $14.9^{+3.02}_{-2.89} \times 10^{4}$ & $>1295$ & $2.29^{+0.08}_{-0.07}$ & $2.6^{+0.2}_{-0.2} \times 10^{47}$  & $\frac{32.52}{32-6}=1.25$\\
CTA 1 & 1.4 & $6.6\pm 0.5$ & 13863 & $4.52 \times 10^{35}$ & $7500^{+7500}_{-2500}$ & $1.4^{+0.4}_{-0.4}$ &$2.7^{+0.2}_{-0.2}$ & $560^{+1500}_{-390}$ & $>420$ & $7.6^{+1.1}_{-1.1}$ & $3.8^{+3.5}_{-2.3} \times 10^{46}$  & $\frac{5.20}{9-6}=1.73$\\
HESS J1418$-$609 & 5.0 & $9.4 \pm 0.9$ & 10376 & $4.93 \times 10^{36}$ & $\sim 8000$ & $0.69^{+0.38}_{-0.37}$ &$2.82^{+0.12}_{-0.13}$ & $220^{+430}_{-120}$ & $>270$ & $2.5^{+0.7}_{-0.6}$ & $2.7^{+2.3}_{-1.6} \times 10^{48}$  & $\frac{6.26}{13-6}=0.89$\\
HESS J1420$-$607 & 5.6 & $7.9 \pm 0.6$ & 12990 & $1.04 \times 10^{37}$ & $\sim 8500$ & $1.2^{+0.2}_{-0.3}$ &$2.89^{+0.08}_{-0.08}$ & $420^{+490}_{-180}$ & $460^{+530}_{-170}$ & $3.1^{+0.5}_{-0.3}$ & $4^{+2}_{-2} \times 10^{48}$  & $\frac{7.78}{20-6}=0.56$\\
HESS J1427$-$608 & 8.0 & 7.0 & 11000 & $6.50 \times 10^{36}$ & $10000^{+0}_{-3600}$ & $1.1^{+0.5}_{-0.6}$ &$2.6^{+0.1}_{-0.1}$ & $77^{+230}_{-51}$ & $103^{+54}_{-27}$ & $3.5^{+1.2}_{-1.0}$ & $2.6^{+2.2}_{-1.4} \times 10^{48}$  & $\frac{6.06}{15-6}=0.67$\\
HESS J1303$-$631 & 6.6 & $20.6 \pm 1.7$ & 11008 & $1.68 \times 10^{36}$ & $\sim 13000$ & $1.3^{+0.3}_{-0.3}$ &$2.5^{+0.1}_{-0.1}$ & $460^{+240}_{-230}$ & $62^{+20}_{-14}$ & $3.0^{+0.8}_{-0.7}$ & $6.5^{+3.5}_{-1.4} \times 10^{47}$  & $\frac{35.9}{15-6}=3.99$\\
  \hline
\end{tabular}
\end{center}
\tablecomments{0.86\textwidth}{$^a$. $R$ represents the 1 $\sigma$ Gaussian extension derived from TeV detection \citep{2018A&A...612A...2H} expect for HESS J1427$-$608, which exhibits a slightly extended morphology consistent with a symmetric Gaussian of radius 
$\sigma \sim 3^{'}$ \citep{2008A&A...477..353A}, corresponding to $R \sim 7$ pc. For VER J2227$+$608, its TeV emission\citep{2009ApJ...703L...6A} can be described by a 1$\sigma$ angular extent of $0.27^o \pm 0.05^o$ along the major axis, and
$0.18^o \pm 0.03^o$ along the minor axis, with $R \sim 3.5$ pc assumed.\\
Information regarding the distance, characteristic age, and spin-down power of the pulsar associated with pulsar wind nebulae (PWNe), as well as the PWN age with errors, can be sourced from \citep{2018A&A...609A.110Z} and \citep{2014JHEAp...1...31T}, except for HESS J1640$-$465, referenced in \citep{2018ApJ...867...55X,2021ApJ...912..158M}, and HESS J1427$-$608, as discussed in \citep{2013ApJ...773..139V,2017ApJ...835...42G}.}
\end{sidewaystable}

\begin{figure}
\centering
   \includegraphics[width=7.0cm, angle=0]{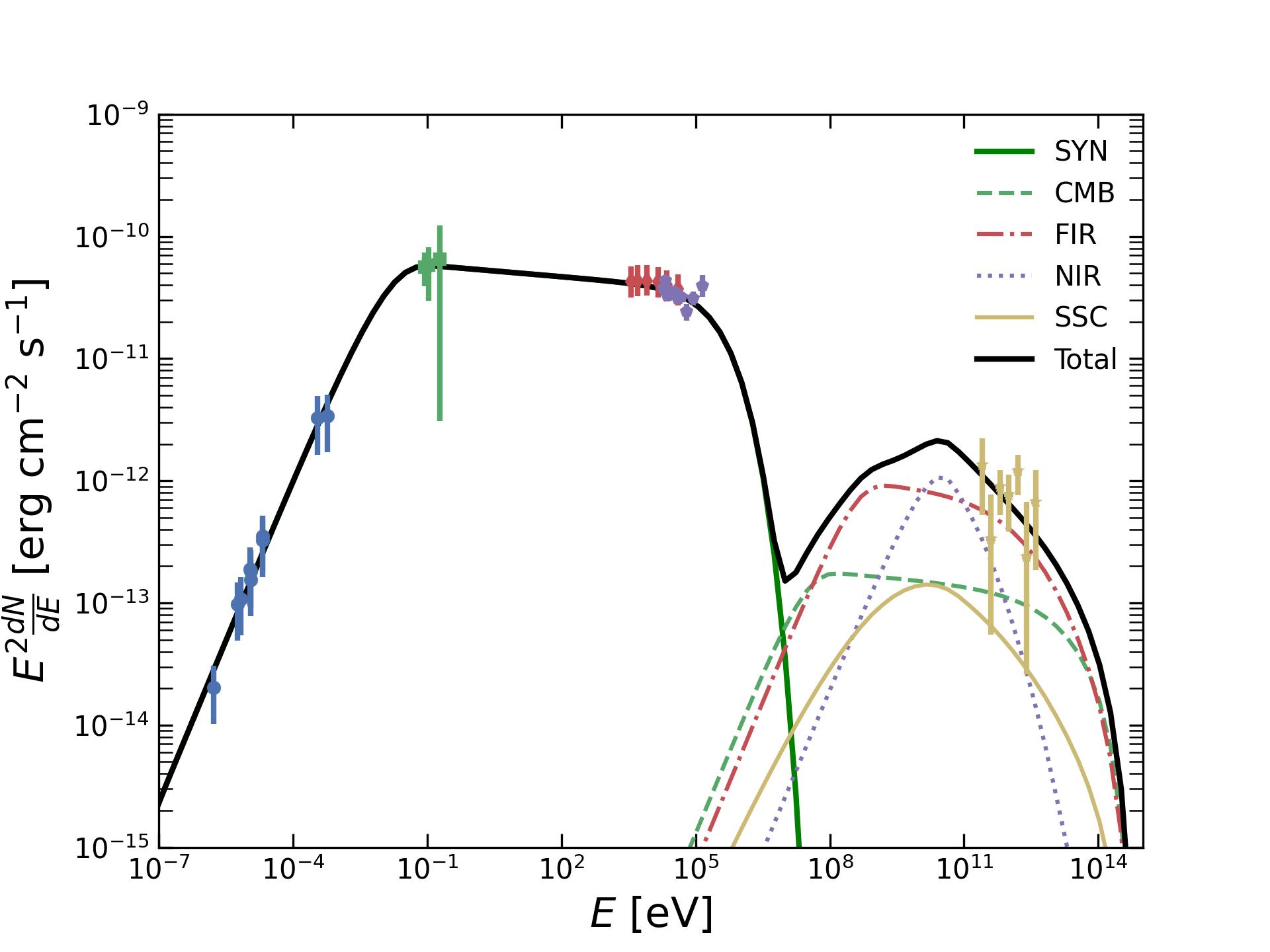}
   \includegraphics[width=6.0cm, angle=0]{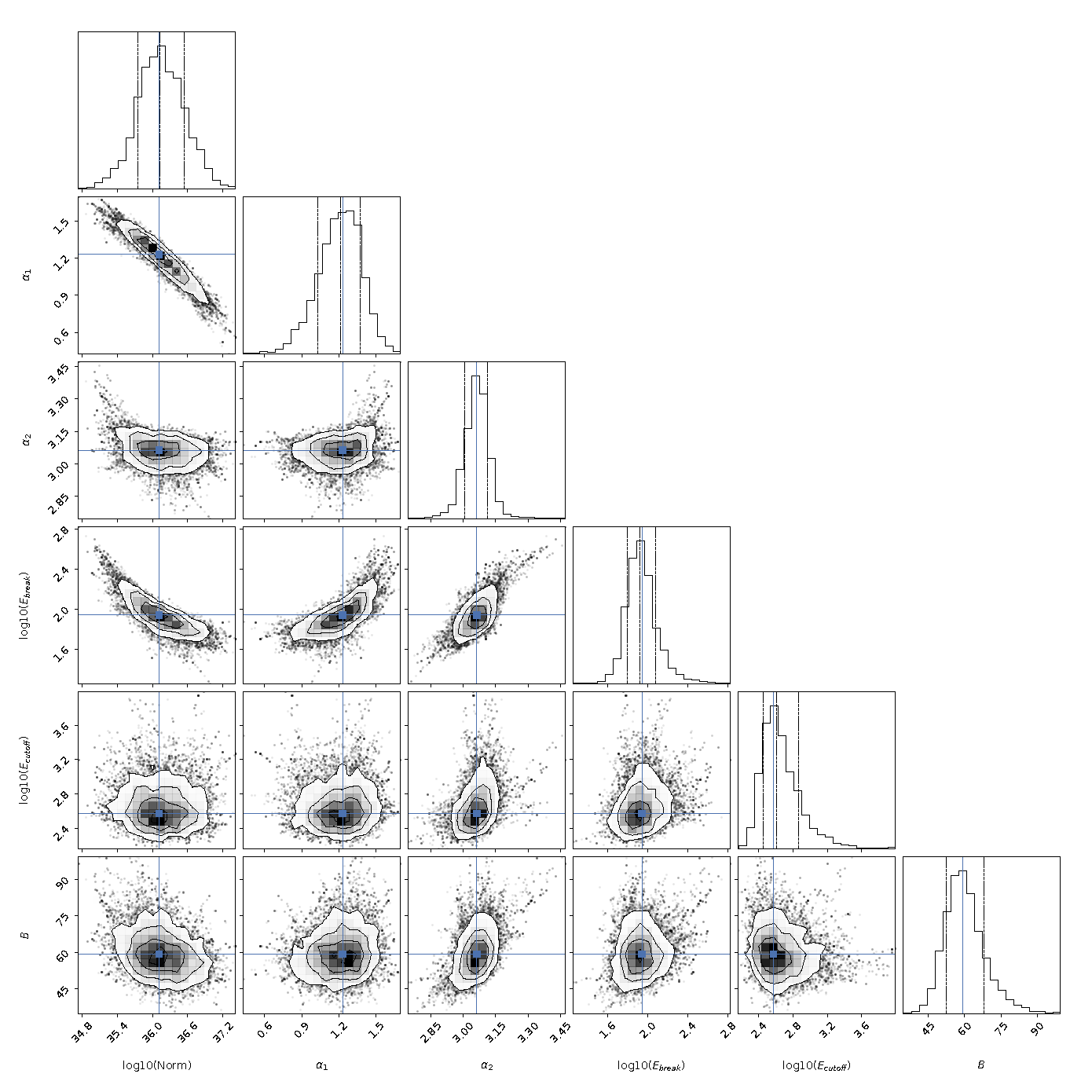}
\caption{The best-fit to the spectral energy distribution (SED) and the posterior probability distribution and two-dimensional(2D) confidence contours of parameters for G21.5$-$0.9. The green thick solid line represents synchrotron radiation, and the green dashed, red dot-dashed, and purple dotted lines are for Inverse-Compton scatterings off CMB, FIR and NIR, respectively. The yellow thick solid line represents synchrotron self-compton radiation. The total SED is displayed by the black thick solid line. The IR background with $T_{\rm FIR} = 35.0$ K and $U_{\rm FIR} = 1.4$ eV cm$^{-3}$ and $T_{\rm NIR} = 3500.0$ K and $U_{\rm NIR} = 5.0$ eV cm$^{-3}$, are used. The radio band data from \cite{1989ApJ...338..171S}, the IR band data from \cite{1998MmSAI..69..963G,1999ESASP.427..313G}, the X-ray band data from \cite{2014ApJ...789...72N}, the TeV $\gamma$-ray band data from \cite{2013ApJ...764...38A}.}
\label{fig:G21.5-0.9}
\end{figure}

\begin{figure}
\includegraphics[width=7.0cm, angle=0]{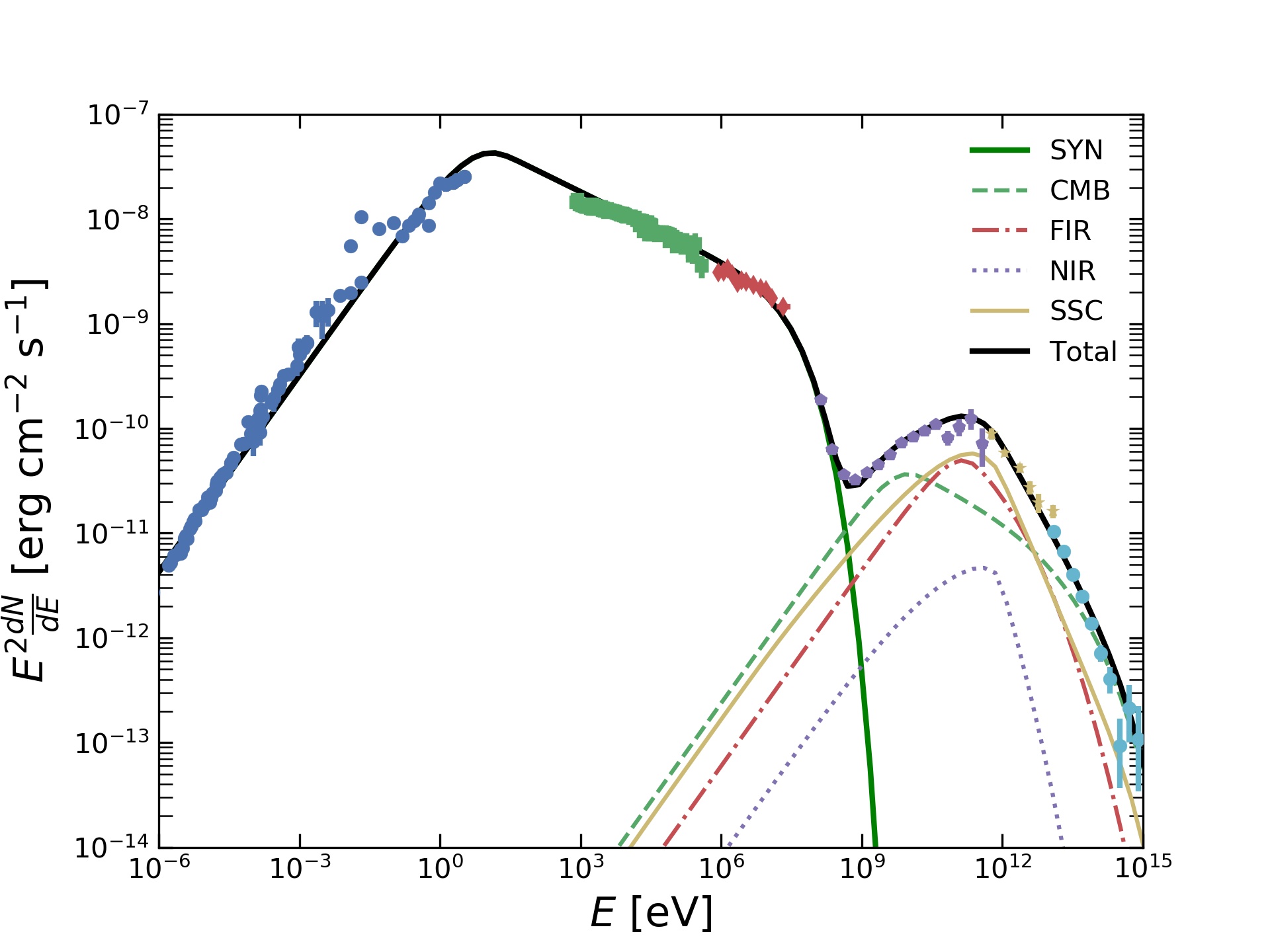}
\includegraphics[width=6.0cm, angle=0]{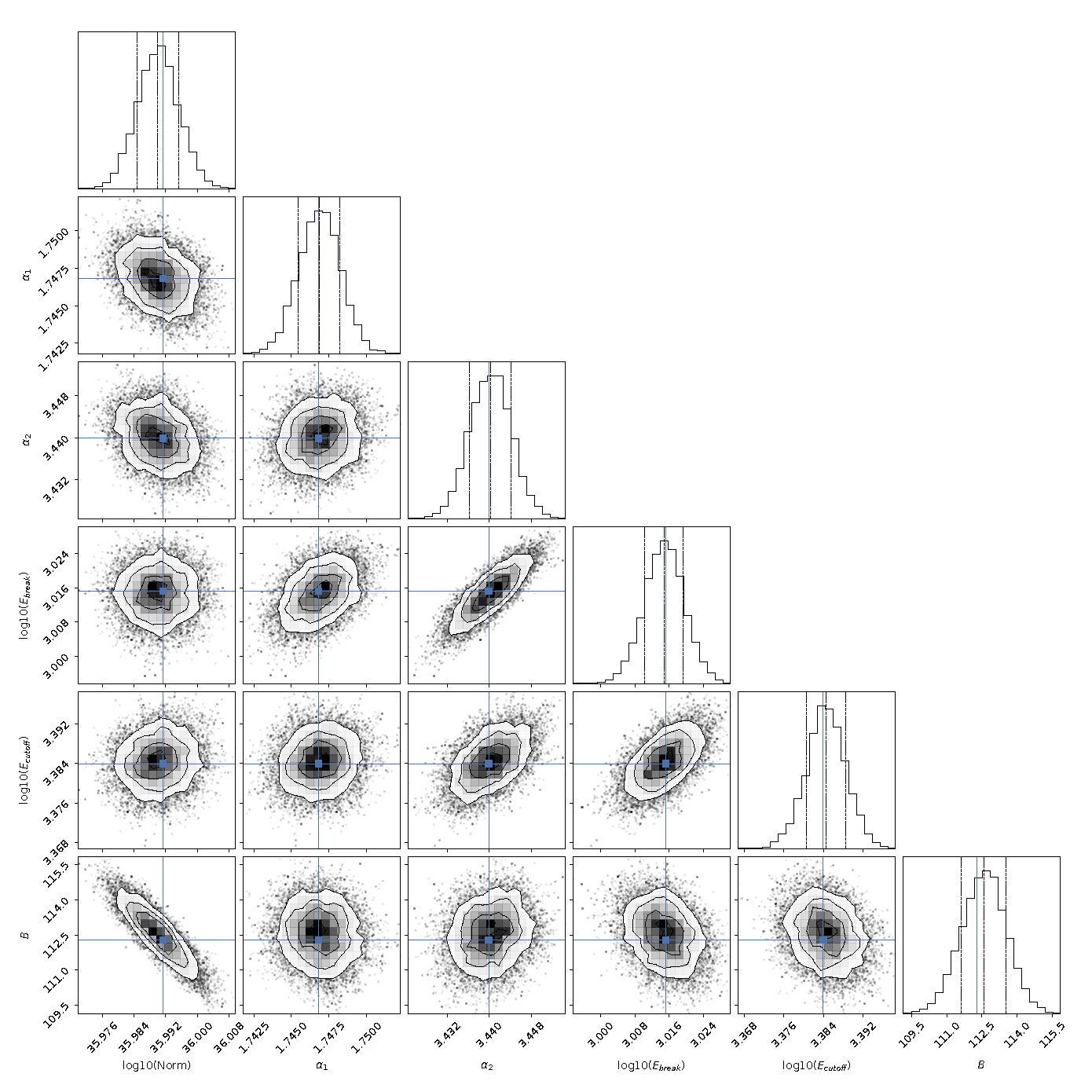}
\caption{The same as Figure \ref{fig:G21.5-0.9} but for Crab nebula. And the energy densities and temperatures of FIR and NIR are $T_{\rm FIR} = 70.0$ K and $U_{\rm FIR} = 0.5$ eV cm$^{-3}$, and $T_{\rm NIR} = 5000.0$ K and $U_{\rm NIR} = 1.0$ eV cm$^{-3}$. The radio band data from \cite{2010ApJ...711..417M}, the IR band data from \cite{2006AJ....132.1610T}, the optical band data from \cite{1993A&A...270..370V}, the X-ray band data from \cite{2001A&A...378..918K}, the GeV data from \cite{2020ApJ...897...33A}  and the TeV $\gamma$-ray band data from \cite{2021Sci...373..425L}.}
\label{fig:Crab}
\end{figure}

\begin{figure}
\includegraphics[width=7.0cm, angle=0]{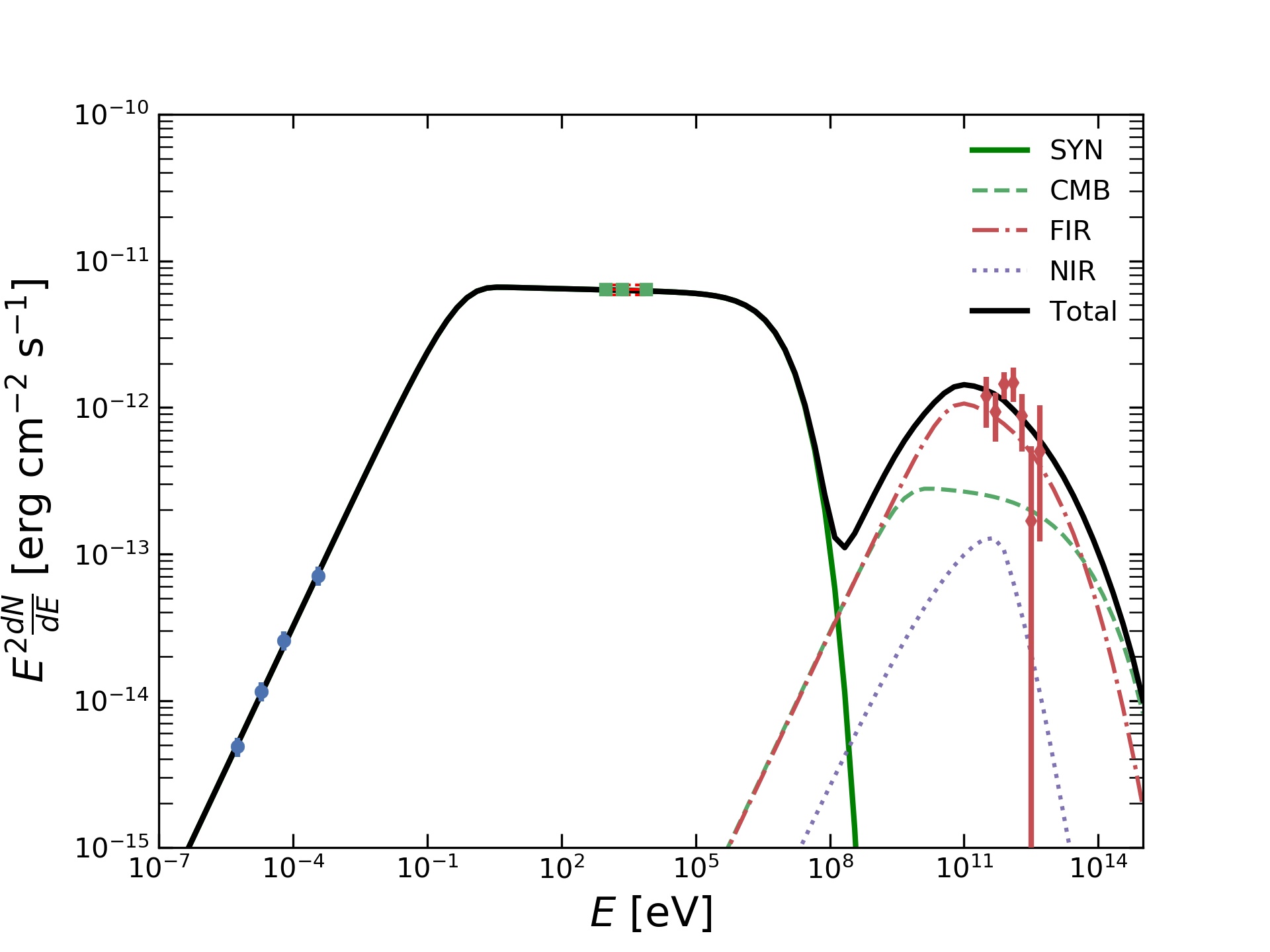}
\includegraphics[width=6.0cm, angle=0]{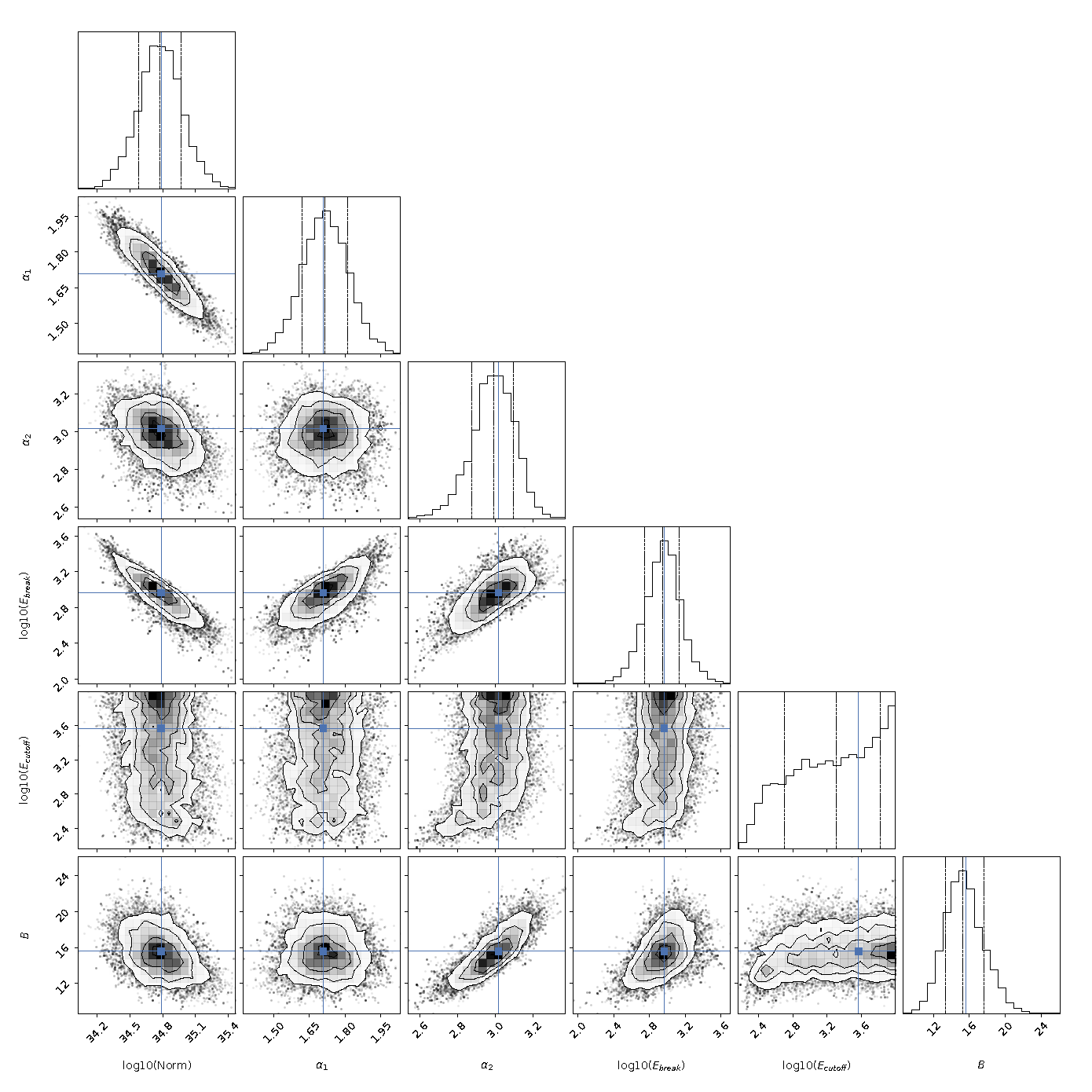}
\caption{The same as Figure \ref{fig:G21.5-0.9} but for Kes 75. And the energy densities and temperatures of FIR and NIR are $T_{\rm FIR} = 30.0$ K and $U_{\rm FIR} = 1.2$ eV cm$^{-3}$, and $T_{\rm NIR} = 3000.0$ K and $U_{\rm NIR} = 2.2$ eV cm$^{-3}$. The radio band data from \cite{1989ApJ...338..171S,2005ApJ...626..343B}, the X-ray band data from \cite{2018ApJ...856..133R} and the TeV $\gamma$-ray band data from \cite{2013ApJ...764...38A}.}
\end{figure}

\begin{figure}
\includegraphics[width=7.0cm, angle=0]{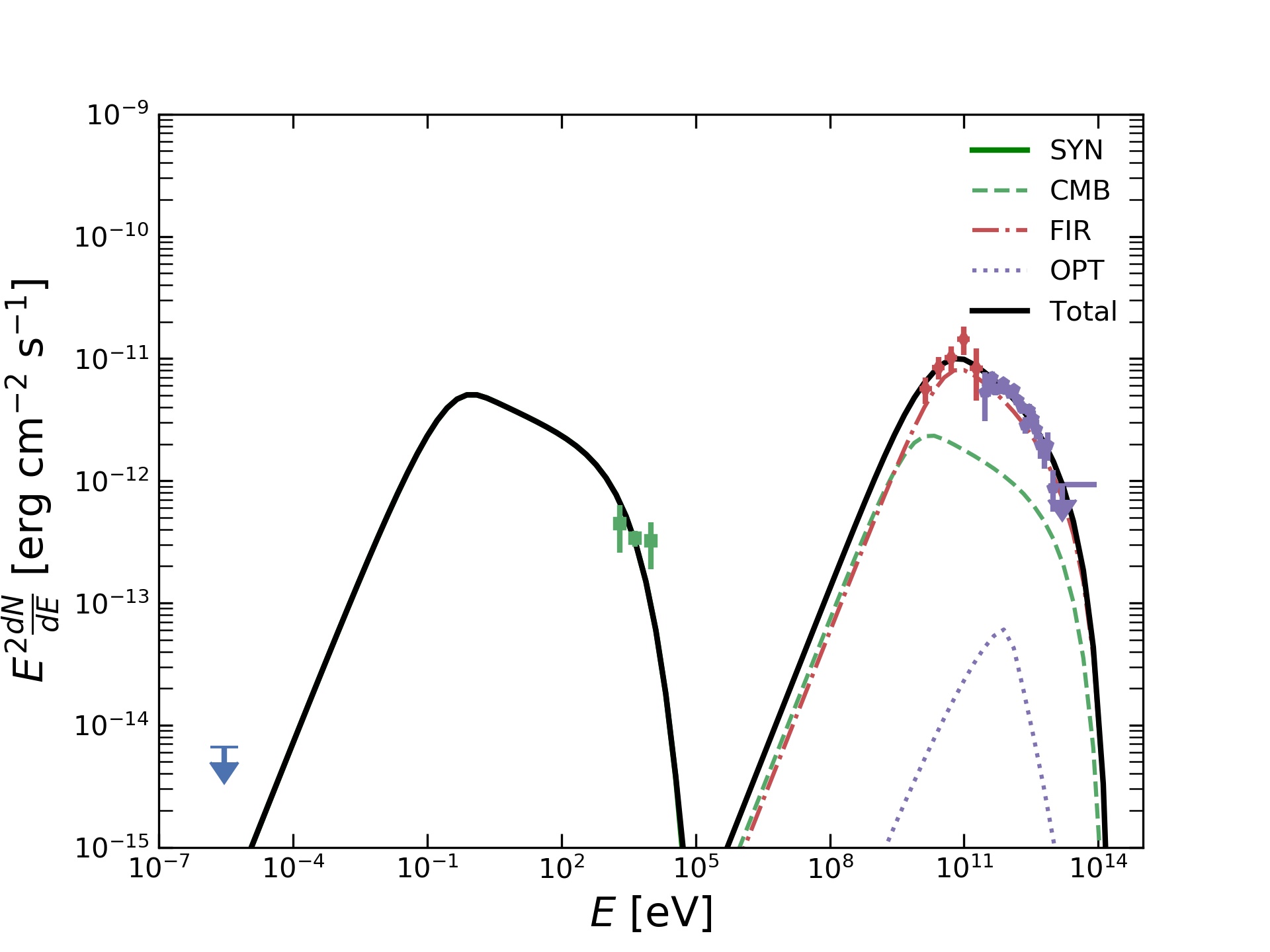}
\includegraphics[width=6.0cm, angle=0]{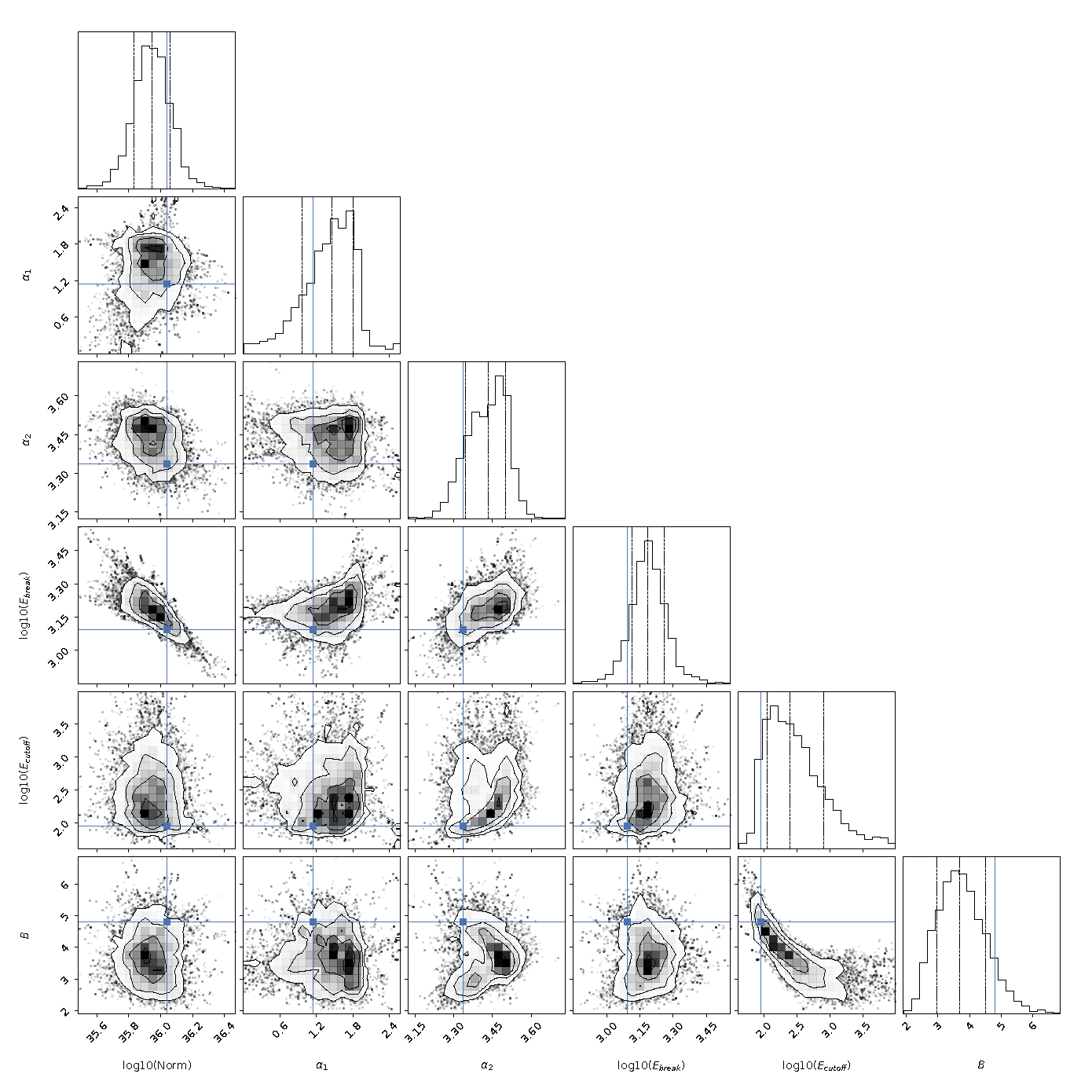}
\caption{The same as Figure \ref{fig:G21.5-0.9} but for HESS J1640$-$465. And the energy densities and temperatures of FIR and OPT are $T_{\rm FIR} = 15.0$ K and $U_{\rm FIR} = 1.0$ eV cm$^{-3}$, and $T_{\rm OPT} = 5000.0$ K and $U_{\rm OPT} = 0.3$ eV cm$^{-3}$. The radio band data from \cite{2014ApJ...788..155G}, the X-ray band data from \cite{2014ApJ...788..155G}, the GeV band data from \cite{2018ApJ...867...55X}, the TeV $\gamma$-ray band data from \cite{2014MNRAS.439.2828A}.}
\end{figure}

\begin{figure}
\includegraphics[width=7.0cm, angle=0]{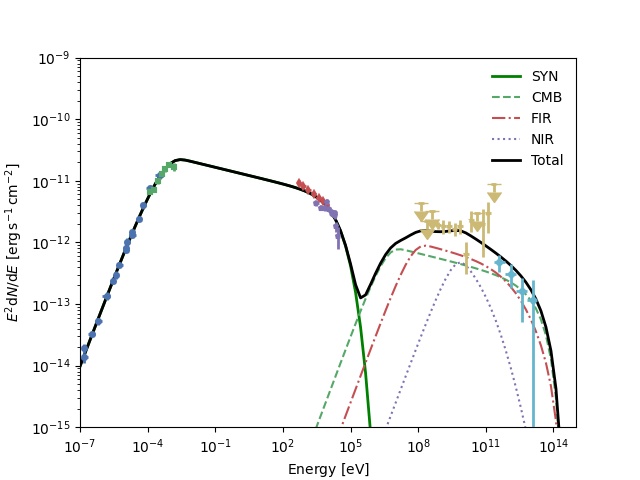}
\includegraphics[width=6.0cm, angle=0]{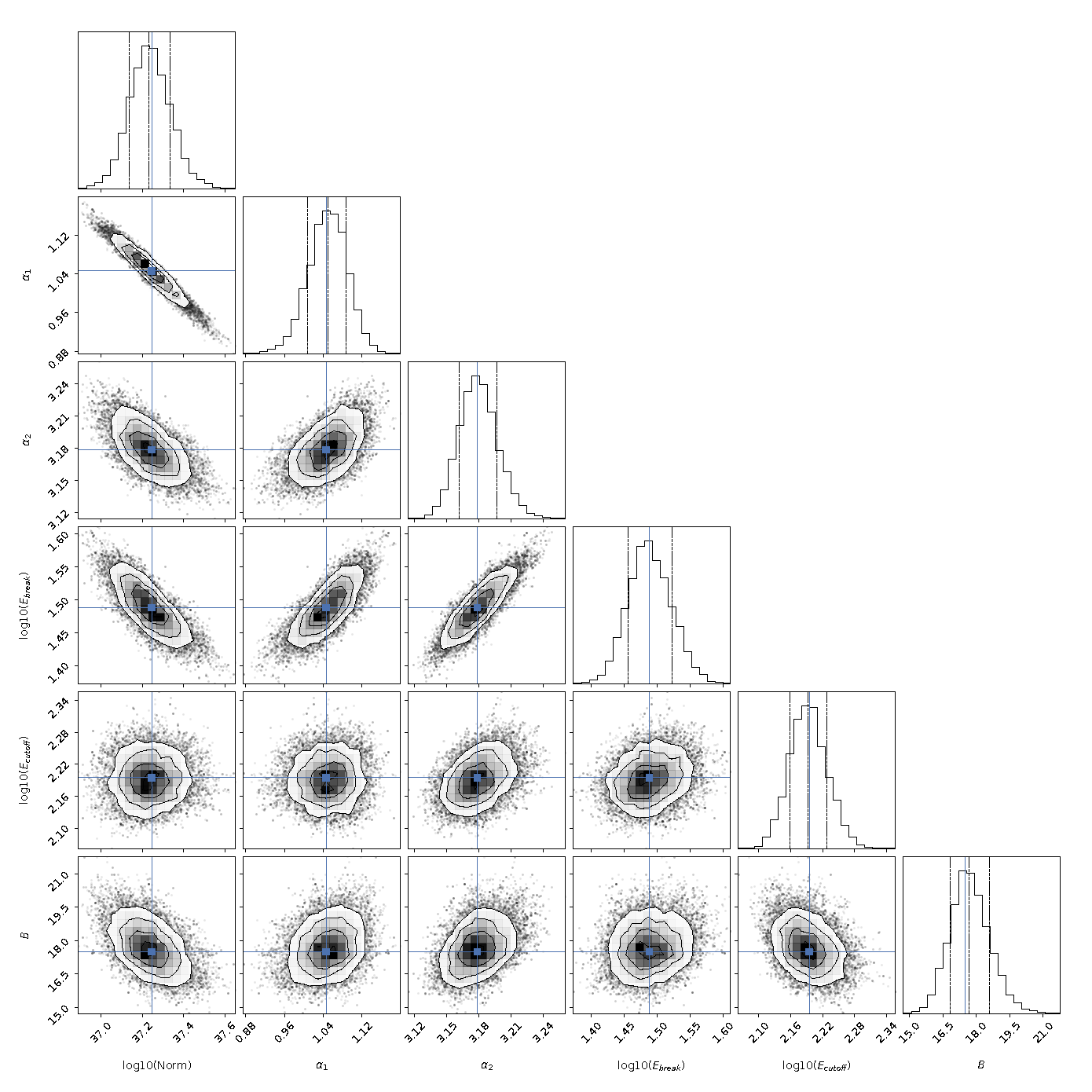}
\caption{The same as Figure \ref{fig:G21.5-0.9} but for 3C 58. The IR background with $T_{\rm FIR} = 40.0$ K and $U_{\rm FIR} = 0.3$ eV cm$^{-3}$ and $T_{\rm NIR} = 4000.0$ K and $U_{\rm NIR} = 0.3$ eV cm$^{-3}$, are used. The radio band data from \cite{1986MNRAS.218..533G,1987A&AS...69..533M,1989ApJ...338..171S}, the Planck data from \cite{2016A&A...586A.134P}, the X-ray band data from \cite{2000PASJ...52..875T,2019ApJ...876..150A}, the GeV band data from \cite{2018ApJ...858...84L} and the TeV $\gamma$-ray band data from \cite{2014A&A...567L...8A}.}
\end{figure}

\begin{figure}
\includegraphics[width=7.0cm, angle=0]{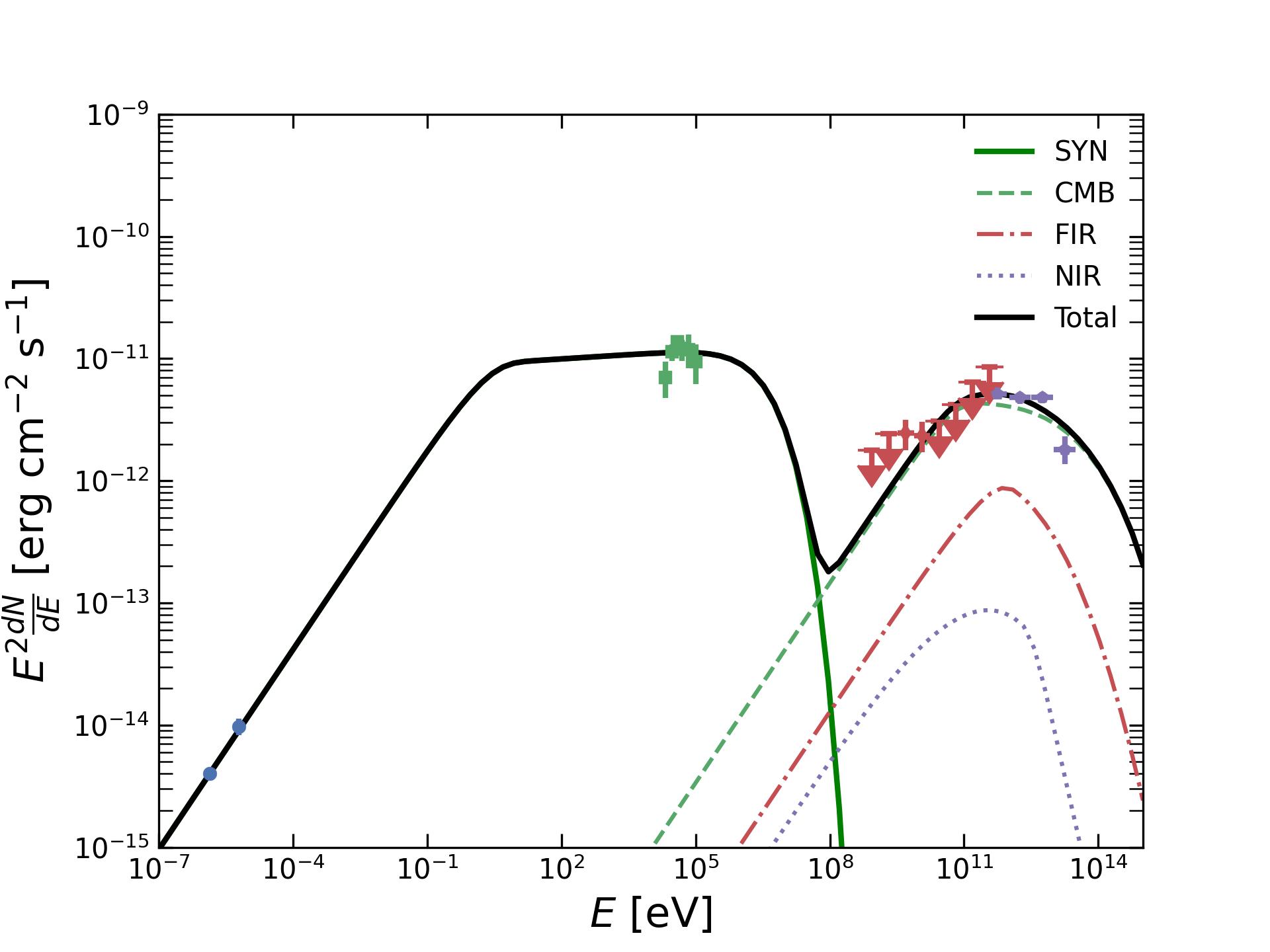}
\includegraphics[width=6.0cm, angle=0]{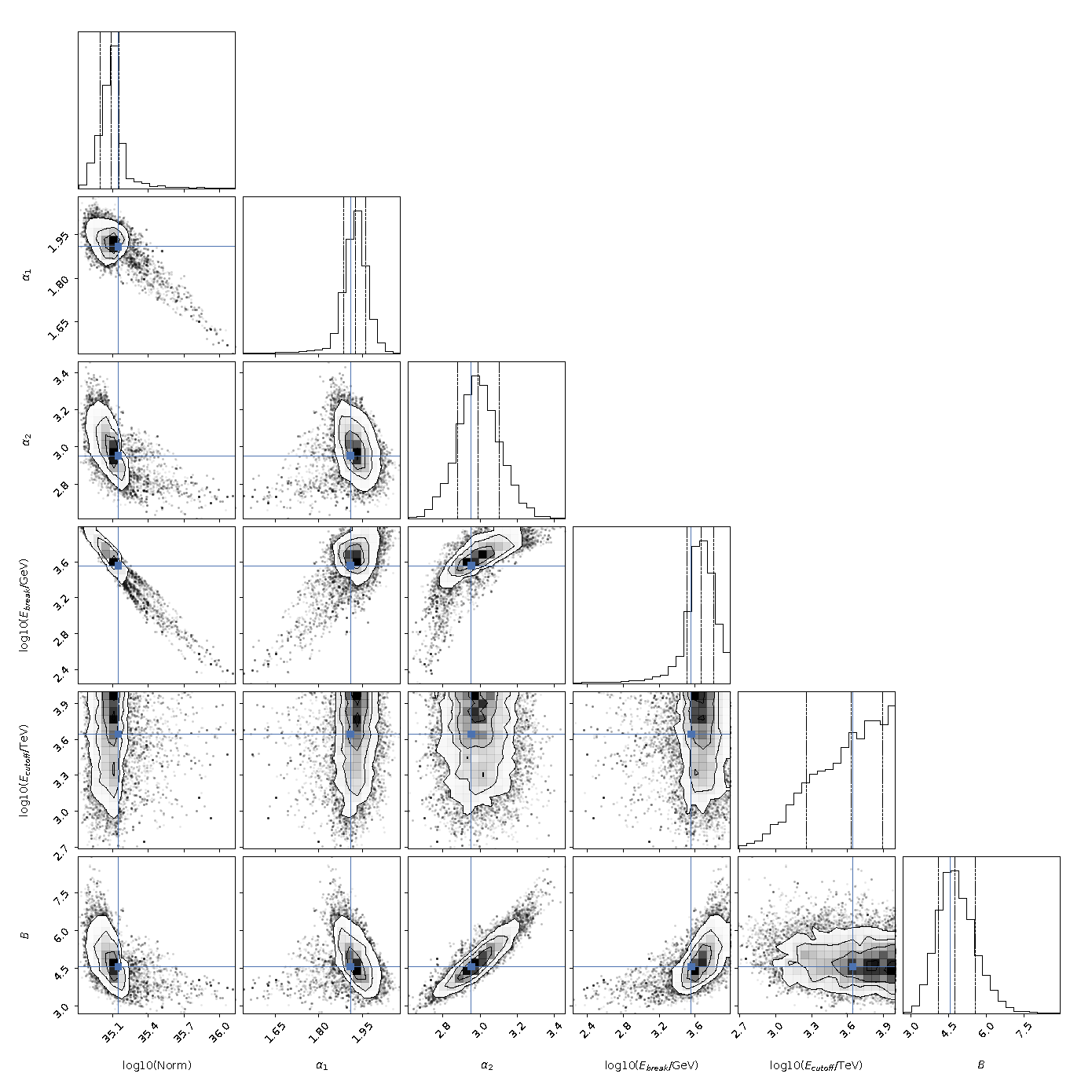}
\caption{The same as Figure \ref{fig:G21.5-0.9} but for HESS J1813$-$178. And the energy densities and temperatures of FIR and NIR are $T_{\rm FIR} = 40.0$ K and $U_{\rm FIR} = 0.1$ eV cm$^{-3}$, and $T_{\rm NIR} = 4000.0$ K and $U_{\rm NIR} = 0.5$ eV cm$^{-3}$. The radio band data from \cite{2005ApJ...629L.105B}, the X-ray band data from \cite{2007A&A...470..249F,2005ApJ...629L.109U}, the GeV band data from \cite{2023arXiv230816717W}, the TeV $\gamma$-ray band data from \cite{2023arXiv230816717W}.}
\end{figure}

\begin{figure}
\includegraphics[width=7.0cm, angle=0]{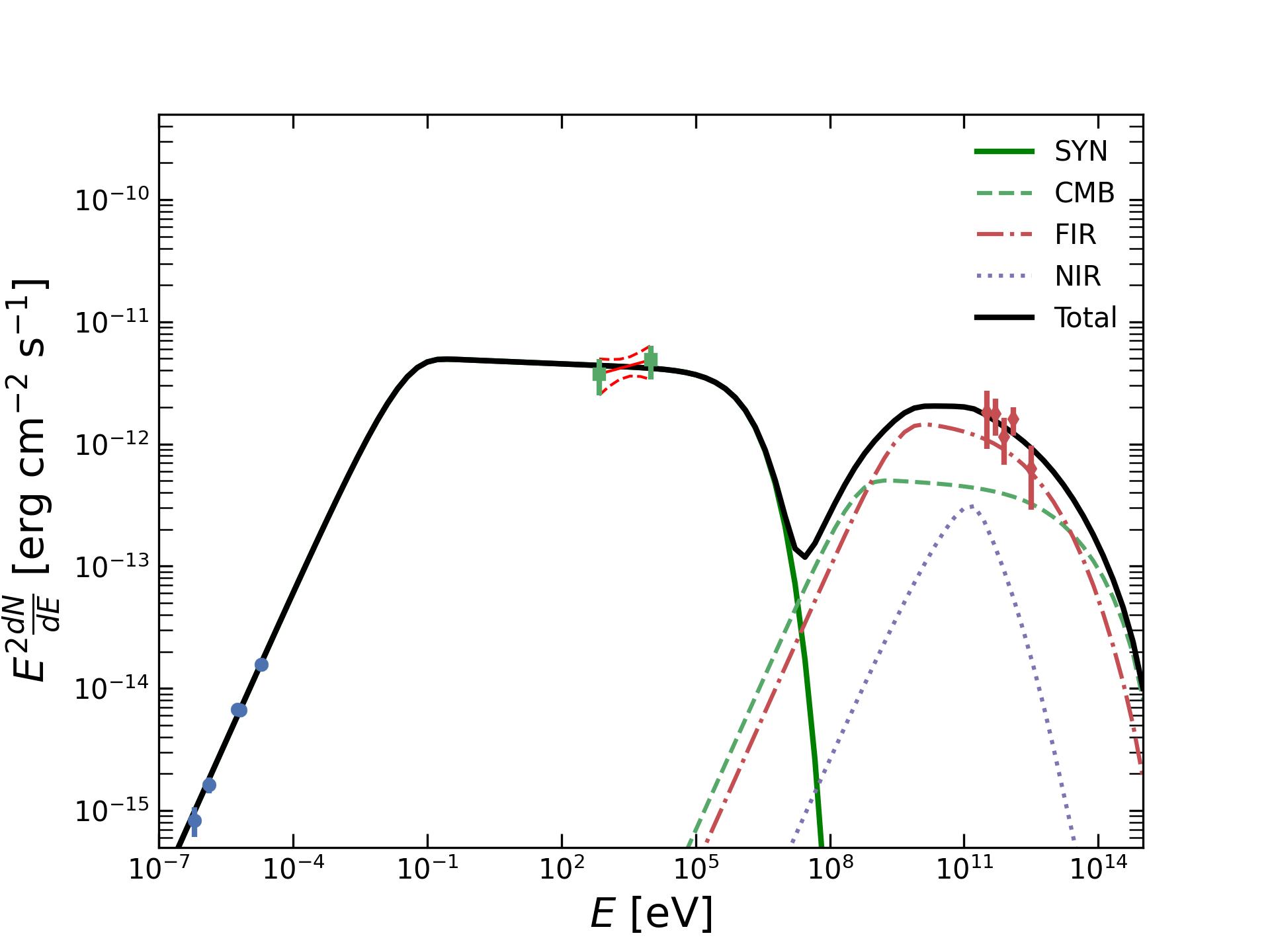}
\includegraphics[width=6.0cm, angle=0]{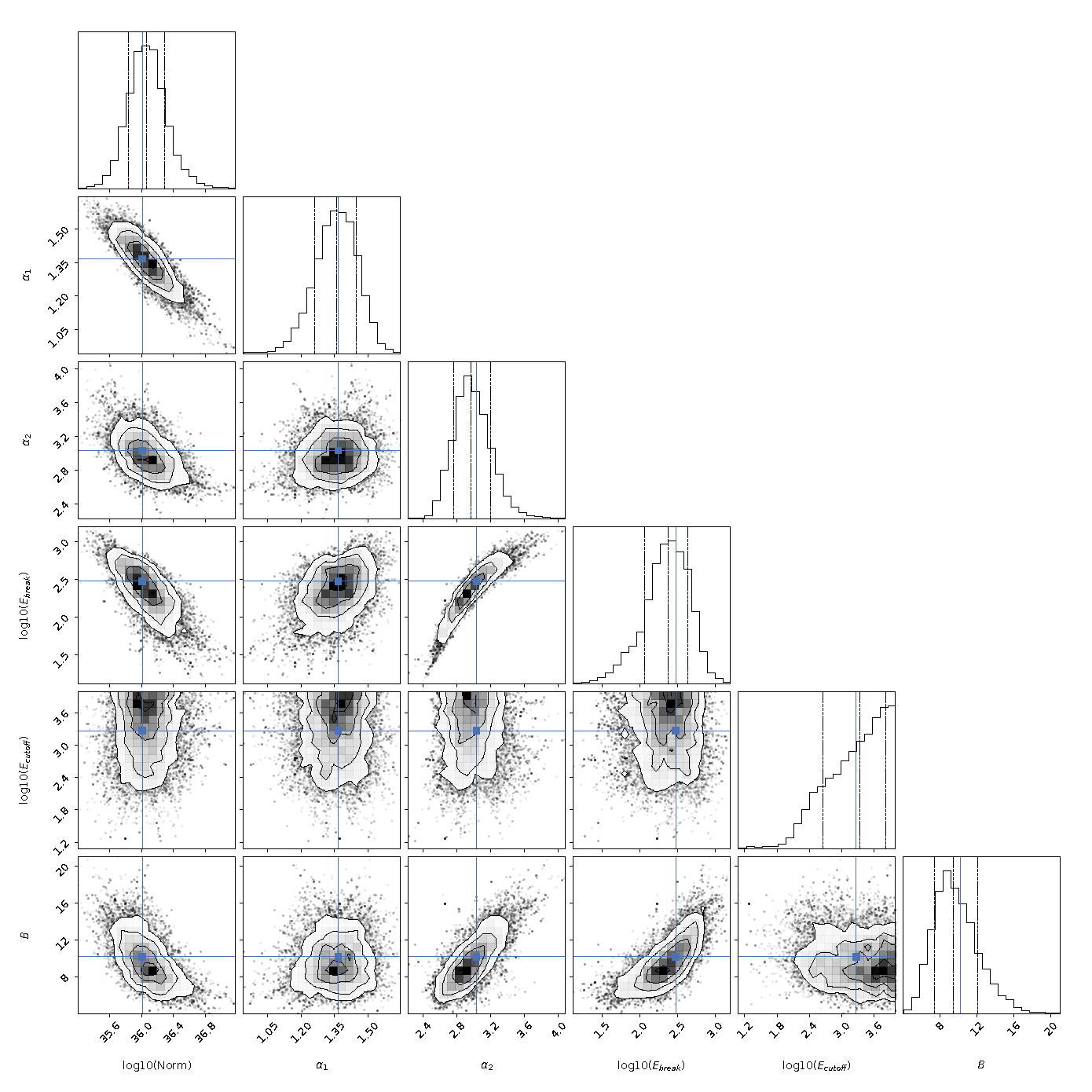}
\caption{The same as Figure \ref{fig:G21.5-0.9} but for G54.1$+$0.3. And the energy densities and temperatures of FIR and NIR are $T_{\rm FIR} = 25.0$ K and $U_{\rm FIR} = 0.8$ eV cm$^{-3}$, and $T_{\rm NIR} = 3000.0$ K and $U_{\rm NIR} = 1.1$ eV cm$^{-3}$. The radio band data from \cite{2010ApJ...709.1125L}, the X-ray band data from \cite{2001A&A...370..570L}, the TeV $\gamma$-ray band data from \cite{2010ApJ...719L..69A}.}
\end{figure}

\begin{figure}
\includegraphics[width=7.0cm, angle=0]{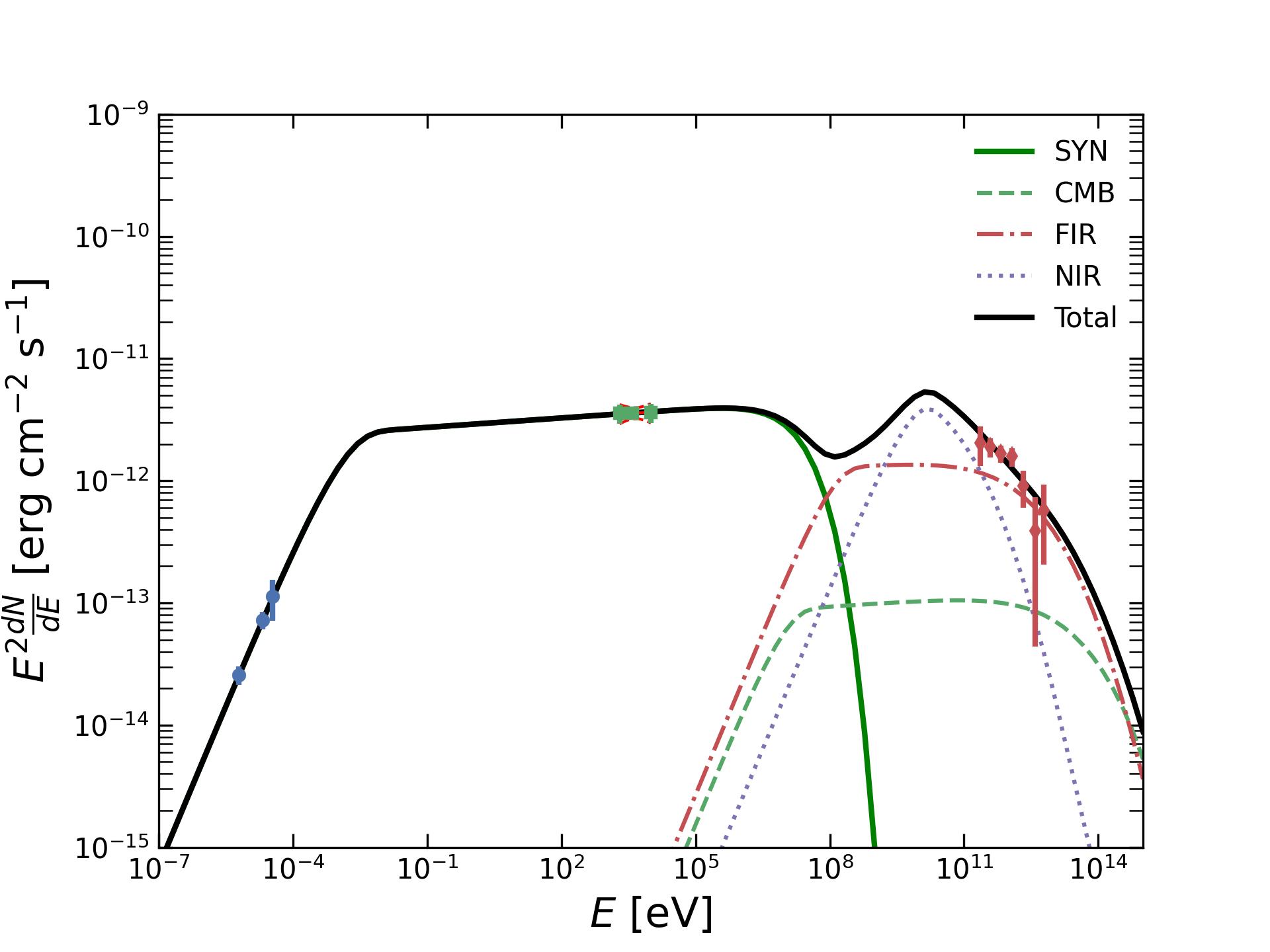}
\includegraphics[width=6.0cm, angle=0]{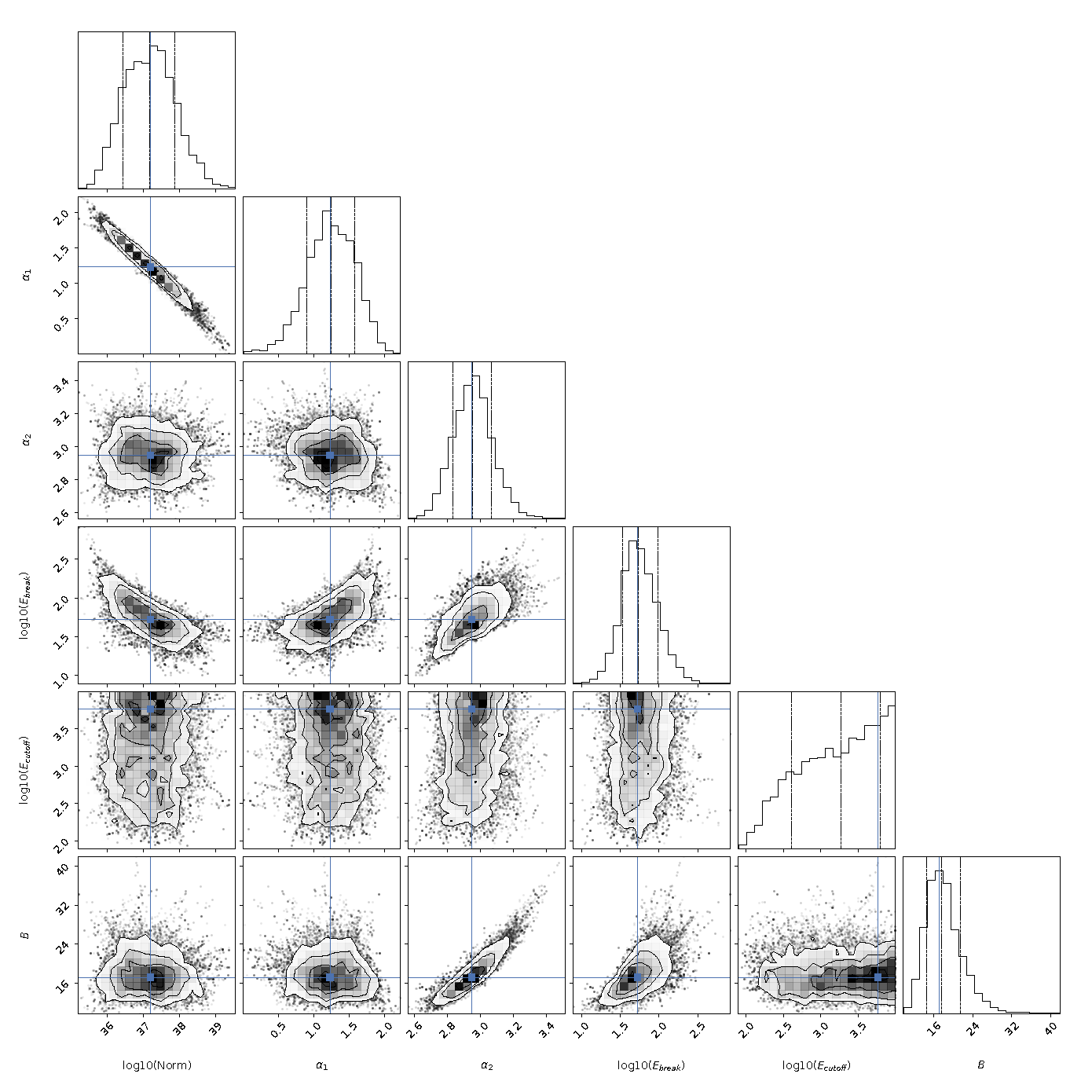}
\caption{The same as Figure \ref{fig:G21.5-0.9} but for G0.9$+$0.1. And the energy densities and temperatures of FIR and NIR are $T_{\rm FIR} = 30.0$ K and $U_{\rm FIR} = 3.8$ eV cm$^{-3}$, and $T_{\rm NIR} = 3000.0$ K and $U_{\rm NIR} = 25.0$ eV cm$^{-3}$. The radio band data from \cite{2008A&A...487.1033D}, the X-ray band data from \cite{2003A&A...401..197P}, the TeV $\gamma$-ray band data from \cite{2005A&A...432L..25A}.}
\end{figure}

\begin{figure}
\includegraphics[width=7.0cm, angle=0]{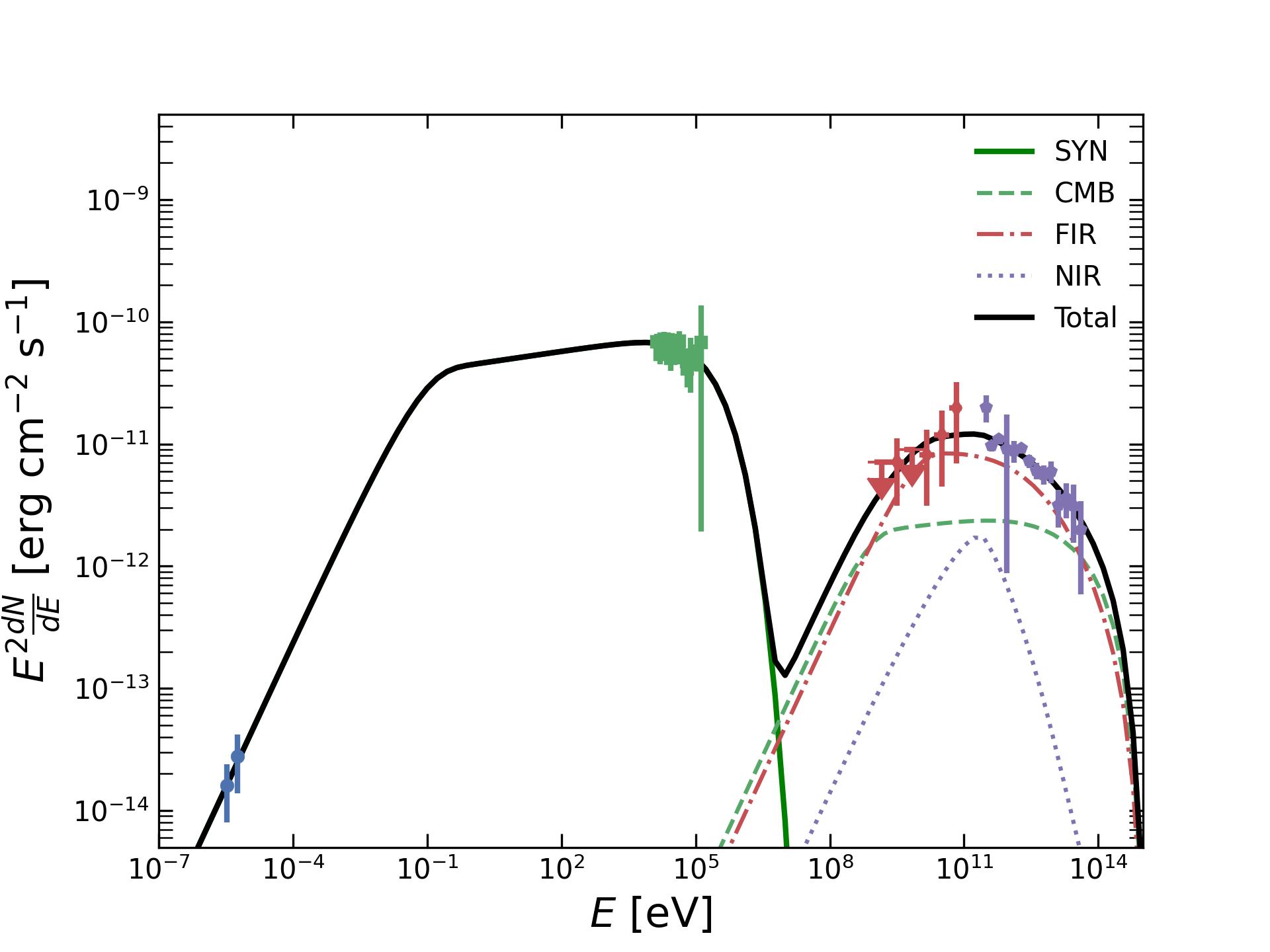}
\includegraphics[width=6.0cm, angle=0]{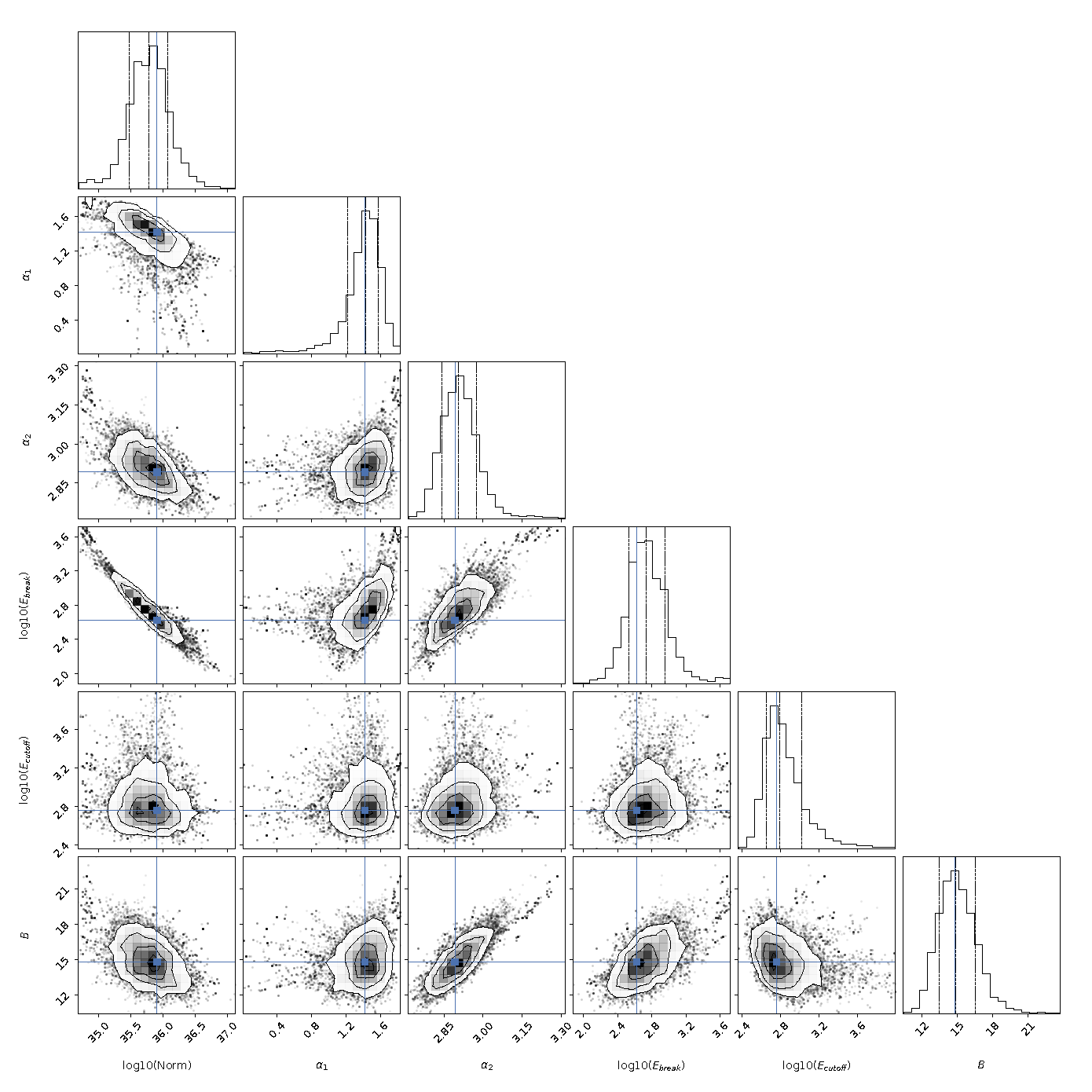}
\caption{The same as Figure \ref{fig:G21.5-0.9} but for MSH 15$-$52. And the energy densities and temperatures of FIR and NIR are $T_{\rm FIR} = 30.0$ K and $U_{\rm FIR} = 1.2$ eV cm$^{-3}$, and $T_{\rm NIR} = 3000.0$ K and $U_{\rm NIR} = 2.2$ eV cm$^{-3}$. The radio band data from \cite{1999MNRAS.305..724G,2002ApJ...569..878G}, the X-ray band data from \cite{2006ApJ...651L..45F}, the GeV band data from \cite{2010ApJ...714..927A}, the TeV $\gamma$-ray band data from \cite{2005A&A...435L..17A}.}
\end{figure}

\begin{figure}
\includegraphics[width=7.0cm, angle=0]{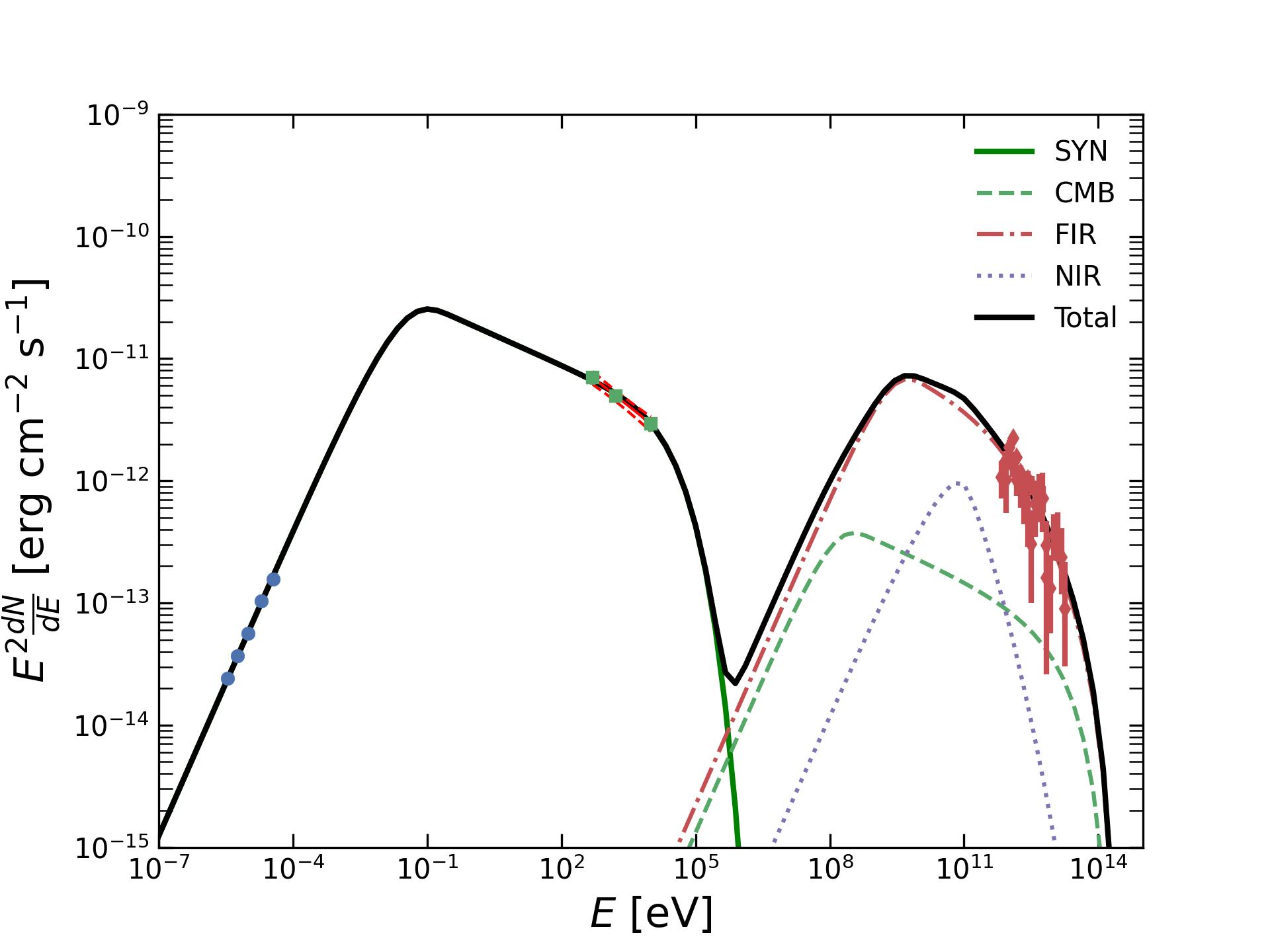}
\includegraphics[width=6.0cm, angle=0]{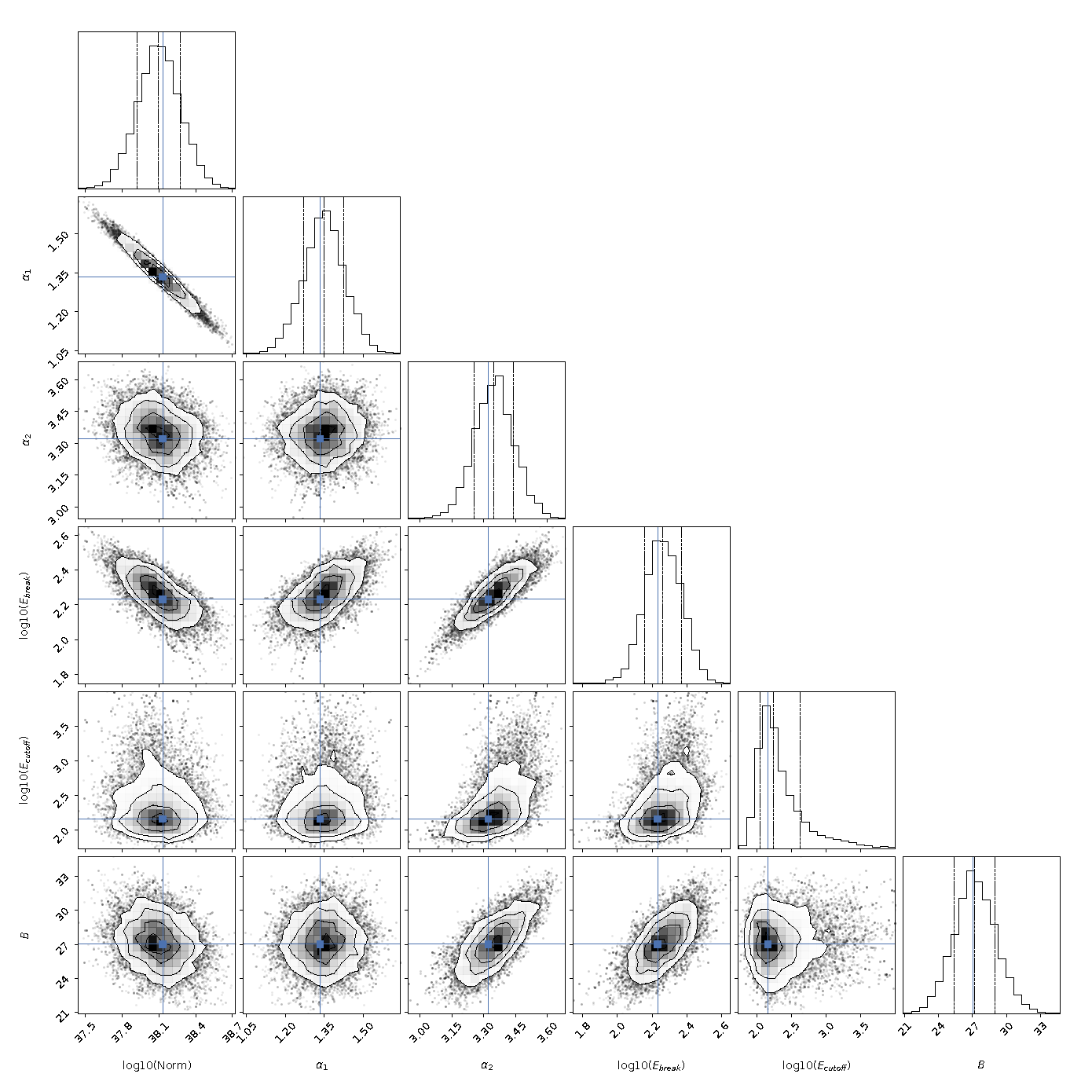}
\caption{The same as Figure \ref{fig:G21.5-0.9} but for N 157B. And the energy densities and temperatures of FIR and NIR are $T_{\rm FIR} = 50.0$ K and $U_{\rm FIR} = 5.0$ eV cm$^{-3}$, and $T_{\rm NIR} = 5000.0$ K and $U_{\rm NIR} = 4.0$ eV cm$^{-3}$. The radio band data from \cite{2000ApJ...540..808L}, the X-ray band data from \cite{2006ApJ...651..237C}, the TeV $\gamma$-ray band data from \cite{2015Sci...347..406H}.}
\end{figure}

\begin{figure}
\includegraphics[width=7.0cm, angle=0]{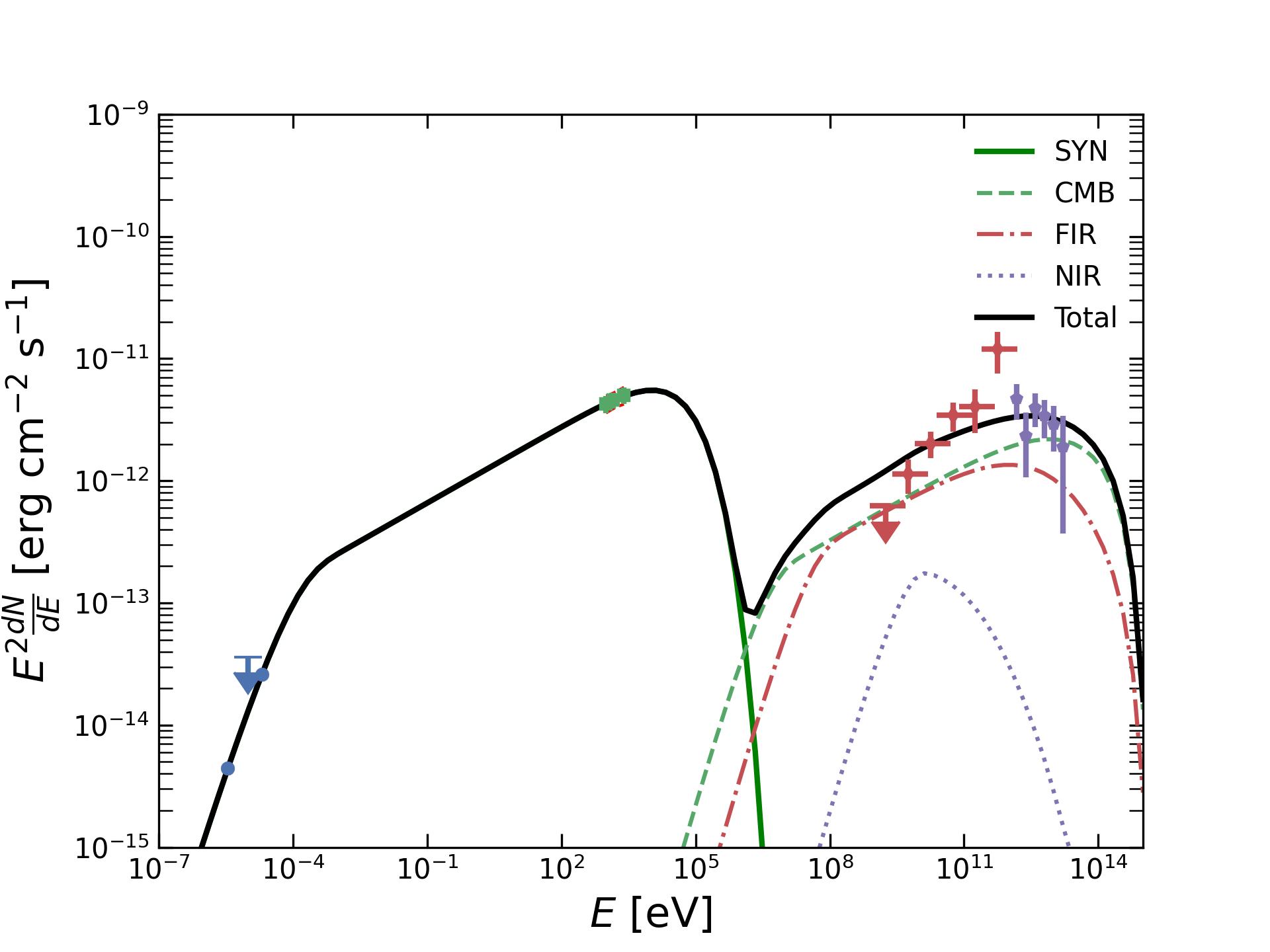}
\includegraphics[width=6.0cm, angle=0]{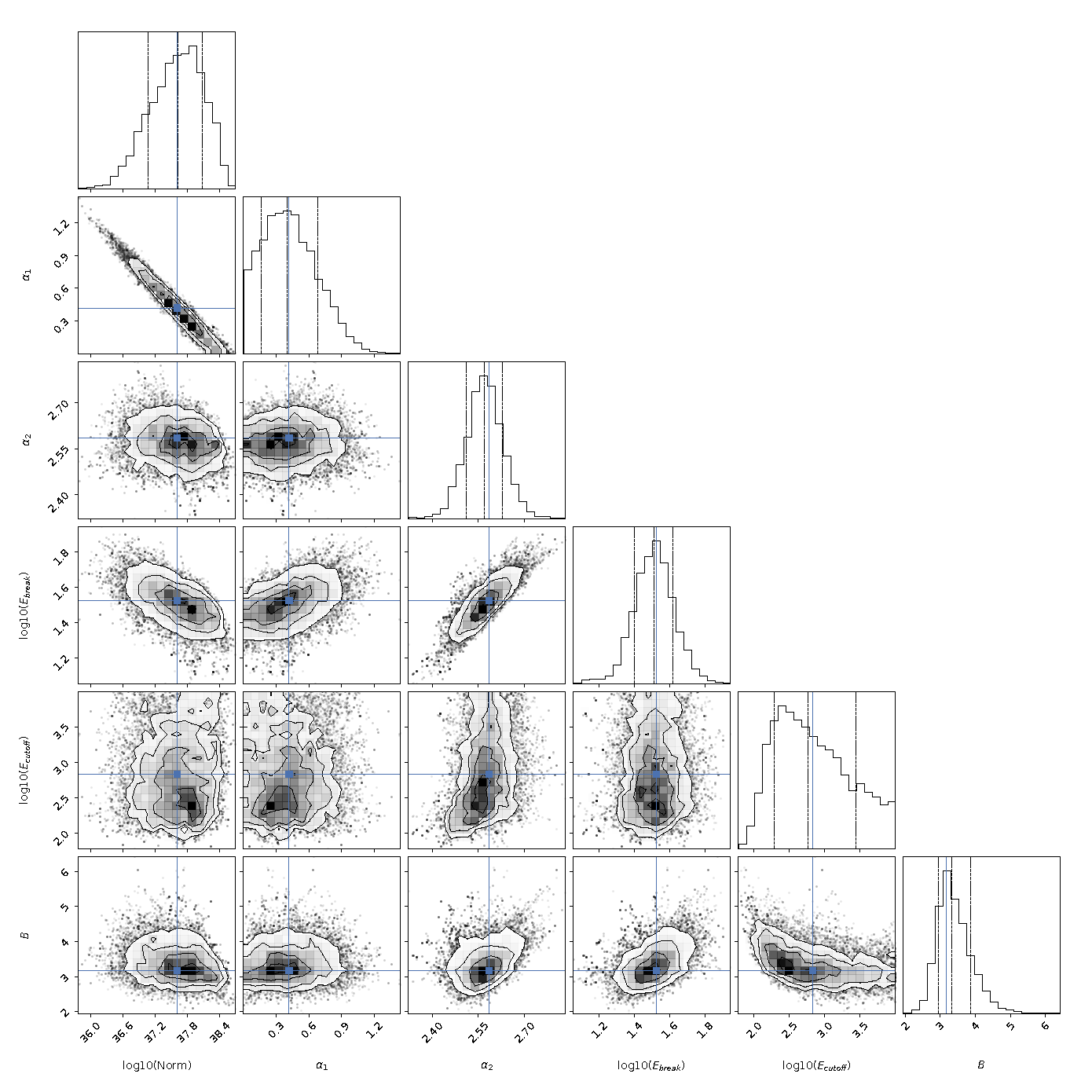}
\caption{The same as Figure \ref{fig:G21.5-0.9} but for HESS J1356$-$645. And the energy densities and temperatures of FIR and NIR are $T_{\rm FIR} = 25.0$ K and $U_{\rm FIR} = 0.4$ eV cm$^{-3}$, and $T_{\rm NIR} = 5000.0$ K and $U_{\rm NIR} = 0.5$ eV cm$^{-3}$. The radio band data from \cite{1995MNRAS.277...36D, 1993AJ....105.1666G, 2007MNRAS.382..382M}, the X-ray band data from \cite{2011A&A...533A.103H}, the GeV band data from \cite{2023ApJ...942..105L}, the TeV $\gamma$-ray band data from \cite{2011A&A...533A.103H}.}
\end{figure}

\begin{figure}
\includegraphics[width=7.0cm, angle=0]{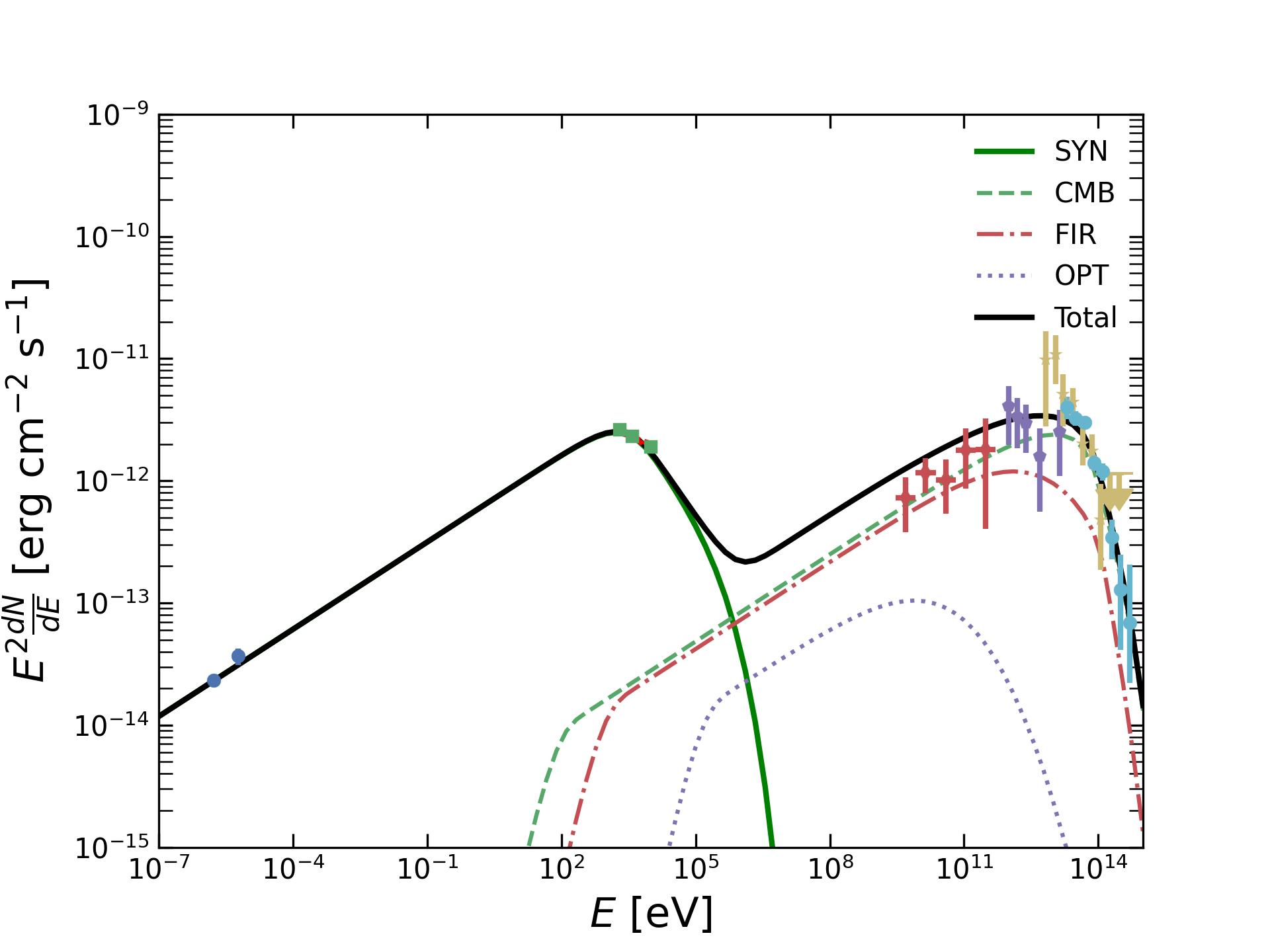}
\includegraphics[width=6.0cm, angle=0]{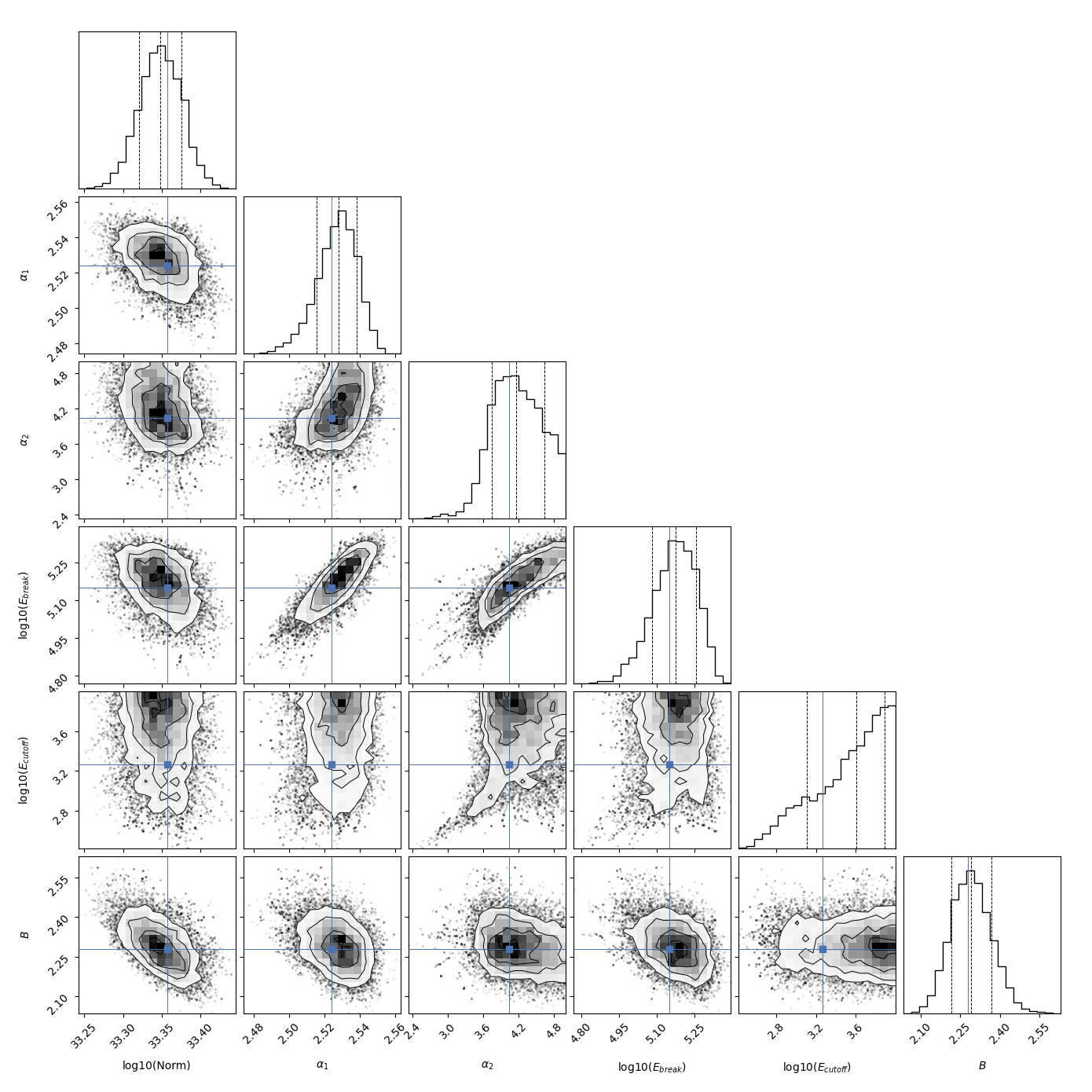}
\caption{The same as Figure \ref{fig:G21.5-0.9} but for VER J2227$+$608. And the energy densities and temperatures of FIR and NIR are $T_{\rm FIR} = 30.0$ K and $U_{\rm FIR} = 0.4$ eV cm$^{-3}$, and $T_{\rm NIR} = 5000.0$ K and $U_{\rm NIR} = 0.4$ eV cm$^{-3}$. The radio band data from \cite{2000AJ....120.3218P}, the X-ray band data from \cite{2021ApJ...912..133F}, the GeV band data from \cite{2019ApJ...885..162X}, the TeV $\gamma$-ray band data from \cite{2009ApJ...703L...6A,2021NatAs...5..460T,2021Natur.594...33C}.}
\end{figure}

\begin{figure}
\includegraphics[width=7.0cm, angle=0]{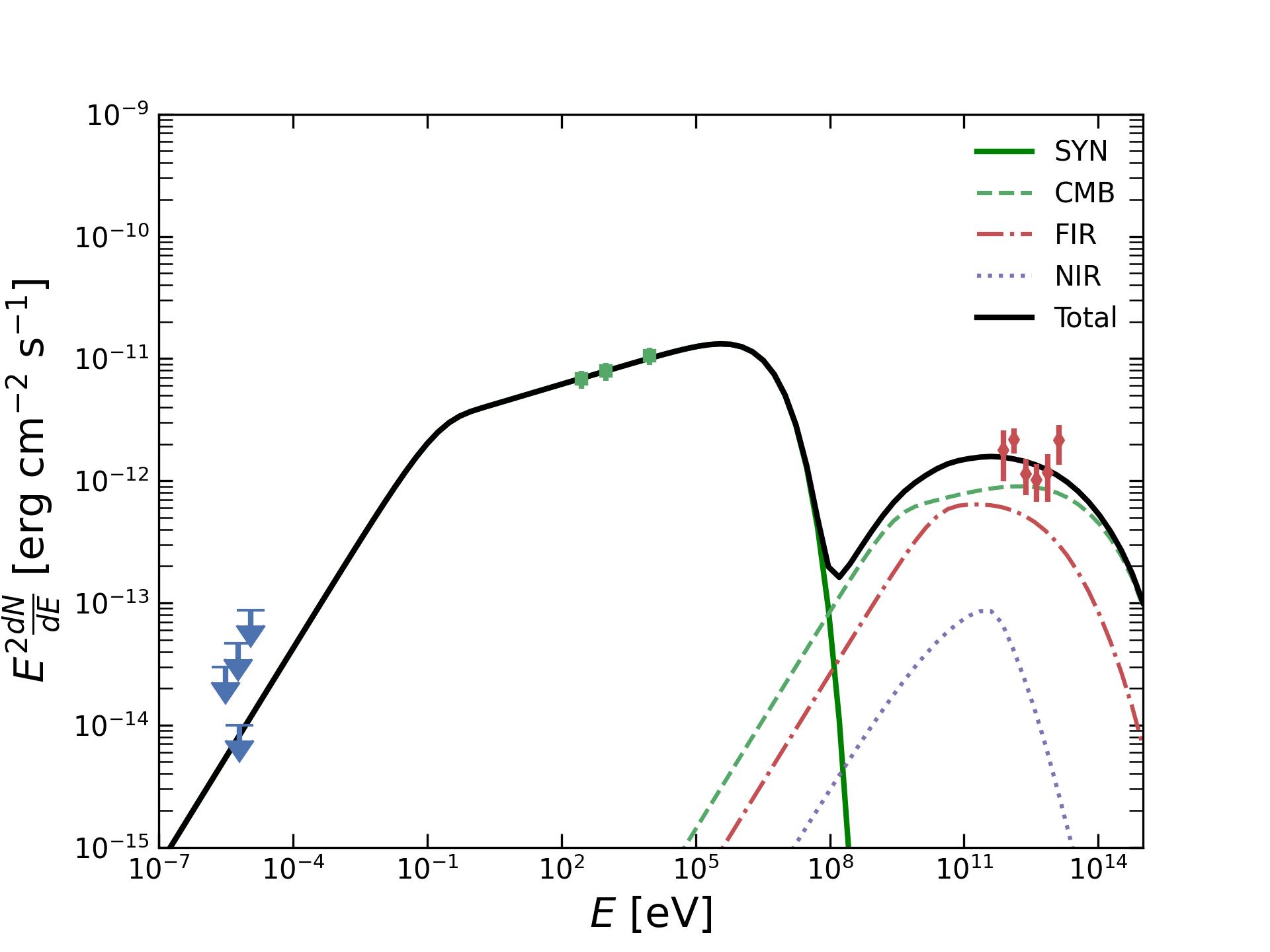}
\includegraphics[width=6.0cm, angle=0]{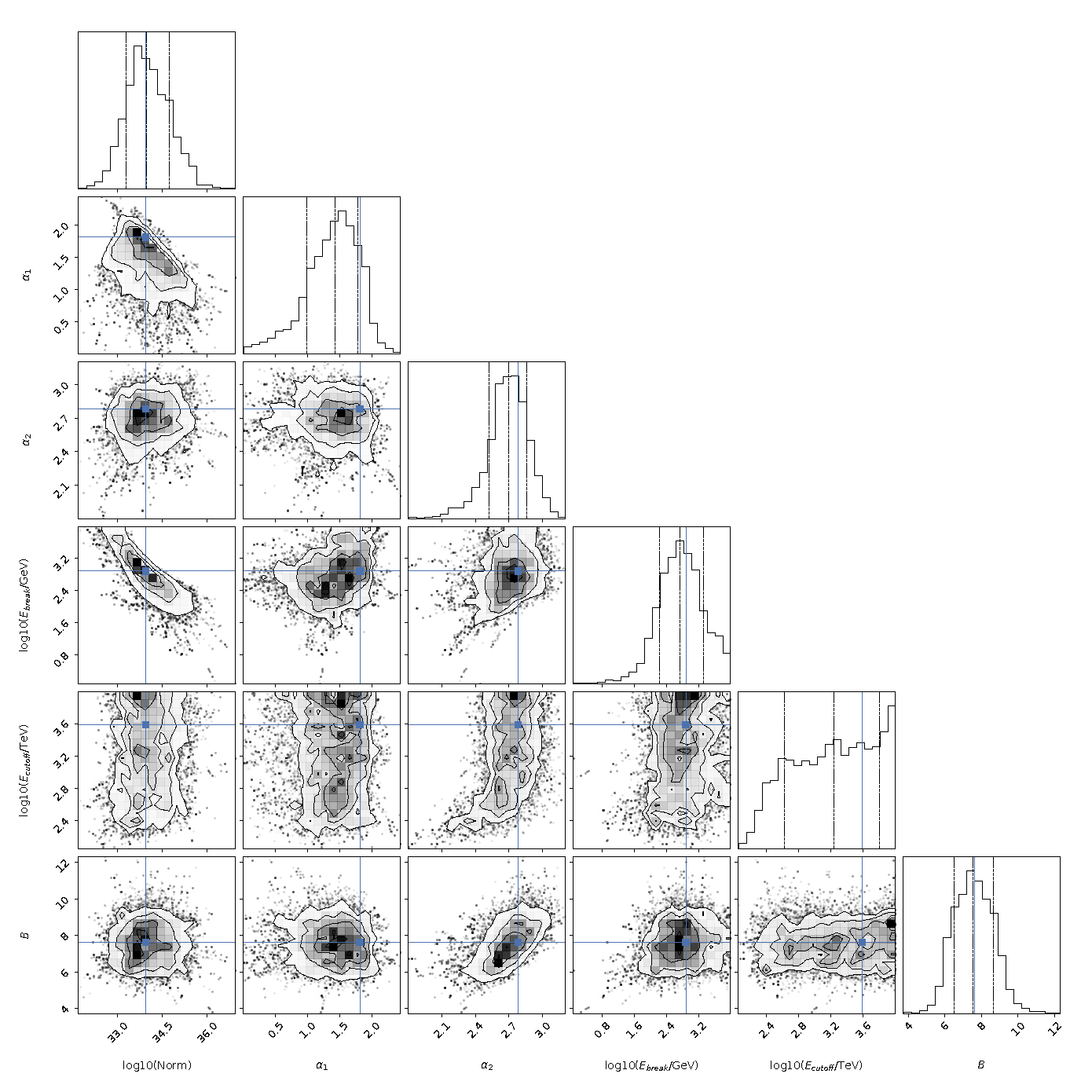}
\caption{The same as Figure \ref{fig:G21.5-0.9} but for CTA 1. And the energy densities and temperatures of FIR and NIR are $T_{\rm FIR} = 25$ K and $U_{\rm FIR} = 0.3$ eV cm$^{-3}$, and $T_{\rm NIR} = 3000$ K and $U_{\rm NIR} = 0.6$ eV cm$^{-3}$. The radio band data from \cite{2013ApJ...764...38A}, the X-ray band data from \cite{2016MNRAS.459.3868M}, the TeV $\gamma$-ray band data from \cite{2013ApJ...764...38A}.}
\end{figure}

\begin{figure}
\includegraphics[width=7.0cm, angle=0]{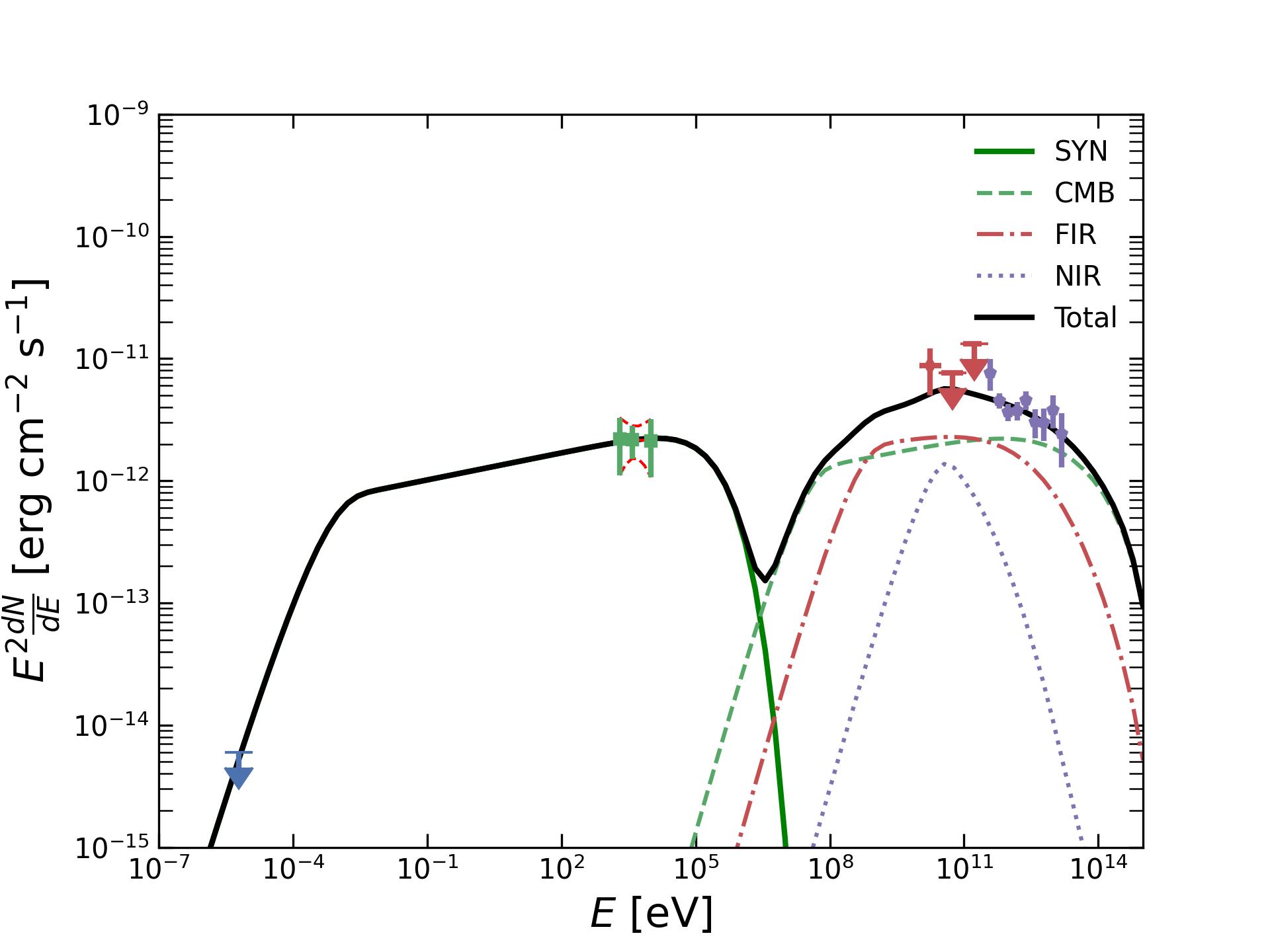}
\includegraphics[width=6.0cm, angle=0]{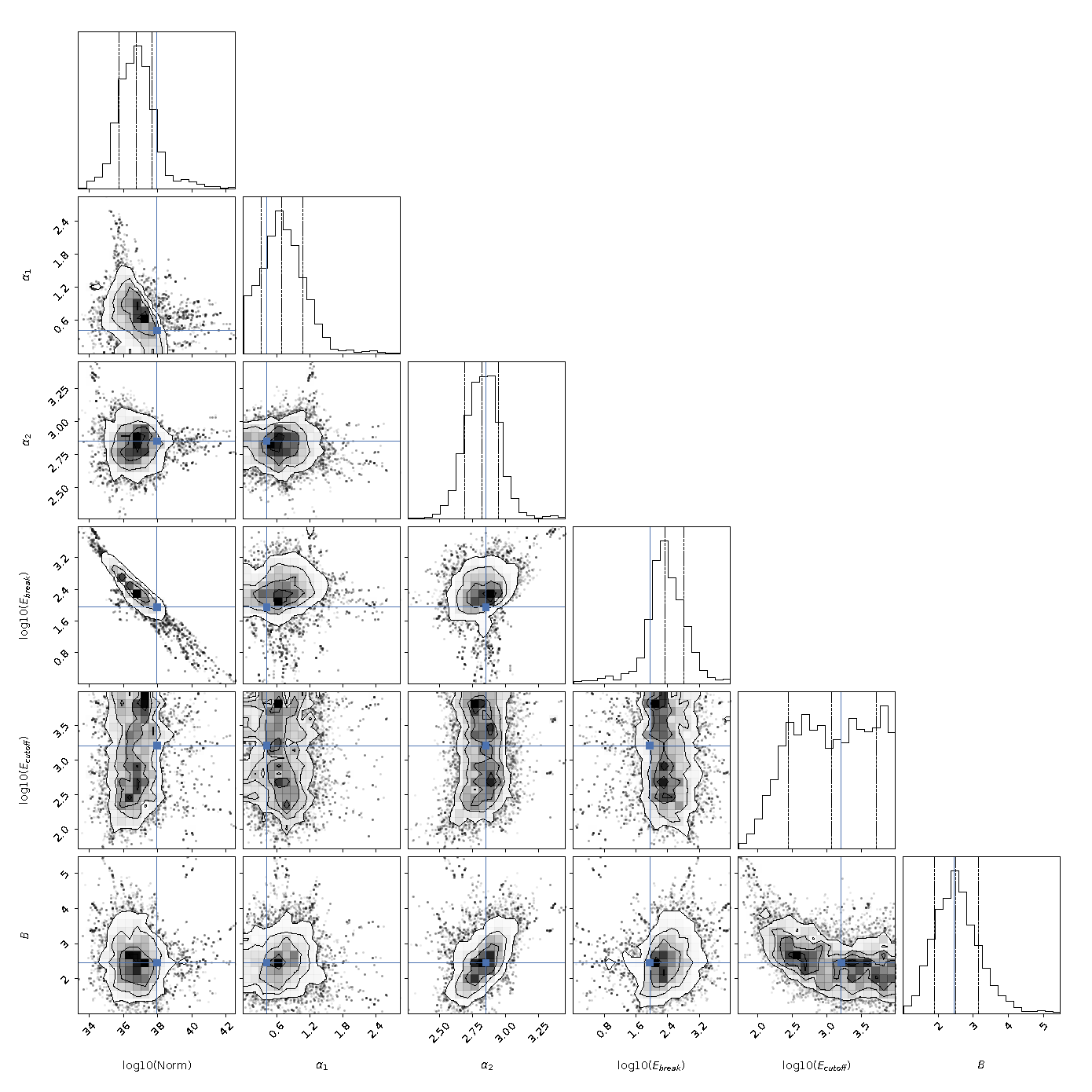}
\caption{The same as Figure \ref{fig:G21.5-0.9} but for HESS J1418$-$609. And the energy densities and temperatures of FIR and NIR are $T_{\rm FIR} = 40.0$ K and $U_{\rm FIR} = 0.4$ eV cm$^{-3}$, and $T_{\rm NIR} = 4000.0$ K and $U_{\rm NIR} = 1.0$ eV cm$^{-3}$. The radio band from \cite{1999ApJ...515..712R}, the X-ray band from \cite{2023ApJ...945...66P}, the GeV band from \cite{2013ApJ...773...77A}, the TeV $\gamma$-ray band from \cite{2006A&A...456..245A}.}
\end{figure}

\begin{figure}
\includegraphics[width=7.0cm, angle=0]{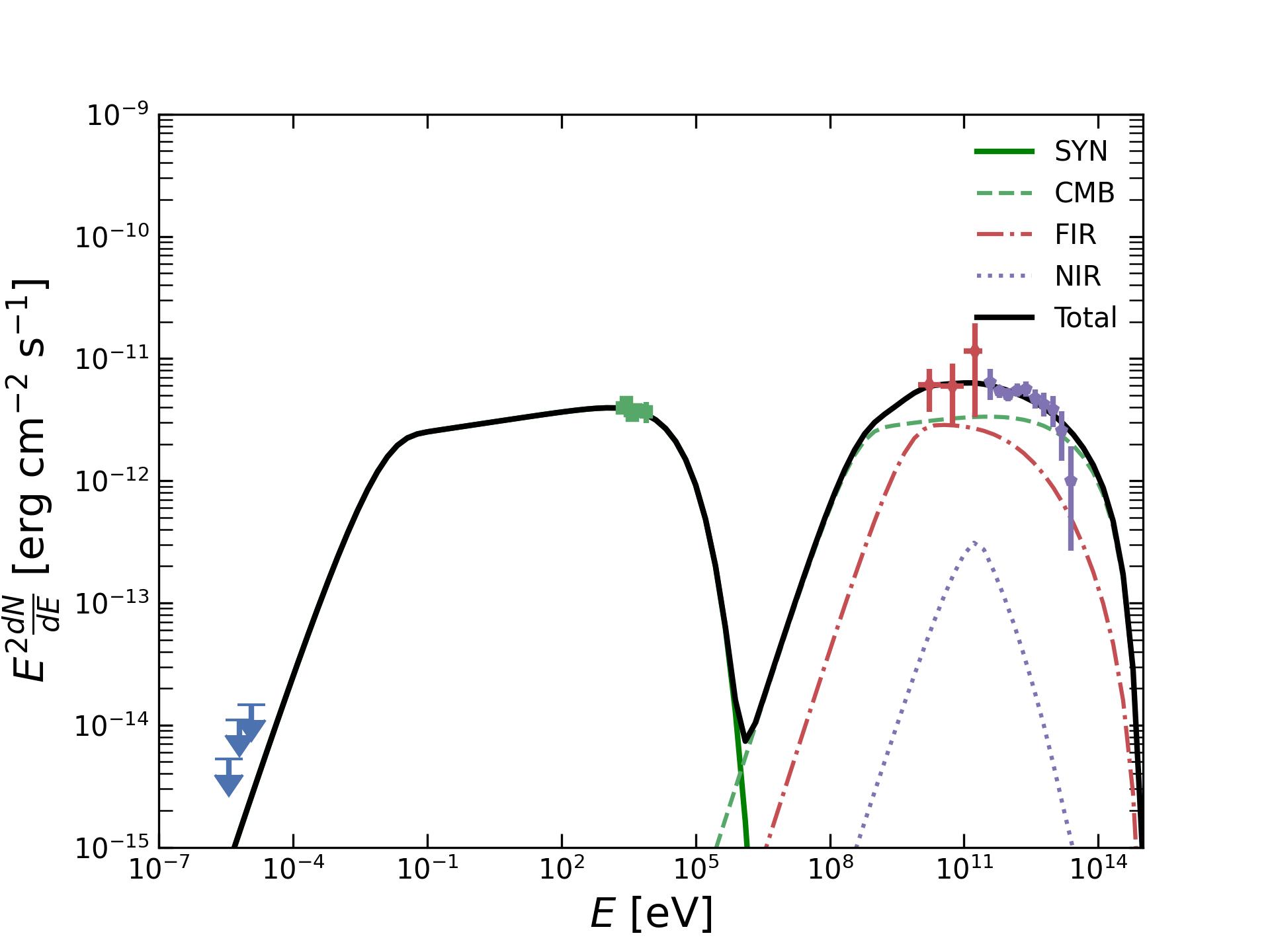}
\includegraphics[width=6.0cm, angle=0]{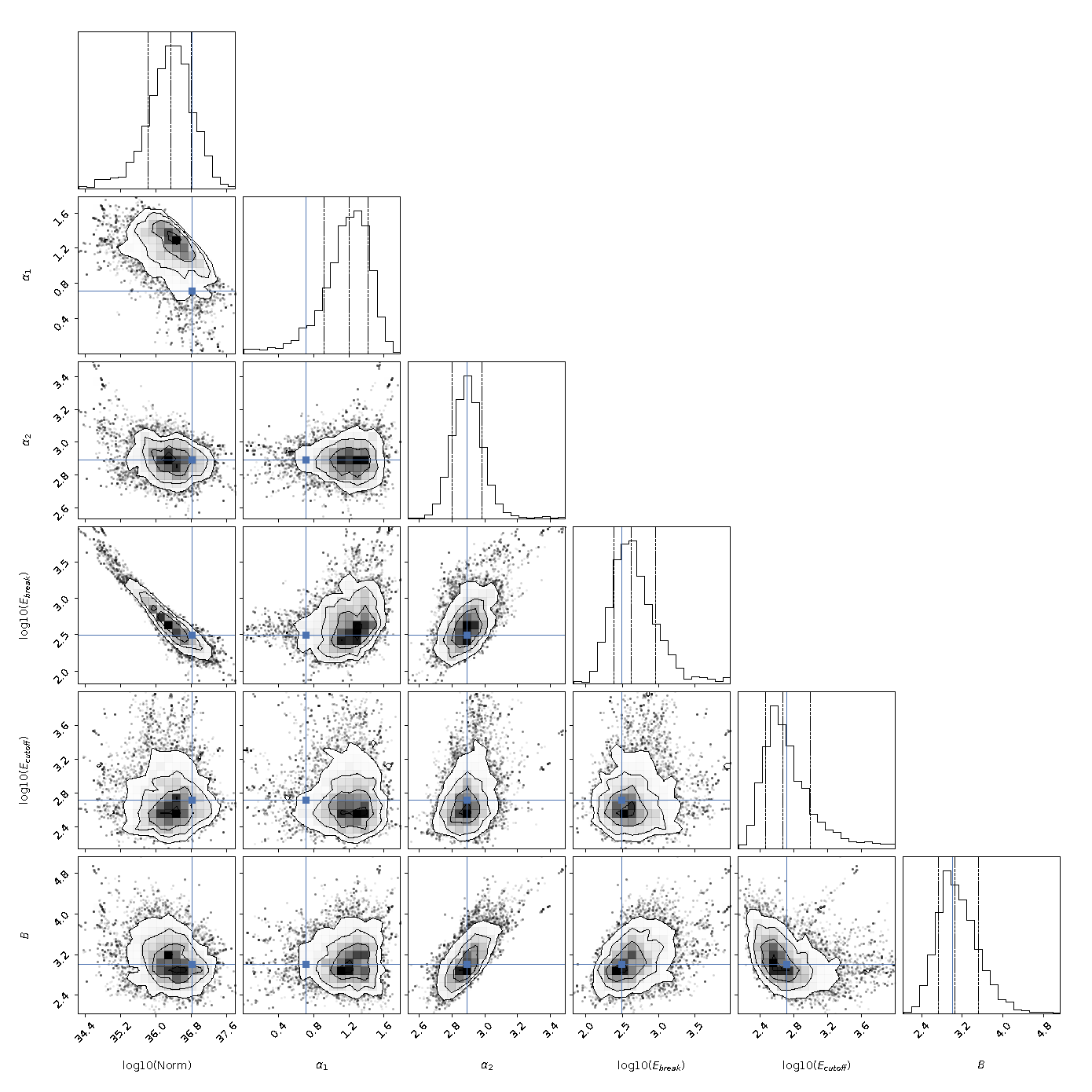}
\caption{The same as Figure \ref{fig:G21.5-0.9} but for HESS J1420-607. And the energy densities and temperatures of FIR and NIR are $T_{\rm FIR} = 40.0$ K and $U_{\rm FIR} = 0.3$ eV cm$^{-3}$, and $T_{\rm NIR} = 4000.0$ K and $U_{\rm NIR} = 0.3$ eV cm$^{-3}$. The radio band from \cite{2010ApJ...711.1168V}, the X-ray band from \cite{2010ApJ...711.1168V}, the GeV $\gamma$-ray band from \cite{2013ApJ...773...77A}, the TeV $\gamma$-ray band from \cite{2006A&A...456..245A}.}
\end{figure}

\begin{figure}
\includegraphics[width=7.0cm, angle=0]{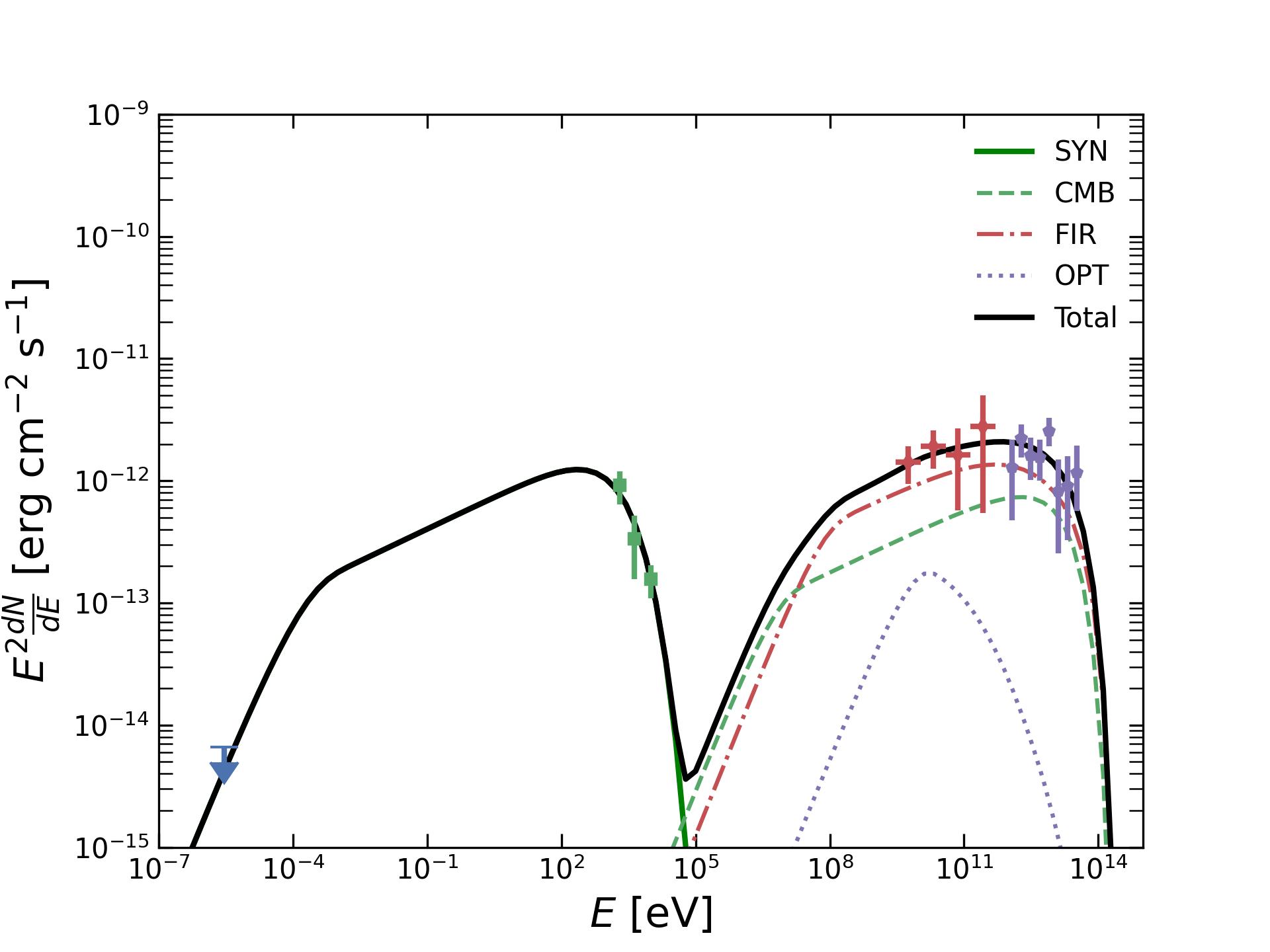}
\includegraphics[width=6.0cm, angle=0]{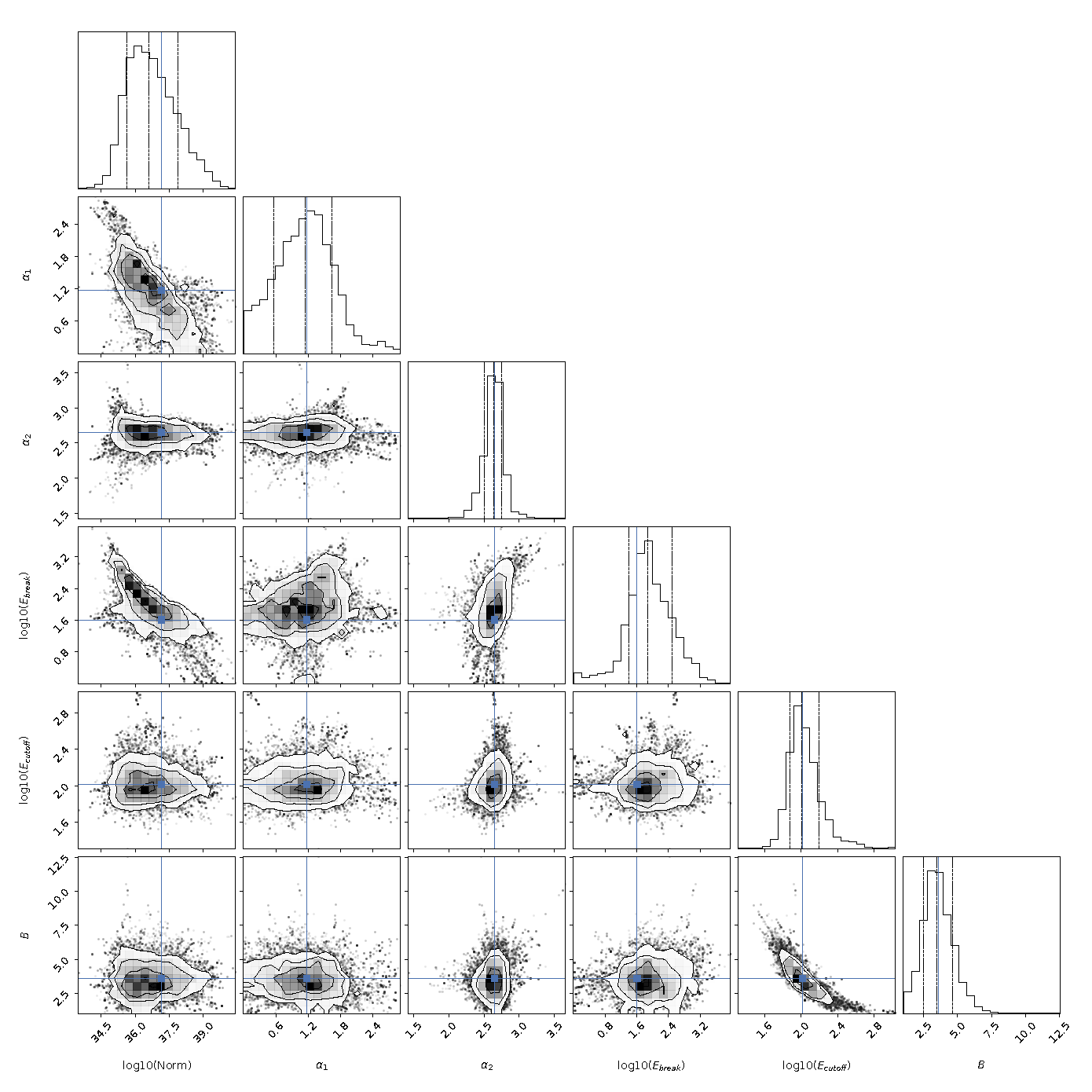}
\caption{The same as Figure \ref{fig:G21.5-0.9} but for HESS J1427-608. And the energy densities and temperatures of FIR and OPT are $T_{\rm FIR} = 30.0$ K and $U_{\rm FIR} = 1.0$ eV cm$^{-3}$, and $T_{\rm OPT} = 6000.0$ K and $U_{\rm OPT} = 1.0$ eV cm$^{-3}$. The radio band from \cite{10.1111/j.1365-2966.2007.12379.x}, the X-ray band from \cite{10.1093/pasj/65.3.61}, the GeV $\gamma$-ray band from \cite{2017ApJ...835...42G},the TeV $\gamma$-ray band from \cite{2008A&A...477..353A}.}
\end{figure}

\begin{figure}
\includegraphics[width=7.0cm, angle=0]{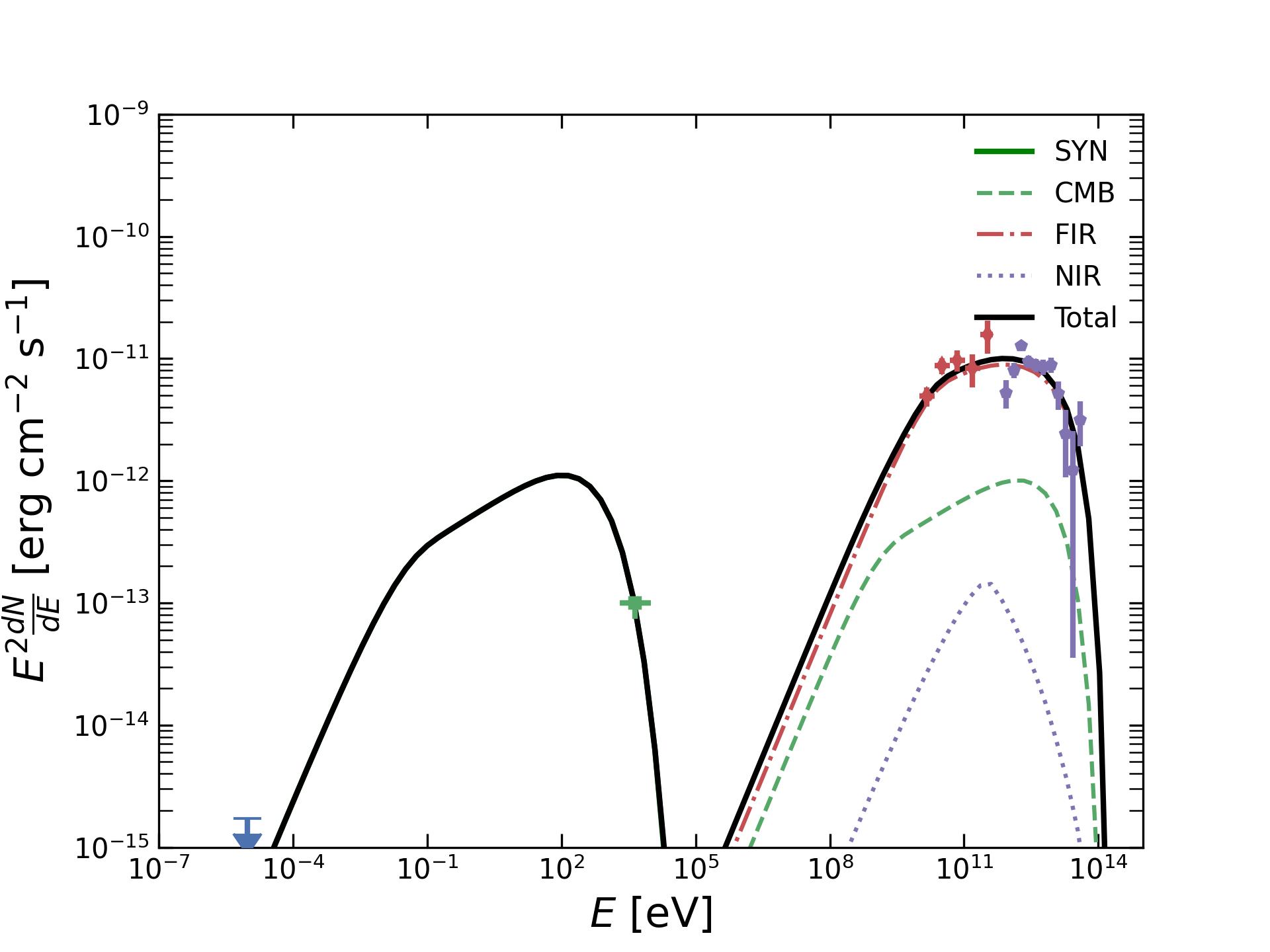}
\includegraphics[width=6.0cm, angle=0]{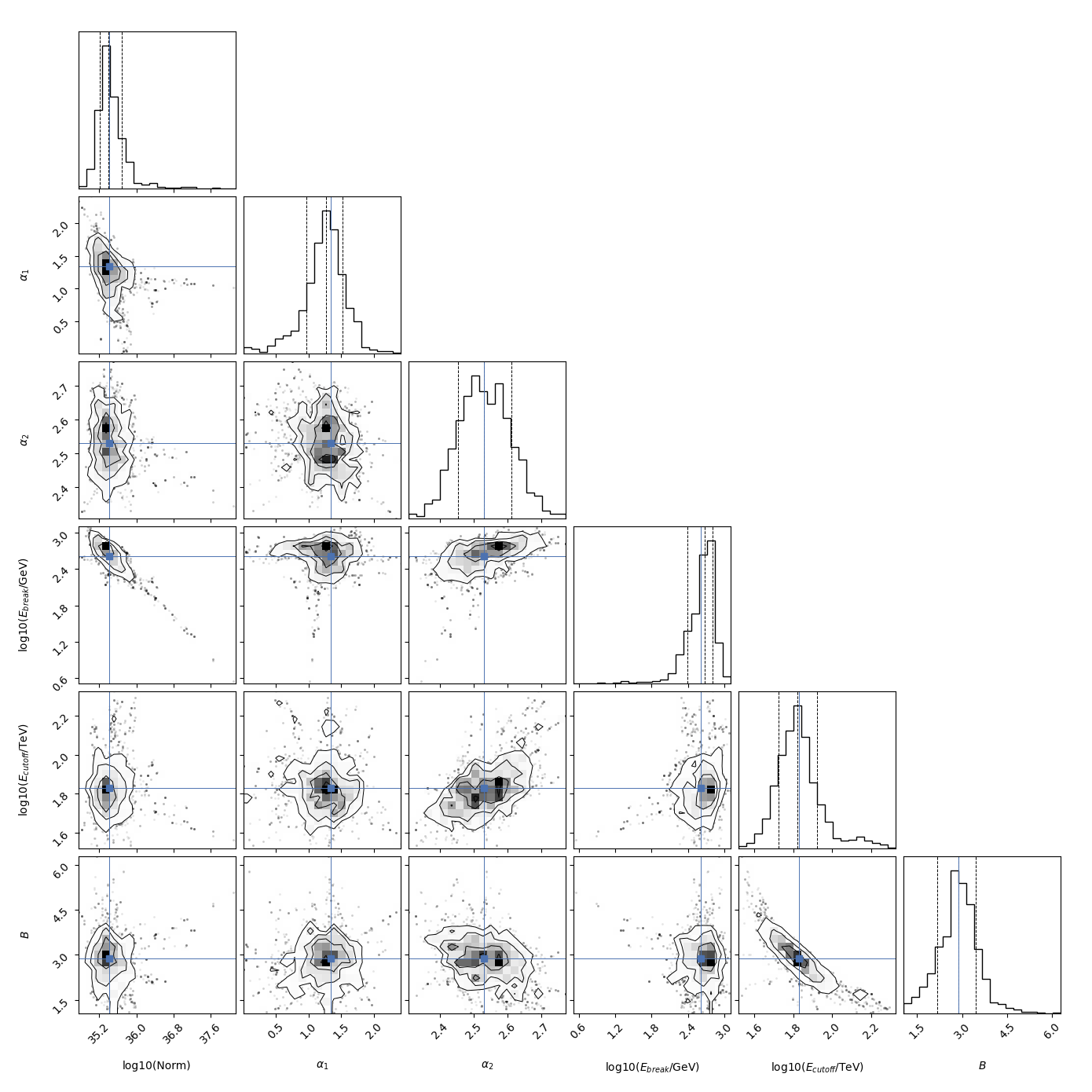}
\caption{The same as Figure \ref{fig:G21.5-0.9} but for HESS J1303-631. And the energy densities and temperatures of FIR and NIR are $T_{\rm FIR} = 40.0$ K and $U_{\rm FIR} = 6.0$ eV cm$^{-3}$, and $T_{\rm NIR} = 4000.0$ K and $U_{\rm NIR} = 2.0$ eV cm$^{-3}$. The radio band data from \cite{1993AJ....106.1095C}, the X-ray band data from \cite{2012A&A...548A..46H}, the GeV $\gamma$-ray band data from \cite{2023RAA....23j5001Z}, the TeV $\gamma$-ray band data from \cite{2012A&A...548A..46H}.}
\end{figure}

\begin{figure*}
\begin{minipage}[t]{0.495\linewidth}
  \centering
\includegraphics[width=7.0cm, angle=0]{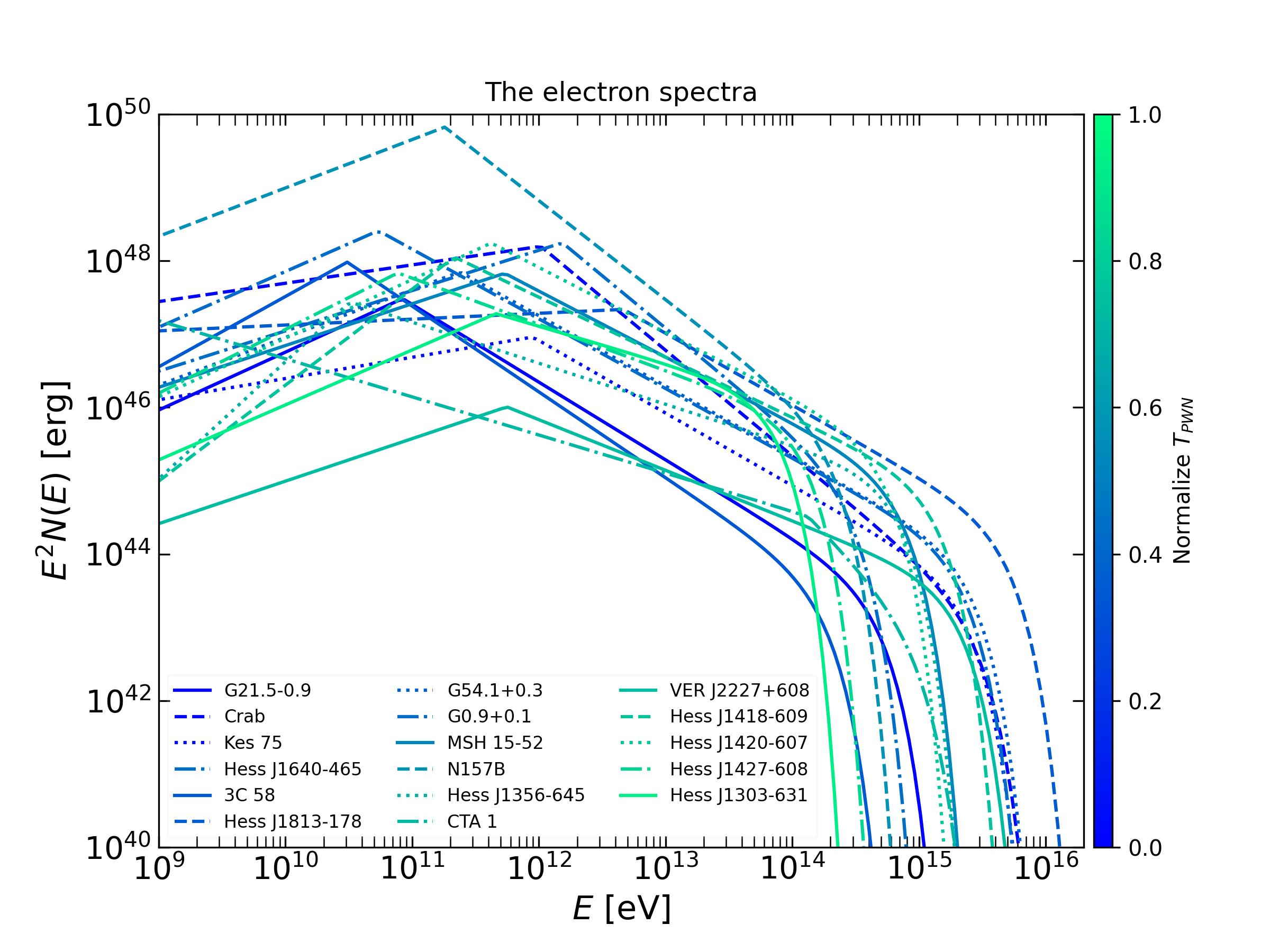}\\
{(a)}
\end{minipage}
\begin{minipage}[t]{0.495\linewidth}
  \centering
\includegraphics[width=7.0cm, angle=0]{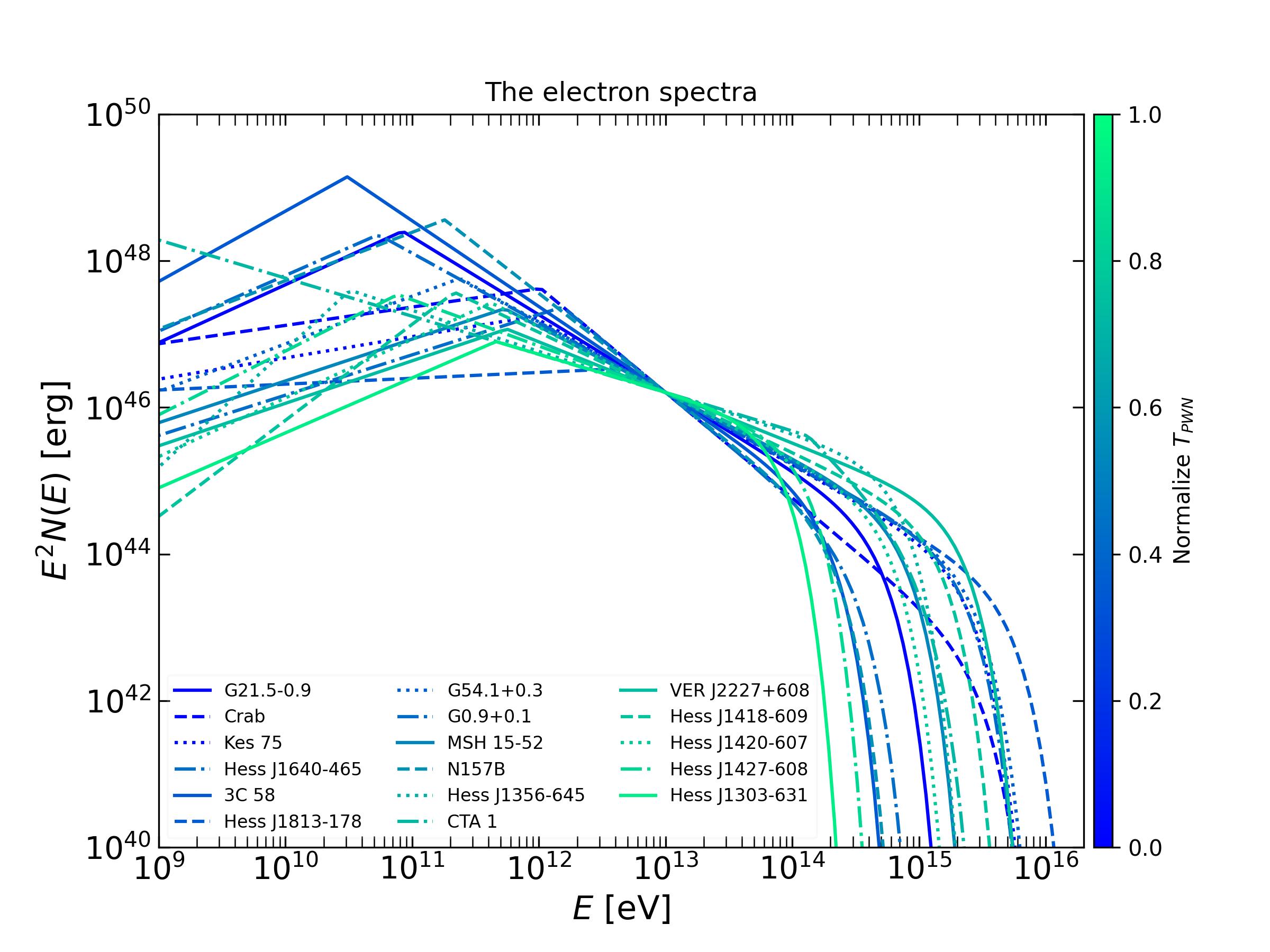}\\
{(b)}
\end{minipage}
\begin{minipage}[t]{0.495\linewidth}
  \centering
\includegraphics[width=7.0cm, angle=0]{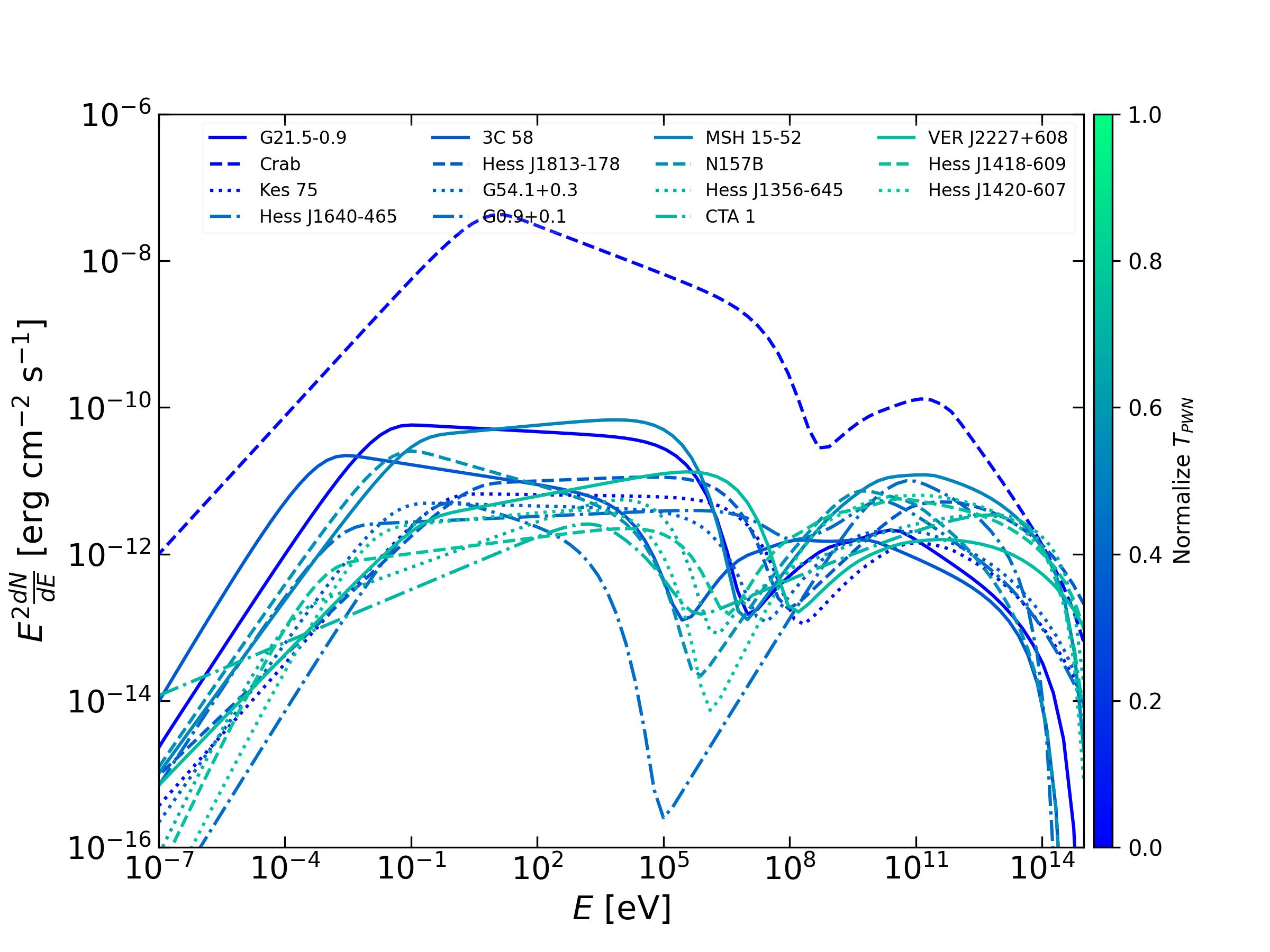}\\
{(c)}
\end{minipage}
\begin{minipage}[t]{0.495\linewidth}
  \centering
\includegraphics[width=7.0cm, angle=0]{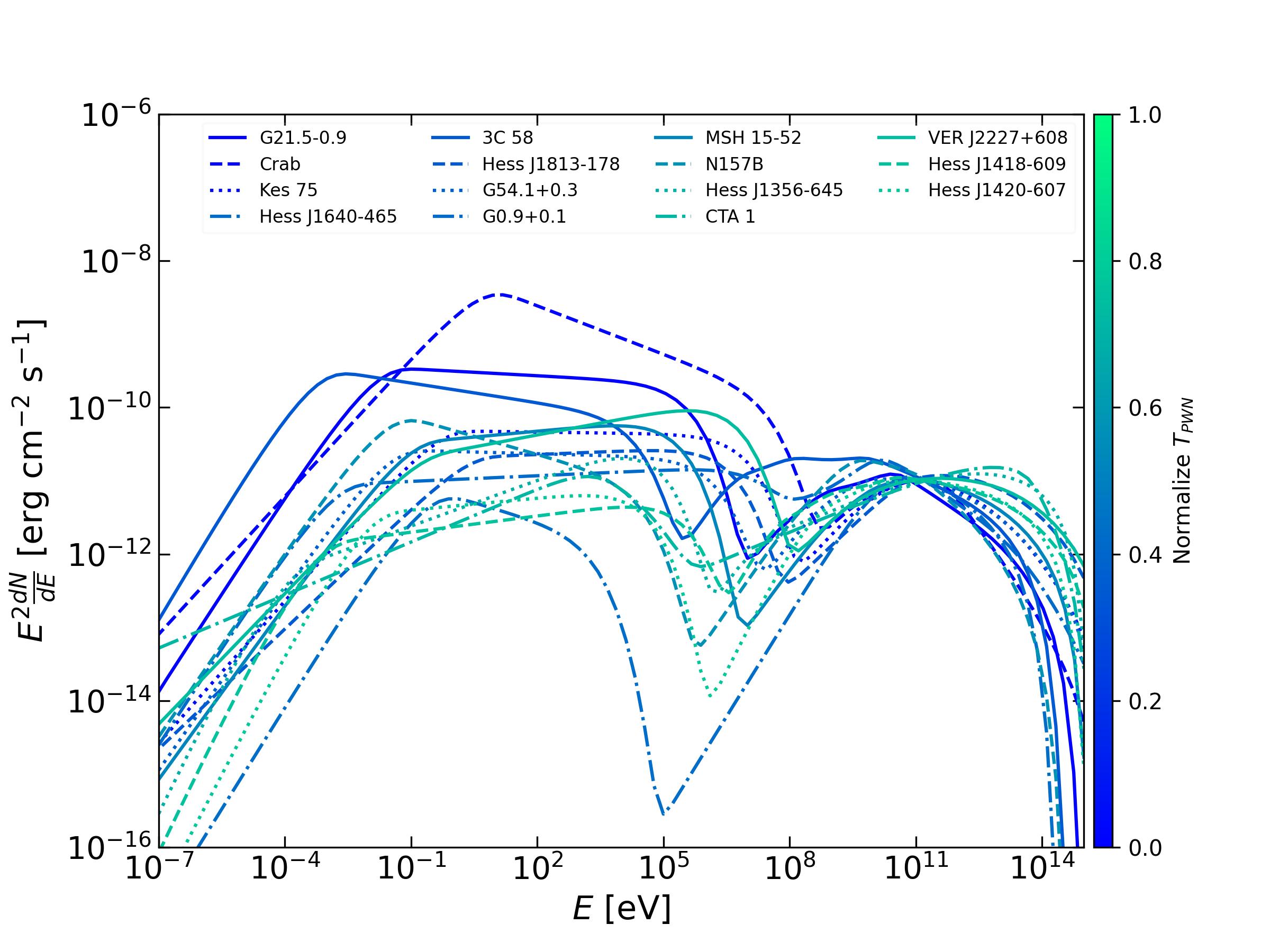}\\
{(d)}
\end{minipage}
\caption{Right: The high-energy electron distribution (a) and SED (c) with the best fit parameters for all PWNe in the sample. Left: the normalized distributions for the high-energy electron distribution (b) and SED (d) normalized at electron energies of 10 TeV and photon energies of 100 GeV, respectively. The color indicates the age of the pulsar wind nebula, from young (blue) to old (green).
\label{fig: eledis and SED}}
\end{figure*}

\begin{figure*}
\begin{minipage}[t]{0.495\linewidth}
  \centering
\includegraphics[width=7.0cm, angle=0]{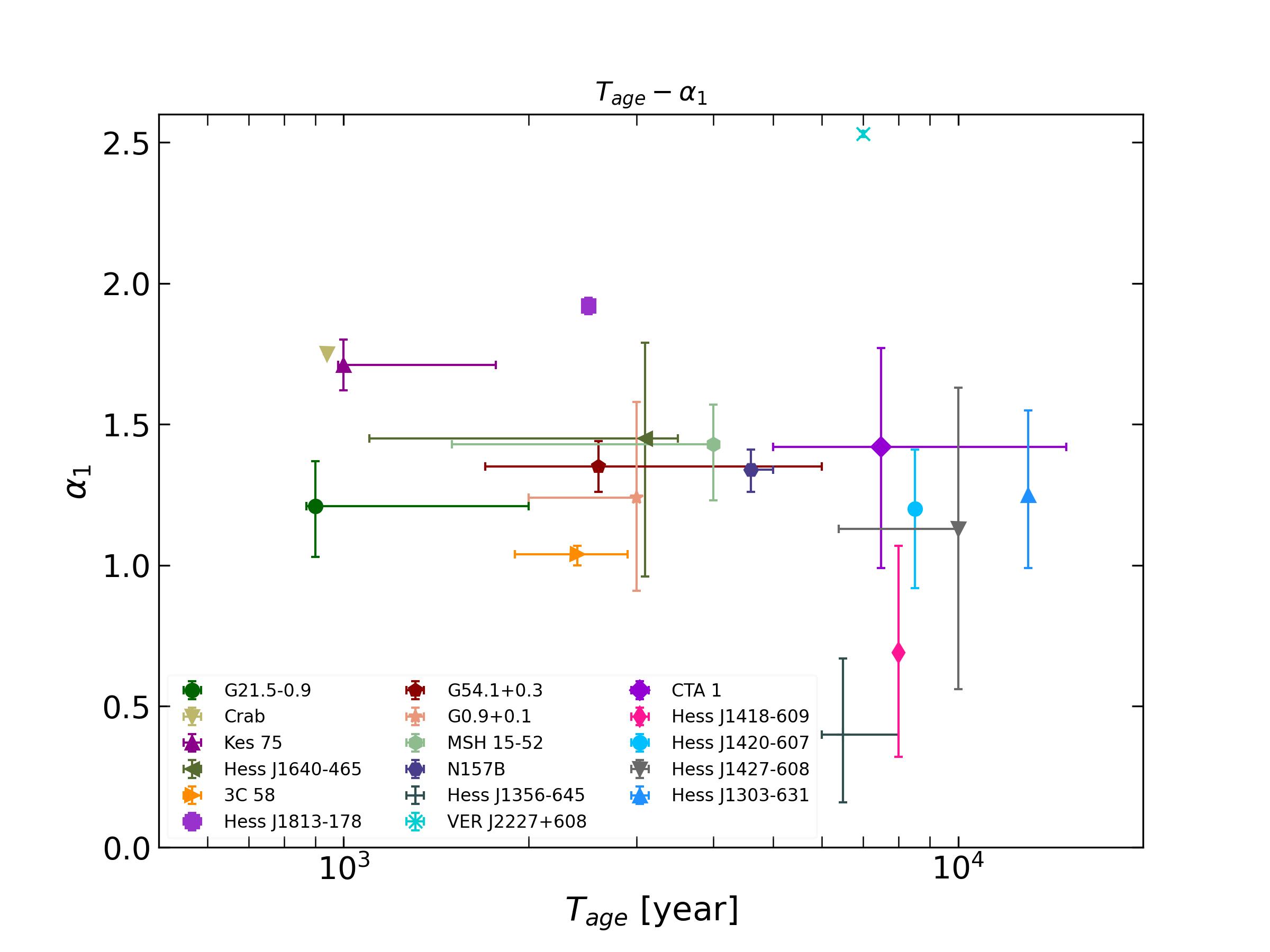}\\
{(a)}
\end{minipage}
\begin{minipage}[t]{0.495\linewidth}
  \centering
\includegraphics[width=7.0cm, angle=0]{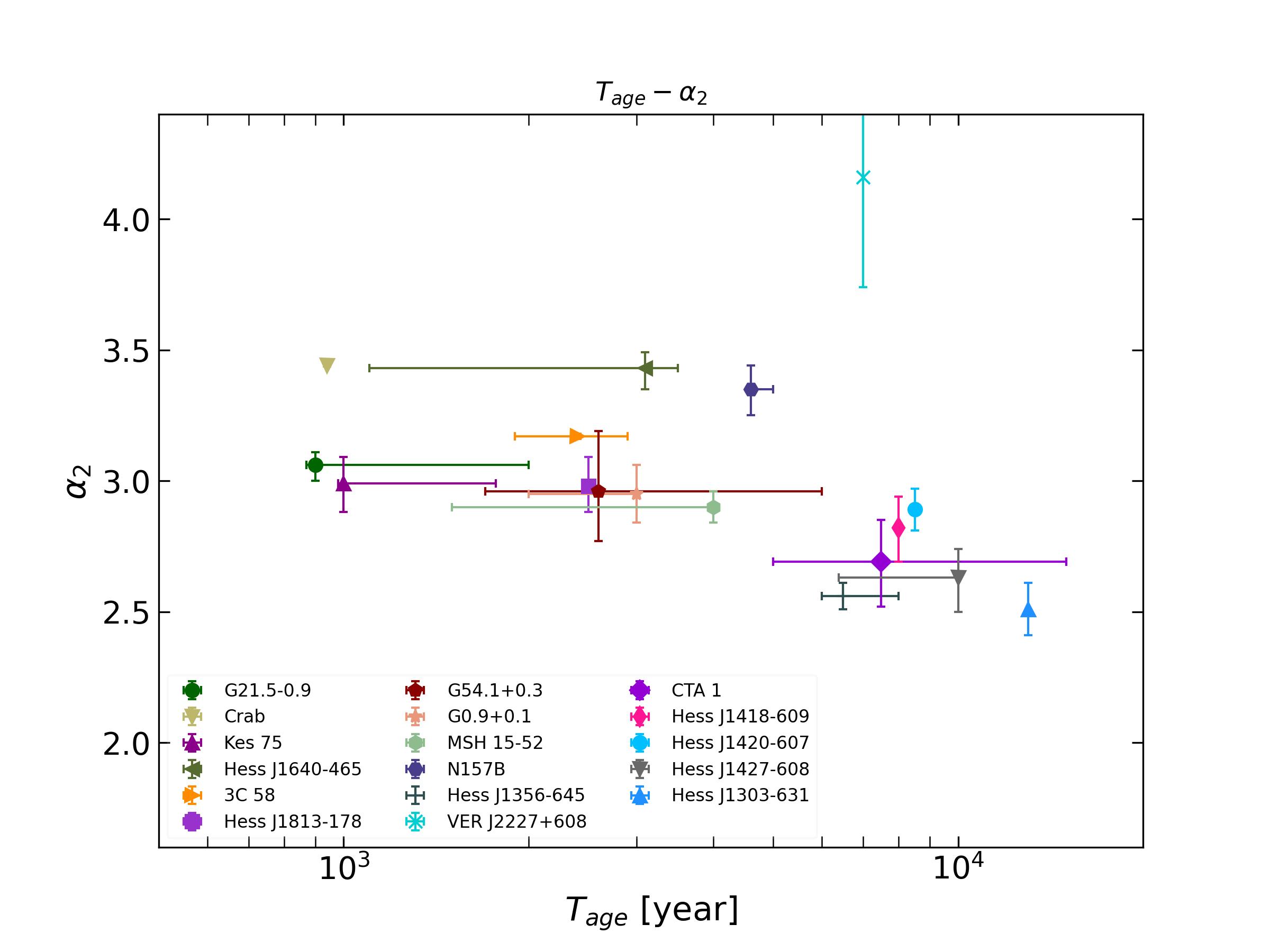}\\
{(b)}
\end{minipage}
\begin{minipage}[t]{1.0\linewidth}
  \centering
\includegraphics[width=7.0cm, angle=0]{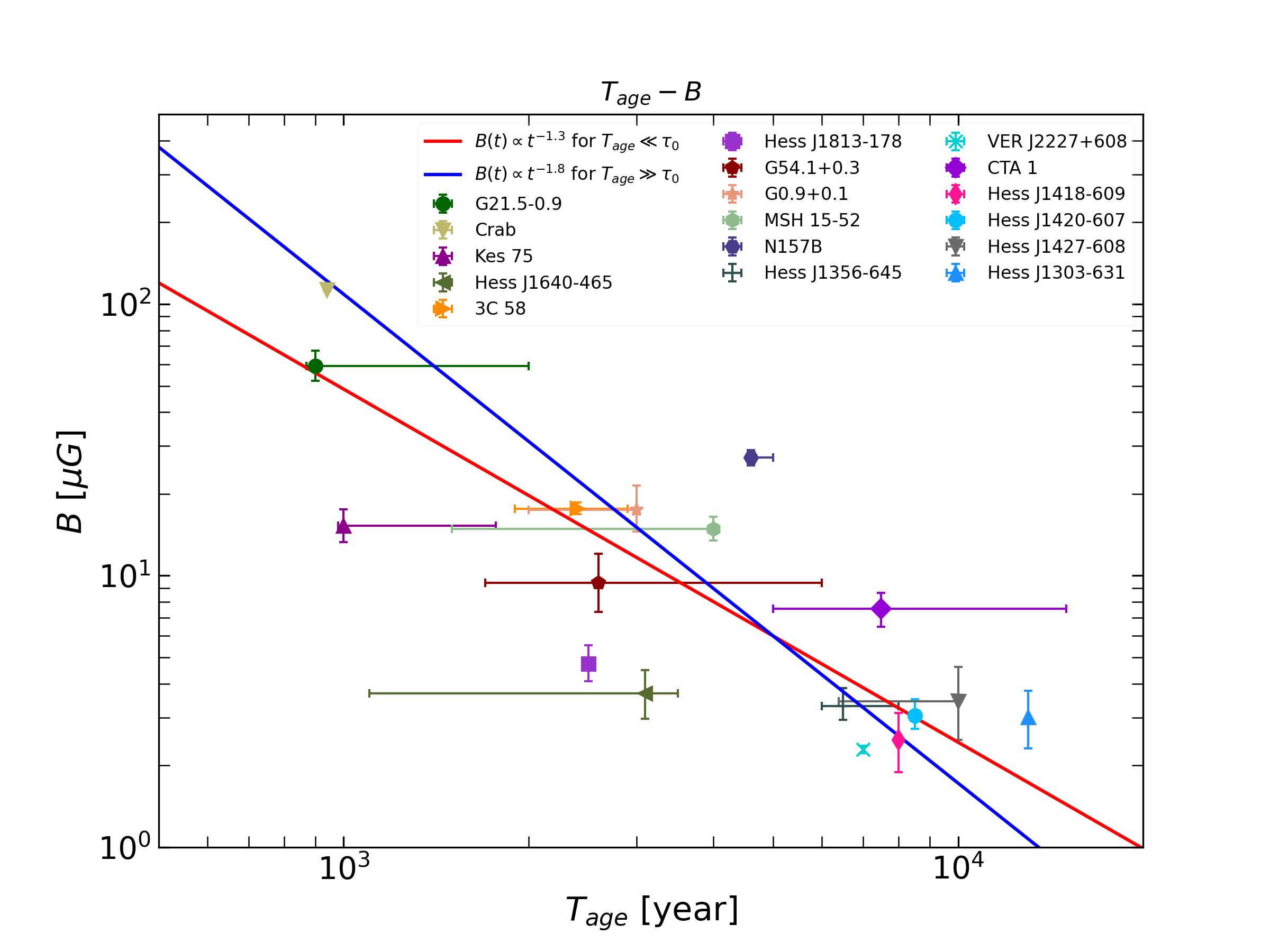}\\
{(c)}
\end{minipage}
\begin{minipage}[t]{0.495\linewidth}
  \centering
\includegraphics[width=7.0cm, angle=0]{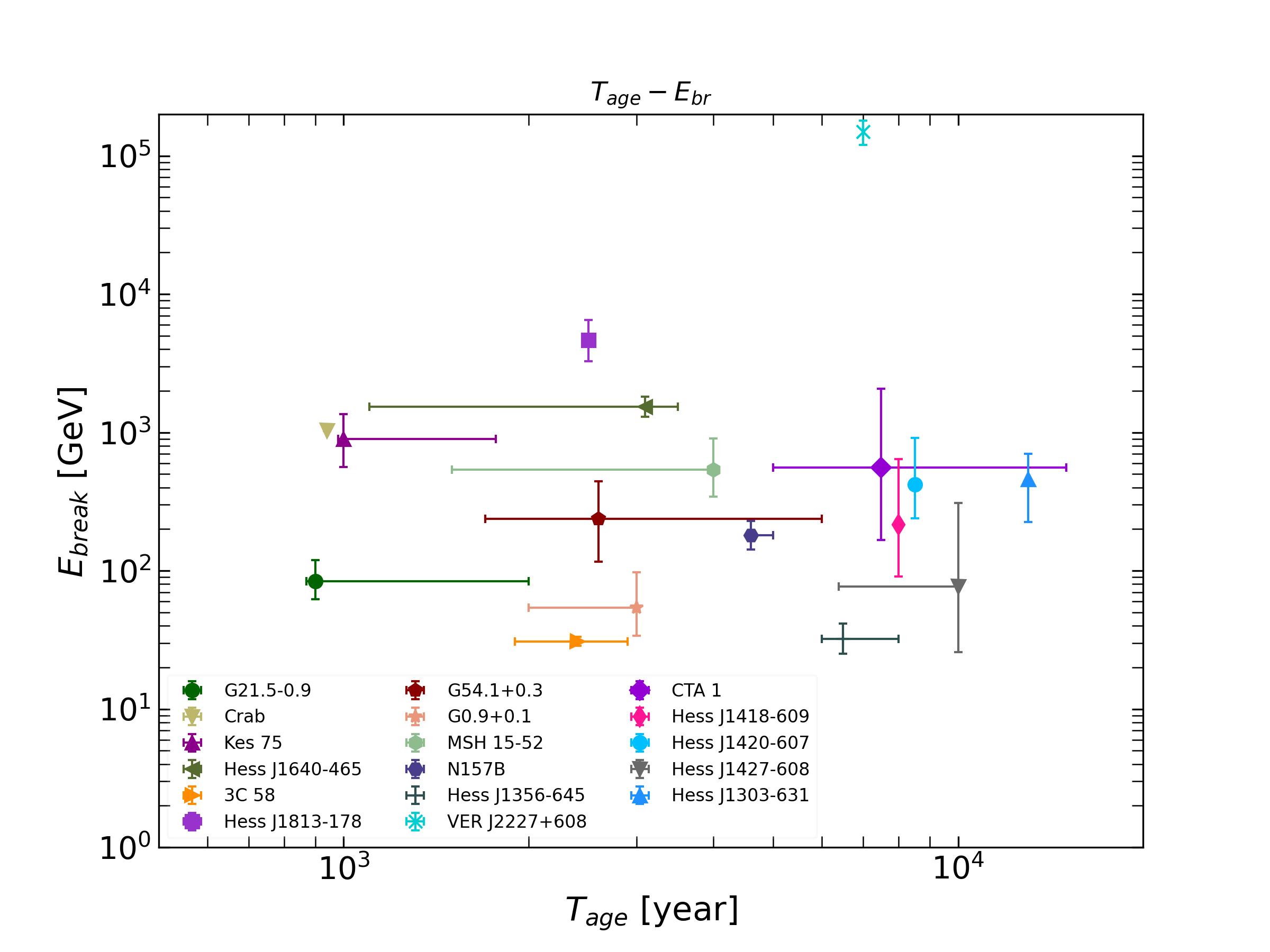}\\
{(d)} 
\end{minipage}
\begin{minipage}[t]{0.495\linewidth}
  \centering
\includegraphics[width=7.0cm, angle=0]{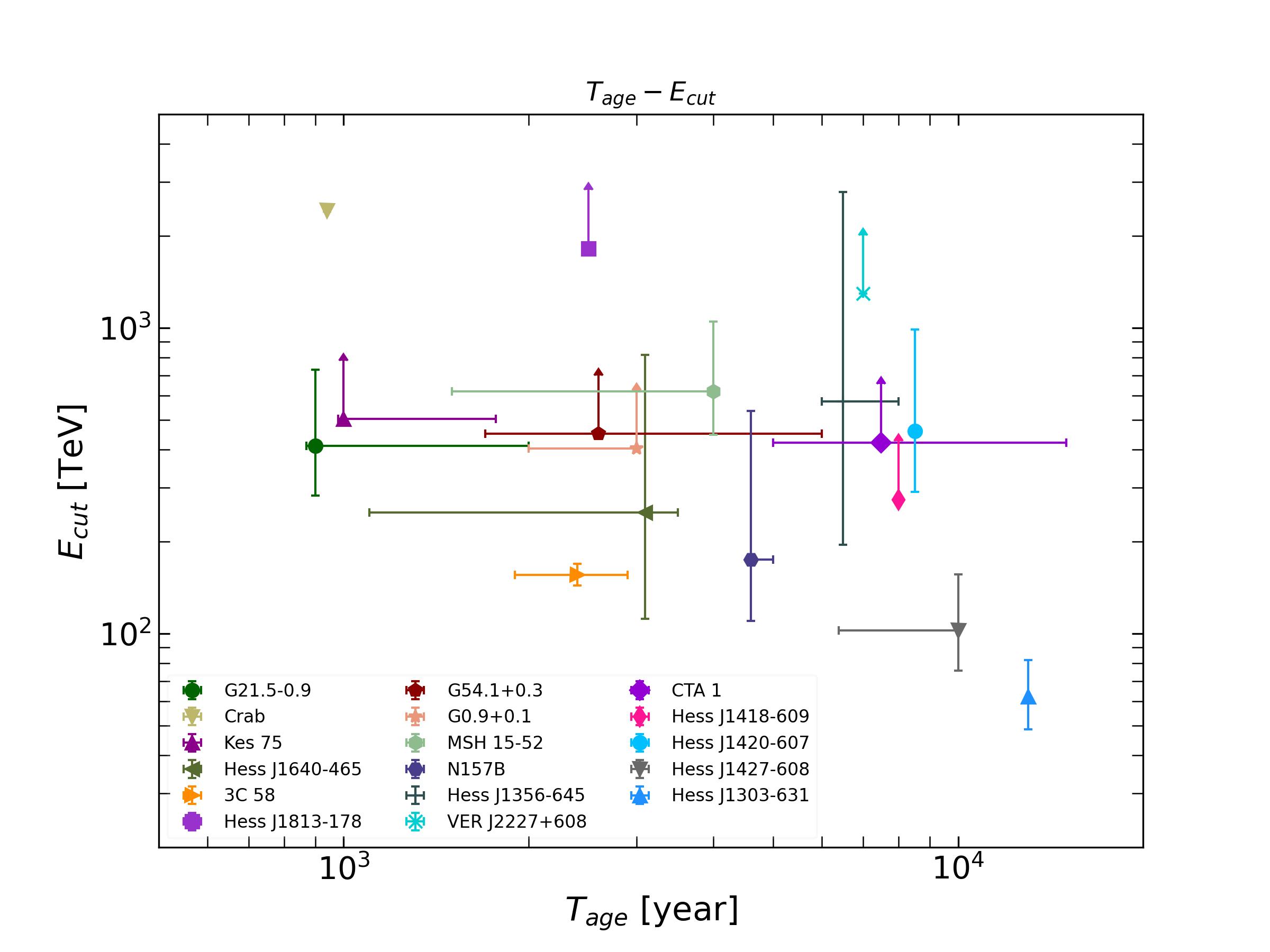}{(e)}\\
{(e)} 
\end{minipage}
\caption{The correlation between the low-end (a) and high-end (b) energy spectral indices, the average magnetic field (c), the break energy (d), and the cutoff energy (e) of the electron distribution and the age of PWNe.
\label{fig: corr1}}
\end{figure*}

\begin{figure*}          
\begin{minipage}[t]{0.495\linewidth}
  \centering
\includegraphics[width=7.0cm, angle=0]{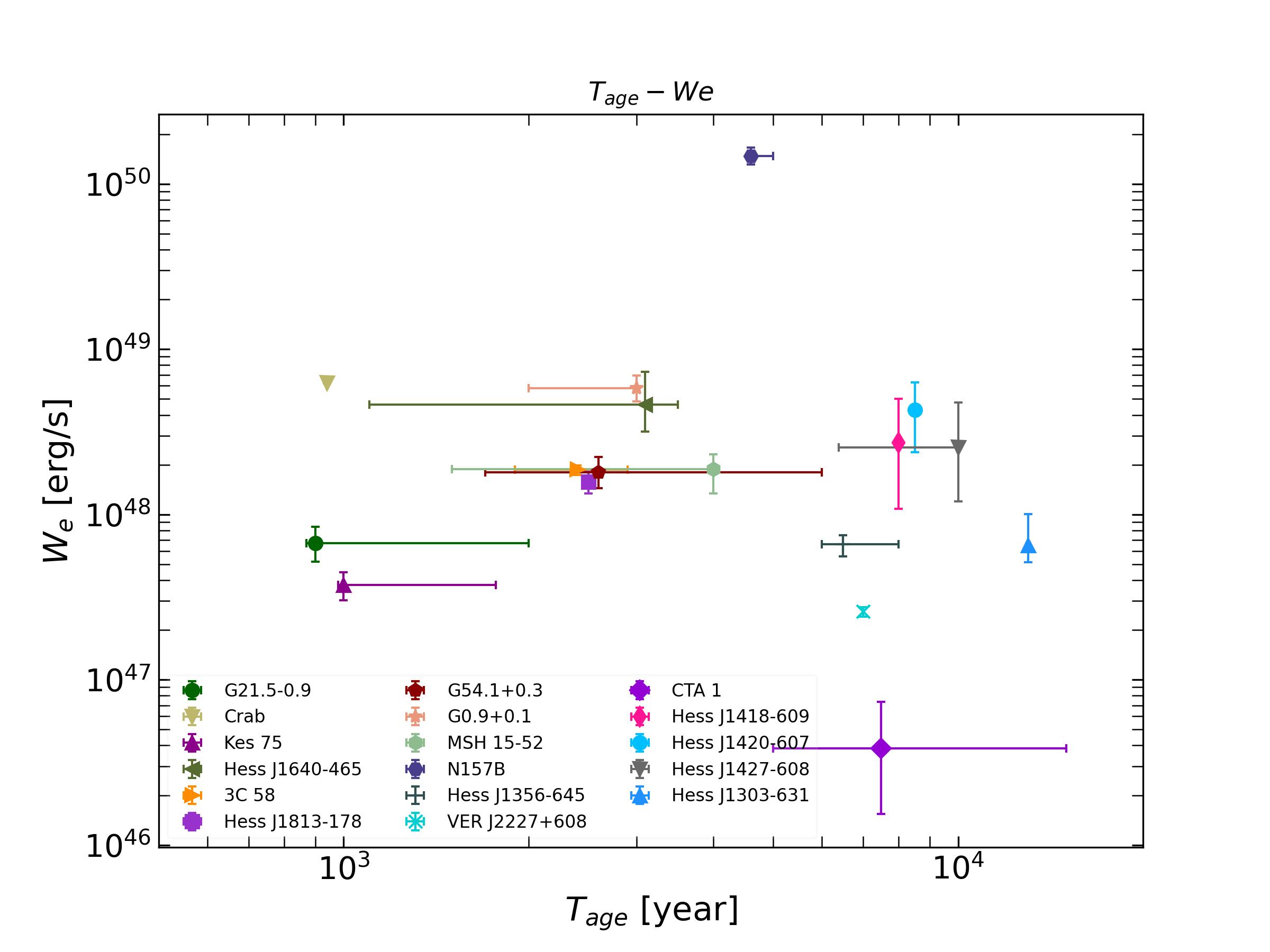}\\
{(f)}
\end{minipage}
\begin{minipage}[t]{0.495\linewidth}
  \centering
\includegraphics[width=7.0cm, angle=0]{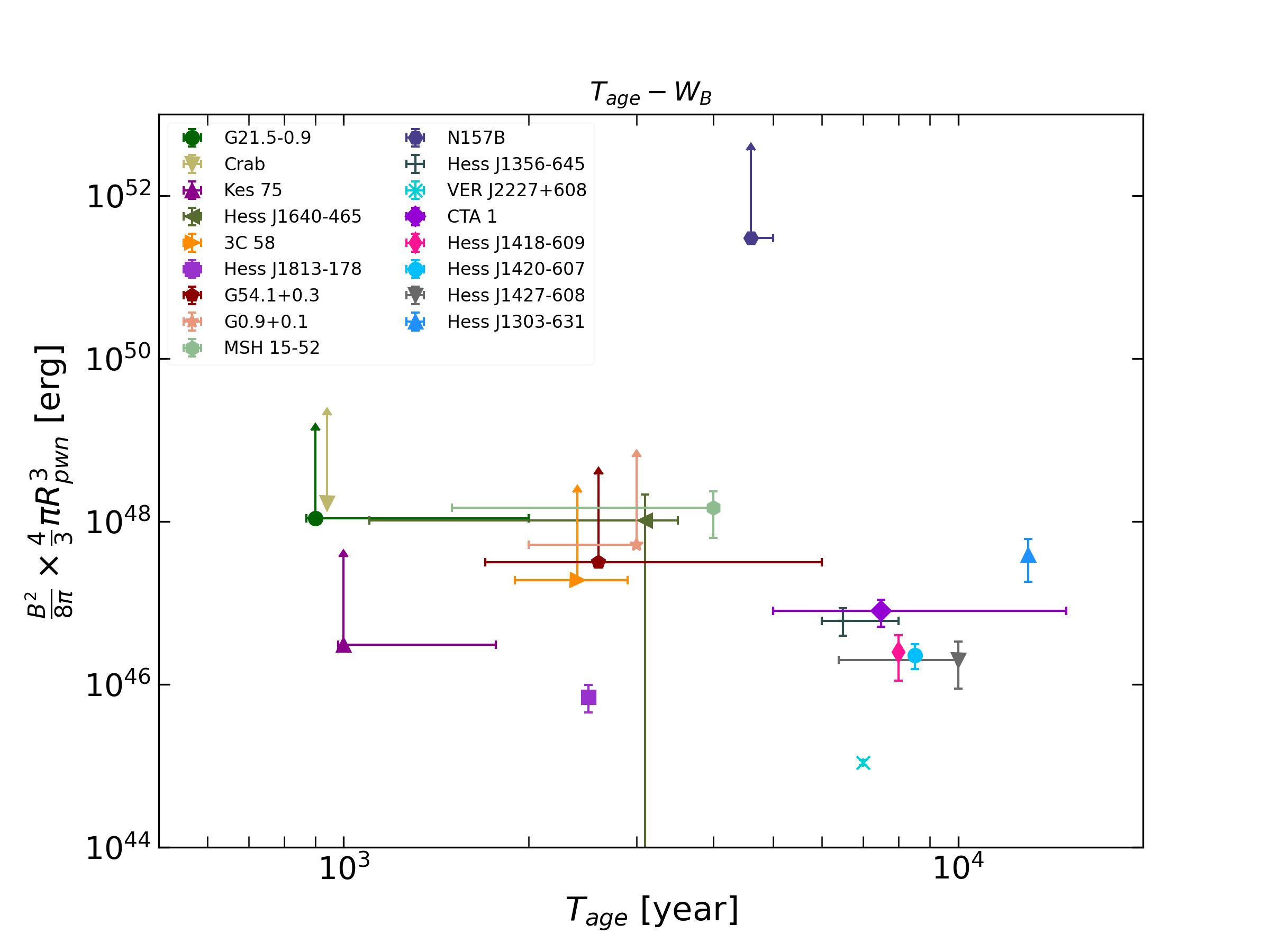}\\
{(g)}
\end{minipage}
\begin{minipage}[t]{0.495\linewidth}
  \centering
\includegraphics[width=7.0cm, angle=0]{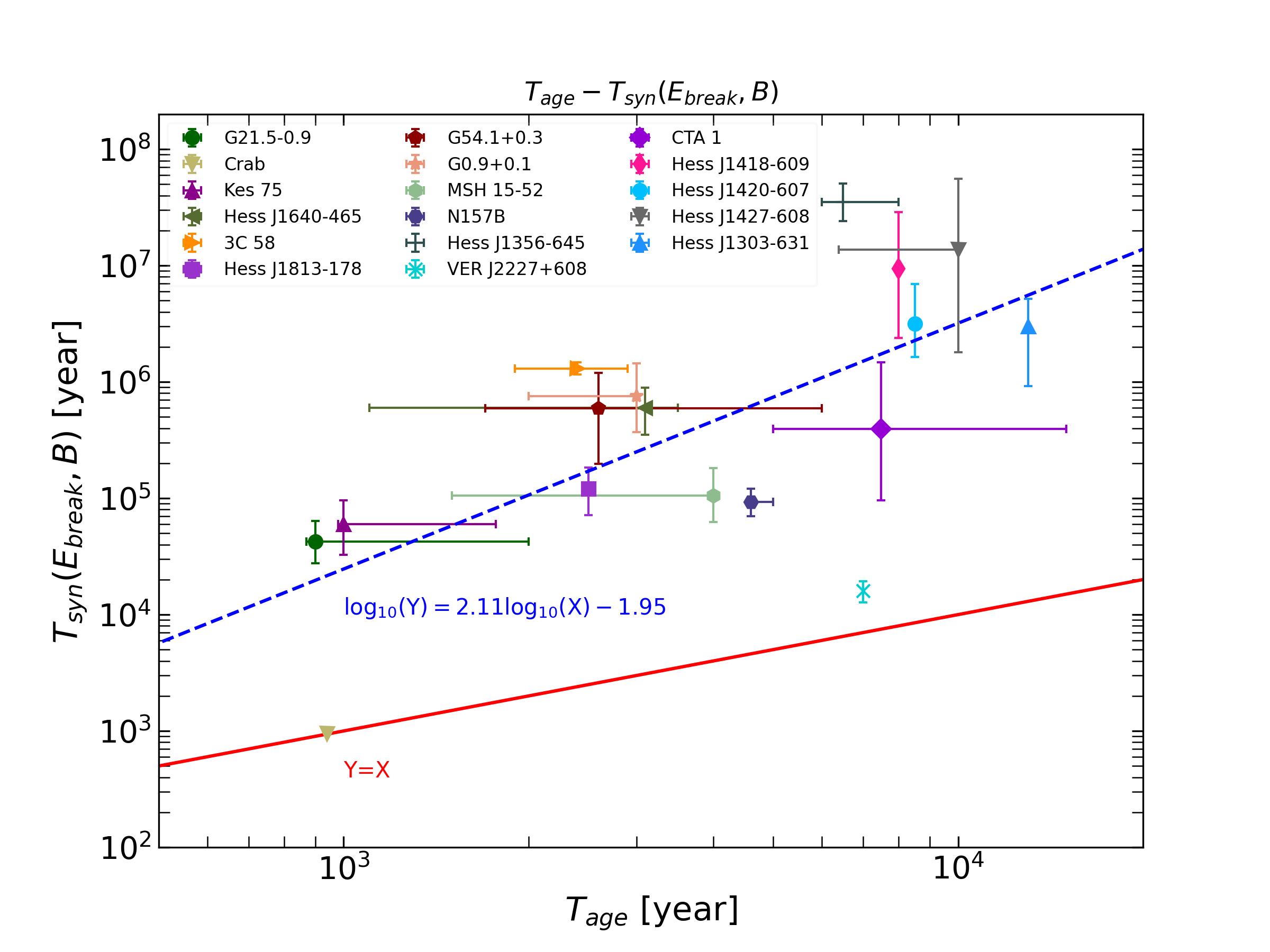}\\
{(h)}
\end{minipage}
\begin{minipage}[t]{0.495\linewidth}
  \centering
\includegraphics[width=7.0cm, angle=0]{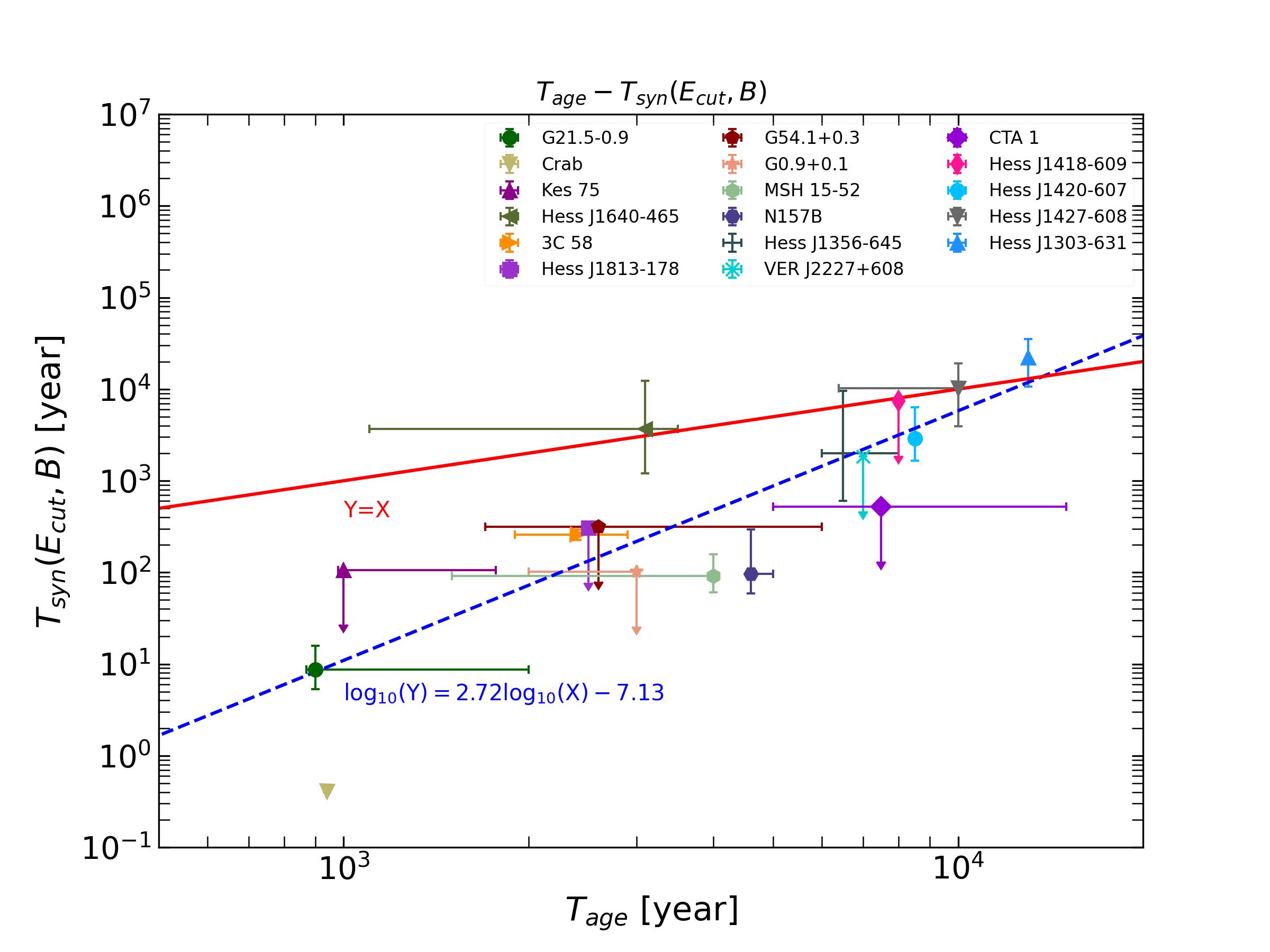}\\
{(j)}
\end{minipage}
\caption{Upper: the correlation between the total electron energy $W_e$ (f) and the total magnetic energy $W_B=\frac{B^2}{8\pi} \frac{4}{3} \pi R_{\rm PWN}^3$ within PWNe (g) and their respective ages. Lower: the correlation between the synchrotron cooling timescale at break energy (h) and at the cutoff energy (j) and the ages of PWNe.
\label{fig: corr2}}
\end{figure*}

Table 1 presents the fundamental physical quantities alongside the best-fit values of the model parameters. It is evident that all spectral fits exhibit a reduced $\chi^2 < 2.0$, except for Crab, and HESS J1303-631.
For PWNe with butterfly-shaped X-ray data, the minimum, maximum, and geometric mean energy values within the energy range will be utilized for participation in the multi-band fitting, such as Kes 75, HESS J1640$-$465, G54.1$+$0.3, G0.9$+$0.1, N157B, HESS J1356$-$645, VER J2227$+$608 and HESS J1418$-$609.
The energy spectrum of the Crab nebula softens from 0.3 at radio wavelength to 0.6 in the optical range, and there is also thermal radiation in the infrared band \citep{2017SSRv..207..175R}, resulting in a poor fit with a single power law. 

The TeV data from HESS observation and the GeV data detected by {\it Fermi}-LAT cannot be well connected. The TeV data of HESS observation \citep{2012A&A...548A..46H} from HESS J1303-631 exhibit uneven fine structures in detail and do not match the GeV data detected by {\it Fermi}-LAT \citep{2023RAA....23j5001Z}, resulting in a relatively large chi-square value ($\chi^2=3.99$) for the fitting results. Generally, the multi-wavelength spectra can be well-fitted by one-zone model described by a broken power law with a high-energy cutoff.

The left panel of Figure \ref{fig: eledis and SED} illustrates the high-energy electron distributions (a) and SEDs (c) of those 17 PWNe with the best-fit parameters. To demonstrate the evolution of PWNe, the right panel depicts the high-energy electron distributions normalized at 10 TeV and SEDs normalized at 100 GeV. It is observed that the high-energy electron spectra gradually harden with the ages of PWNe. Indeed, younger PWNe exhibit softer spectra in the gamma-ray band, while older PWNe show harder GeV spectra with a high-energy cutoff in the TeV band, which is opposite to the evolution of supernova remnants \citep{2019ApJ...874...50Z}.

Figure \ref{fig: corr1}(a) and \ref{fig: corr1}(b) display the correlation between the low-end and high-end energy spectral indices and the ages of PWNe. The results indicate that the low-end spectral index remains relatively constant with age, mainly between 1.0 and 1.5, while the high-end spectral index decreases with age. The origin of these electron spectra within PWNe is still under debate. \cite{1996MNRAS.278..525A} suggested the existence of two distinct populations of relativistic electrons within PWNe. The low-energy electrons with an index of $\sim 1.5$ are believed to be generated within the light cylinder of the pulsar, reflecting the history of the pulsar and the nebula, and primarily contributing to radio photons by synchrotron radiation, thus called ``radio electrons''. Therefore, this process has not significantly changed over time, resulting in the constancy of the low-end spectral index. On the other hand, the high-energy electrons are freshly accelerated ultra-high-energy electrons with $E>E_b$, mainly produced at the wind termination shock location beyond the light cylinder, via Fermi 
acceleration, and form the high-energy part with an index $\geq 2$, also called as the ``wind electrons'' component. When the termination shock propagates outward, the magnetic field $B(r)$ will also decay (also see Figure \ref{fig: corr1}(c)), and the cooling of high-energy electrons will slow down rapidly, resulting in a hardening of the energy spectrum. Many time-dependent models considering such magnetic field evolution exhibit the characteristic of hardening in the high-end energy spectrum (\cite[e.g.][]{2014JHEAp...1...31T,2018A&A...609A.110Z}). Additionally, the radio electron possibly originates from relativistic magnetic reconnection, and its spectral slope is in the range of $1.0-2.0$ \cite[e.g.][]{2011ApJ...741...39S,2012CS&D....5a4014S,2014ApJ...783L..21S}. The present of \cite{2021MNRAS.504.4952A} suggests a sub-population with a hard spectrum (radio photon spectral indices of $\alpha = 0.01 \pm 0.06$) near the termination shock and softer spectra elsewhere, possibly due to a recent evacuation of the shock surroundings. Note that the spectral index of the source VER J2227$+$608 deviates significantly from the clustering distribution of the sample sources, 
as the fitting data provided in this paper mainly come from the tail region of the radio morphology, which is generally considered to be a supernova remnant.

Figure \ref{fig: corr1}(c), (d) and (e) illustrates the correlations between the average magnetic field, the break energy, and the cutoff energy of the electron distribution and the age of PWNe. It is observed that the magnetic field gradually decreases with the evolution of PWNe. Two theoretical lines are provided in the figure based on the dynamical evolution of PWNe (the conversion of the pulsar's rotational energy into magnetic energy)\citep{2014JHEAp...1...31T}: when $T_{age} \ll \tau_0$(where $\tau_0$ is the initial spin-down timescale), the magnetic field follows $B \propto t^{-1.3}$, and when $T_{age} \gg \tau_0$, $B \propto t^{-1.8}$. 
From Figure \ref{fig: corr1}(d), it can be seen that the break energy of the electron distribution shows no significant correlation with the age of PWNe. In traditional time-dependent evolution models, the magnitude of the break energy of the injected spectral lies between 10 GeV and 1.0 TeV \citep[e.g][]{2014JHEAp...1...31T,2018A&A...609A.110Z}, which is basely consistent with the break energy in our non-time-dependent model, although its physical mechanism remains unclear \citep{2011MNRAS.410..381B,2011ApJ...741...40T,2014JHEAp...1...31T,2018A&A...609A.110Z}. If the magnetic energy is entirely transferred to particles, either in the pulsar wind or at the termination shock, the break energy is believed to be proportional to the product of the Lorentz factor $\gamma_w$ and the wind magnetization strength $\sigma_w$
\citep{2019MNRAS.489.2403L,2020ApJ...896..147L}, i.e., $E_b \propto \gamma_w \sigma_w m_{\pm e}c^2$. Although most studies suggest $\gamma_w$ ranges between $10^2$ and $10^6$, there is significant variation in magnetization strength. Theoretical models of pulsar magnetosphere and wind predict $\sigma_w$ to be much greater than 1.0 \citep[e.g.][]{2012SSRv..173..341A}, while simple one-dimensional models of PWNe require $\sigma_w$ to be much less than 1.0 \citep[e.g.][]{1984ApJ...283..694K}, a discrepancy known as the "$\sigma$-problem".  
Figure \ref{fig: corr1}(e) also shows no apparent correlation between the cutoff energy and the age of PWNe. This lack of correlation may stem from the fact that the fitting of most PWNe only provides lower limits. However, the decrease in the cutoff energy for the two oldest PWNe  suggests a potential negative correlation between them.

Figure \ref{fig: corr1}(f) and (g) depicts the correlation between the total electron energy $W_e$, and the total magnetic energy $W_B=\frac{B^2}{8\pi} \frac{4}{3} \pi R_{\rm PWN}^3$ within PWNe and their respective ages. Except for PWN N157B in the Large Magellanic Cloud and PWN CTA 1, the total electron energy of most PWNe ranges from $3 \times 10^{47}-10^{49}$ erg, show no significant variation with the evolution of PWNe. However, the total magnetic energy appears to be inversely correlated with age. As the evolution of PWNe progresses, there is a conversion of magnetic energy into electron energy, resulting in an overall decrease in magnetic energy. 

Figure \ref{fig: corr1}(h) and (j) illustrates the correlation between the synchrotron cooling timescale at break energy (h), and at the cutoff energy (j) and the ages of PWNe. A strong positive correlation is observed between the synchrotron cooling timescale and the ages of the corresponding PWNe, likely attributable to the relationship between magnetic fields and PWN ages. The red solid line in the figure represents when the cooling timescale equals the age, and the blue dashed line is a linear fit without error. It is notable that except for the Crab Nebula PWN, the cooling timescale of the break energies exceeds the age of PWNe and a slope between them greater than 1.0, indicating a gradual increase in the cooling timescale as the magnetic field weakens. The cooling timescale of the cutoff energy timescales are mostly smaller than or equal to the ages of the corresponding PWNe, implying that the termination shock can efficient accelerate electrons to higher energies. Simultaneously, with the evolution of PWNe, the wind speed and magnetic field of PWNe weaken, resulting in a decrease in the maximum energy attained by accelerated electrons.

\section{Summary and Discussion}
Multiwavelength observations of PWNe provide a unique opportunity to study evolution of high-energy electron distribution within PWNes, which is closely related to the acceleration, radiation, escape, and transport of high-energy particles inside them \citep[e.g][]{2008ApJ...676.1210Z,2018ApJ...867..141I,2019A&A...621A.116H,2020MNRAS.498.1911L,2021A&A...655A..41Z,Zhu_2023}, and the the origin of Galactic electrons and positrons \citep[e.g][]{2014JCAP...04..006D,2020PhRvD.102b3015M,2021PhRvD.104h3012D} and the contributions to the gamma-ray sky \citep[e.g][]{Pagliaroli:2021sQ,2022A&A...666A...7M}. In this paper, we select 16 PWNe as sample sources with observational data in at least three bands between radio, X-ray, GeV, and TeV. We fit the SED of each PWN source using a simplified time-independent single-zone model,  and then reveal the evolution of high-energy electron distributions in PWNe by analyzing the correlation of the model parameters.
Below we give a brief summary of our conclusion.

\begin{itemize}
    \item \textit{The electron distribution in PWNe:} 
    Due to observational limitations, multi-band information for most PWNe is still somewhat lacking, with the exception of the Crab Nebula. Currently, a simplified double power law with a super-exponential cutoff can be used to describe the electron distribution inside observed PWNe.
    \item \textit{The evolution of the electron distribution:}
    With the evolution of PWNe, the electron spectrum becomes harder at higher energies, transitioning from approximately 3.5 to around 2.5 for the power-law index, while maintaining a power-law index around 1.0-1.5 at low energies. The break energy of the electron distribution is not correlated with the age of the PWN, and there is no significant correlation between the cutoff energy and the age of the PWN.
    \item \textit{The evolution of the magnetic field:}
    The mean magnetic field decreases with the age of the respective PWNe, resulting in a positive correlation between the synchrotron cooling timescales of the break energy and the cutoff energy of the electron distribution with the age of the PWN.
    \item \textit{The total energy:}
    The total energy of electrons inside PWNe remains relatively constant at approximately $2.0 \times 10^{48}$ erg, while the total magnetic energy decreases with the age of the corresponding PWNe.
\end{itemize}

\normalem
\begin{acknowledgements}
This work is supported by the National Natural Science Foundation of China (No. 12220101003, No. 12273114, No. U1931204, No. 12103040, No. 12147208 and No. U2031111), the Project for Young Scientists in Basic Research of Chinese Academy of Sciences (No. YSBR-061) and the Program for Innovative Talents and Entrepreneur in Jiangsu.

\end{acknowledgements}
  
\bibliographystyle{raa}

\begin{thebibliography}{116}
\providecommand\natexlab[1]{#1}
\providecommand\JournalTitle[1]{#1}

\bibitem[{Abdo} {et~al.}(2010)]{2010ApJ...714..927A}
{Abdo}, A.~A., {Ackermann}, M., {Ajello}, M., {et~al.} 2010, \apj, 714, 927

\bibitem[{Abdollahi} {et~al.}(2022)]{2022ApJS..260...53A}
{Abdollahi}, S., {Acero}, F., {Baldini}, L., {et~al.} 2022, \apjs, 260, 53

\bibitem[{Abramowski} {et~al.}(2014)]{2014MNRAS.439.2828A}
{Abramowski}, A., {Aharonian}, F., {Benkhali}, F.~A., {et~al.} 2014, \mnras,
  439, 2828

\bibitem[{Acciari} {et~al.}(2009)]{2009ApJ...703L...6A}
{Acciari}, V.~A., {Aliu}, E., {Arlen}, T., {et~al.} 2009, \apjl, 703, L6

\bibitem[{Acciari} {et~al.}(2010)]{2010ApJ...719L..69A}
{Acciari}, V.~A., {Aliu}, E., {Arlen}, T., {et~al.} 2010, \apjl, 719, L69

\bibitem[{Acero} {et~al.}(2013)]{2013ApJ...773...77A}
{Acero}, F., {Ackermann}, M., {Ajello}, M., {et~al.} 2013, \apj, 773, 77

\bibitem[{Aharonian} {et~al.}(2005{\natexlab{a}})]{2005A&A...435L..17A}
{Aharonian}, F., {Akhperjanian}, A.~G., {Aye}, K.~M., {et~al.}
  2005{\natexlab{a}}, \aap, 435, L17

\bibitem[{Aharonian} {et~al.}(2005{\natexlab{b}})]{2005A&A...432L..25A}
{Aharonian}, F., {Akhperjanian}, A.~G., {Aye}, K.~M., {et~al.}
  2005{\natexlab{b}}, \aap, 432, L25

\bibitem[{Aharonian} {et~al.}(2006)]{2006A&A...456..245A}
{Aharonian}, F., {Akhperjanian}, A.~G., {Bazer-Bachi}, A.~R., {et~al.} 2006,
  \aap, 456, 245

\bibitem[{Aharonian} {et~al.}(2008)]{2008A&A...477..353A}
{Aharonian}, F., {Akhperjanian}, A.~G., {Barres de Almeida}, U., {et~al.} 2008,
  \aap, 477, 353

\bibitem[{Aleksi{\'c}} {et~al.}(2014)]{2014A&A...567L...8A}
{Aleksi{\'c}}, J., {Ansoldi}, S., {Antonelli}, L.~A., {et~al.} 2014, \aap, 567,
  L8

\bibitem[{Aliu} {et~al.}(2013)]{2013ApJ...764...38A}
{Aliu}, E., {Archambault}, S., {Arlen}, T., {et~al.} 2013, \apj, 764, 38

\bibitem[{Amato}(2024)]{2024arXiv240210912A}
{Amato}, E. 2024, arXiv e-prints, arXiv:2402.10912

\bibitem[{Amato} {et~al.}(2003)]{2003A&A...402..827A}
{Amato}, E., {Guetta}, D., \& {Blasi}, P. 2003, \aap, 402, 827

\bibitem[{An}(2019)]{2019ApJ...876..150A}
{An}, H. 2019, \apj, 876, 150

\bibitem[{Arad} {et~al.}(2021)]{2021MNRAS.504.4952A}
{Arad}, O., {Lavi}, A., \& {Keshet}, U. 2021, \mnras, 504, 4952

\bibitem[{Arakawa} {et~al.}(2020)]{2020ApJ...897...33A}
{Arakawa}, M., {Hayashida}, M., {Khangulyan}, D., \& {Uchiyama}, Y. 2020, \apj,
  897, 33

\bibitem[{Arons}(2012)]{2012SSRv..173..341A}
{Arons}, J. 2012, ssr, 173, 341

\bibitem[{Atoyan} \& {Aharonian}(1996)]{1996MNRAS.278..525A}
{Atoyan}, A.~M., \& {Aharonian}, F.~A. 1996, \mnras, 278, 525

\bibitem[{Bednarek} \& {Bartosik}(2003)]{2003A&A...405..689B}
{Bednarek}, W., \& {Bartosik}, M. 2003, \aap, 405, 689

\bibitem[{Bednarek} \& {Protheroe}(1997)]{1997PhRvL..79.2616B}
{Bednarek}, W., \& {Protheroe}, R.~J. 1997, \prl, 79, 2616

\bibitem[{Bock} \& {Gaensler}(2005)]{2005ApJ...626..343B}
{Bock}, D.~C.~J., \& {Gaensler}, B.~M. 2005, \apj, 626, 343

\bibitem[{Brogan} {et~al.}(2005)]{2005ApJ...629L.105B}
{Brogan}, C.~L., {Gaensler}, B.~M., {Gelfand}, J.~D., {et~al.} 2005, \apjl,
  629, L105

\bibitem[{Bucciantini} {et~al.}(2011)]{2011MNRAS.410..381B}
{Bucciantini}, N., {Arons}, J., \& {Amato}, E. 2011, \mnras, 410, 381

\bibitem[{Cao} {et~al.}(2021)]{2021Natur.594...33C}
{Cao}, Z., {Aharonian}, F.~A., {An}, Q., {et~al.} 2021, \nat, 594, 33

\bibitem[{Cao} {et~al.}(2024)]{2024ApJS..271...25C}
{Cao}, Z., {Aharonian}, F., {An}, Q., {et~al.} 2024, \apjs, 271, 25

\bibitem[{Chen} {et~al.}(2006)]{2006ApJ...651..237C}
{Chen}, Y., {Wang}, Q.~D., {Gotthelf}, E.~V., {et~al.} 2006, \apj, 651, 237

\bibitem[Collaboration*† {et~al.}(2021)]{doi:10.1126/science.abg5137}
Collaboration*†, T.~L., Cao, Z., Aharonian, F., {et~al.} 2021, Science, 373,
  425

\bibitem[{Condon} {et~al.}(1993)]{1993AJ....106.1095C}
{Condon}, J.~J., {Griffith}, M.~R., \& {Wright}, A.~E. 1993, \aj, 106, 1095

\bibitem[{de Jager} {et~al.}(2008)]{2008ApJ...689L.125D}
{de Jager}, O.~C., {Slane}, P.~O., \& {LaMassa}, S. 2008, \apjl, 689, L125

\bibitem[{Di Mauro} {et~al.}(2014)]{2014JCAP...04..006D}
{Di Mauro}, M., {Donato}, F., {Fornengo}, N., {Lineros}, R., \& {Vittino}, A.
  2014, \jcap, 2014, 006

\bibitem[{Di Mauro} {et~al.}(2021)]{2021PhRvD.104h3012D}
{Di Mauro}, M., {Donato}, F., \& {Manconi}, S. 2021, \prd, 104, 083012

\bibitem[{Dubner} {et~al.}(2008)]{2008A&A...487.1033D}
{Dubner}, G., {Giacani}, E., \& {Decourchelle}, A. 2008, \aap, 487, 1033

\bibitem[{Duncan} {et~al.}(1995)]{1995MNRAS.277...36D}
{Duncan}, A.~R., {Stewart}, R.~T., {Haynes}, R.~F., \& {Jones}, K.~L. 1995,
  \mnras, 277, 36

\bibitem[{Eagle}(2022)]{2022arXiv220911855E}
{Eagle}, J.~L. 2022, arXiv e-prints, arXiv:2209.11855

\bibitem[{Fang} \& {Zhang}(2010)]{2010A&A...515A..20F}
{Fang}, J., \& {Zhang}, L. 2010, \aap, 515, A20

\bibitem[{Ferrand} \& {Safi-Harb}(2012)]{2012AdSpR..49.1313F}
{Ferrand}, G., \& {Safi-Harb}, S. 2012, Advances in Space Research, 49, 1313

\bibitem[{Fiori} {et~al.}(2022)]{2022MNRAS.511.1439F}
{Fiori}, M., {Olmi}, B., {Amato}, E., {et~al.} 2022, \mnras, 511, 1439

\bibitem[{Forot} {et~al.}(2006)]{2006ApJ...651L..45F}
{Forot}, M., {Hermsen}, W., {Renaud}, M., {et~al.} 2006, \apjl, 651, L45

\bibitem[Fujinaga {et~al.}(2013)]{10.1093/pasj/65.3.61}
Fujinaga, T., Mori, K., Bamba, A., {et~al.} 2013, Publications of the
  Astronomical Society of Japan, 65, 61

\bibitem[{Fujita} {et~al.}(2021)]{2021ApJ...912..133F}
{Fujita}, Y., {Bamba}, A., {Nobukawa}, K.~K., \& {Matsumoto}, H. 2021, \apj,
  912, 133

\bibitem[{Funk} {et~al.}(2007)]{2007A&A...470..249F}
{Funk}, S., {Hinton}, J.~A., {Moriguchi}, Y., {et~al.} 2007, \aap, 470, 249

\bibitem[{Gaensler} {et~al.}(2002)]{2002ApJ...569..878G}
{Gaensler}, B.~M., {Arons}, J., {Kaspi}, V.~M., {et~al.} 2002, \apj, 569, 878

\bibitem[{Gaensler} {et~al.}(1999)]{1999MNRAS.305..724G}
{Gaensler}, B.~M., {Brazier}, K.~T.~S., {Manchester}, R.~N., {Johnston}, S., \&
  {Green}, A.~J. 1999, \mnras, 305, 724

\bibitem[{Gallant} \& {Tuffs}(1998)]{1998MmSAI..69..963G}
{Gallant}, Y.~A., \& {Tuffs}, R.~J. 1998, \memsai, 69, 963

\bibitem[{Gallant} \& {Tuffs}(1999)]{1999ESASP.427..313G}
{Gallant}, Y.~A., \& {Tuffs}, R.~J. 1999, in ESA Special Publication, Vol. 427,
  The Universe as Seen by ISO, ed. P.~{Cox} \& M.~{Kessler}, 313

\bibitem[{Ge} {et~al.}(2021)]{2021Innov...200118G}
{Ge}, C., {Liu}, R.-Y., {Niu}, S., {Chen}, Y., \& {Wang}, X.-Y. 2021, The
  Innovation, 2, 100118

\bibitem[{Gelfand} {et~al.}(2009)]{2009ApJ...703.2051G}
{Gelfand}, J.~D., {Slane}, P.~O., \& {Zhang}, W. 2009, \apj, 703, 2051

\bibitem[{Gotthelf} {et~al.}(2014)]{2014ApJ...788..155G}
{Gotthelf}, E.~V., {Tomsick}, J.~A., {Halpern}, J.~P., {et~al.} 2014, \apj,
  788, 155

\bibitem[{Green}(1986)]{1986MNRAS.218..533G}
{Green}, D.~A. 1986, \mnras, 218, 533

\bibitem[{Griffith} \& {Wright}(1993)]{1993AJ....105.1666G}
{Griffith}, M.~R., \& {Wright}, A.~E. 1993, \aj, 105, 1666

\bibitem[{Guo} {et~al.}(2017)]{2017ApJ...835...42G}
{Guo}, X.-L., {Xin}, Y.-L., {Liao}, N.-H., {et~al.} 2017, \apj, 835, 42

\bibitem[{H.~E.~S.~S. Collaboration} {et~al.}(2011)]{2011A&A...533A.103H}
{H.~E.~S.~S. Collaboration}, {Abramowski}, A., {Acero}, F., {et~al.} 2011,
  \aap, 533, A103

\bibitem[{H.~E.~S.~S. Collaboration} {et~al.}(2012)]{2012A&A...548A..46H}
{H.~E.~S.~S. Collaboration}, {Abramowski}, A., {Acero}, F., {et~al.} 2012,
  \aap, 548, A46

\bibitem[{H.~E.~S.~S. Collaboration} {et~al.}(2015)]{2015Sci...347..406H}
{H.~E.~S.~S. Collaboration}, {Abramowski}, A., {Aharonian}, F., {et~al.} 2015,
  Science, 347, 406

\bibitem[{H.~E.~S.~S. Collaboration} {et~al.}(2018)]{2018A&A...612A...2H}
{H.~E.~S.~S. Collaboration}, {Abdalla}, H., {Abramowski}, A., {et~al.} 2018,
  \aap, 612, A2

\bibitem[{H.~E.~S.~S. Collaboration} {et~al.}(2019)]{2019A&A...621A.116H}
{H.~E.~S.~S. Collaboration}, {Abdalla}, H., {Aharonian}, F., {et~al.} 2019,
  \aap, 621, A116

\bibitem[{Ishizaki} {et~al.}(2018)]{2018ApJ...867..141I}
{Ishizaki}, W., {Asano}, K., \& {Kawaguchi}, K. 2018, \apj, 867, 141

\bibitem[{Kargaltsev} \& {Pavlov}(2008)]{2008AIPC..983..171K}
{Kargaltsev}, O., \& {Pavlov}, G.~G. 2008, in American Institute of Physics
  Conference Series, Vol. 983, 40 Years of Pulsars: Millisecond Pulsars,
  Magnetars and More, ed. C.~{Bassa}, Z.~{Wang}, A.~{Cumming}, \& V.~M.
  {Kaspi}, 171

\bibitem[{Kargaltsev} {et~al.}(2013)]{2013uean.book..359K}
{Kargaltsev}, O., {Rangelov}, B., \& {Pavlov}, G. 2013, in The Universe
  Evolution: Astrophysical and Nuclear Aspects. Edited by I. Strakovsky and L.
  Blokhintsev. Nova Science Publishers, 359

\bibitem[{Kennel} \& {Coroniti}(1984)]{1984ApJ...283..694K}
{Kennel}, C.~F., \& {Coroniti}, F.~V. 1984, \apj, 283, 694

\bibitem[{Kuiper} {et~al.}(2001)]{2001A&A...378..918K}
{Kuiper}, L., {Hermsen}, W., {Cusumano}, G., {et~al.} 2001, \aap, 378, 918

\bibitem[{Lang} {et~al.}(2010)]{2010ApJ...709.1125L}
{Lang}, C.~C., {Wang}, Q.~D., {Lu}, F., \& {Clubb}, K.~I. 2010, \apj, 709, 1125

\bibitem[{Lazendic} {et~al.}(2000)]{2000ApJ...540..808L}
{Lazendic}, J.~S., {Dickel}, J.~R., {Haynes}, R.~F., {Jones}, P.~A., \&
  {White}, G.~L. 2000, \apj, 540, 808

\bibitem[{Lhaaso Collaboration} {et~al.}(2021)]{2021Sci...373..425L}
{Lhaaso Collaboration}, {Cao}, Z., {Aharonian}, F., {et~al.} 2021, Science,
  373, 425

\bibitem[{Li} {et~al.}(2018)]{2018ApJ...858...84L}
{Li}, J., {Torres}, D.~F., {Lin}, T.~T., {et~al.} 2018, \apj, 858, 84

\bibitem[{Liu} {et~al.}(2023)]{2023ApJ...942..105L}
{Liu}, X., {Guo}, X., {Xin}, Y., {Zhu}, F., \& {Liu}, S. 2023, \apj, 942, 105

\bibitem[{Lu} {et~al.}(2001)]{2001A&A...370..570L}
{Lu}, F.~J., {Aschenbach}, B., \& {Song}, L.~M. 2001, \aap, 370, 570

\bibitem[{Lu} {et~al.}(2020)]{2020MNRAS.498.1911L}
{Lu}, F.-W., {Gao}, Q.-G., \& {Zhang}, L. 2020, \mnras, 498, 1911

\bibitem[{Luo} {et~al.}(2020)]{2020ApJ...896..147L}
{Luo}, Y., {Lyutikov}, M., {Temim}, T., \& {Comisso}, L. 2020, apj, 896, 147

\bibitem[{Lyutikov} {et~al.}(2019)]{2019MNRAS.489.2403L}
{Lyutikov}, M., {Temim}, T., {Komissarov}, S., {et~al.} 2019, \mnras, 489, 2403

\bibitem[{Mac{\'\i}as-P{\'e}rez} {et~al.}(2010)]{2010ApJ...711..417M}
{Mac{\'\i}as-P{\'e}rez}, J.~F., {Mayet}, F., {Aumont}, J., \& {D{\'e}sert},
  F.~X. 2010, \apj, 711, 417

\bibitem[{Manconi} {et~al.}(2020)]{2020PhRvD.102b3015M}
{Manconi}, S., {Di Mauro}, M., \& {Donato}, F. 2020, \prd, 102, 023015

\bibitem[{Mares} {et~al.}(2021)]{2021ApJ...912..158M}
{Mares}, A., {Lemoine-Goumard}, M., {Acero}, F., {et~al.} 2021, \apj, 912, 158

\bibitem[{Mart{\'\i}n} {et~al.}(2016)]{2016MNRAS.459.3868M}
{Mart{\'\i}n}, J., {Torres}, D.~F., \& {Pedaletti}, G. 2016, \mnras, 459, 3868

\bibitem[{Mart{\'\i}n} {et~al.}(2012)]{2012MNRAS.427..415M}
{Mart{\'\i}n}, J., {Torres}, D.~F., \& {Rea}, N. 2012, \mnras, 427, 415

\bibitem[{Martin} {et~al.}(2022)]{2022A&A...666A...7M}
{Martin}, P., {Tibaldo}, L., {Marcowith}, A., \& {Abdollahi}, S. 2022, \aap,
  666, A7

\bibitem[{Morsi} \& {Reich}(1987)]{1987A&AS...69..533M}
{Morsi}, H.~W., \& {Reich}, W. 1987, \aaps, 69, 533

\bibitem[{Murphy} {et~al.}(2007)]{2007MNRAS.382..382M}
{Murphy}, T., {Mauch}, T., {Green}, A., {et~al.} 2007, \mnras, 382, 382

\bibitem[Murphy {et~al.}(2007)]{10.1111/j.1365-2966.2007.12379.x}
Murphy, T., Mauch, T., Green, A., {et~al.} 2007, Monthly Notices of the Royal
  Astronomical Society, 382, 382

\bibitem[{Nynka} {et~al.}(2014)]{2014ApJ...789...72N}
{Nynka}, M., {Hailey}, C.~J., {Reynolds}, S.~P., {et~al.} 2014, \apj, 789, 72

\bibitem[Pagliaroli {et~al.}(2021)]{Pagliaroli:2021sQ}
Pagliaroli, G., Vecchiotti, V., \& Villante, F. 2021, PoS, ICRC2021, 671

\bibitem[{Park} {et~al.}(2023)]{2023ApJ...945...66P}
{Park}, J., {Kim}, C., {Woo}, J., {et~al.} 2023, \apj, 945, 66

\bibitem[{Pineault} \& {Joncas}(2000)]{2000AJ....120.3218P}
{Pineault}, S., \& {Joncas}, G. 2000, \aj, 120, 3218

\bibitem[{Planck Collaboration} {et~al.}(2016)]{2016A&A...586A.134P}
{Planck Collaboration}, {Arnaud}, M., {Ashdown}, M., {et~al.} 2016, \aap, 586,
  A134

\bibitem[{Porquet} {et~al.}(2003)]{2003A&A...401..197P}
{Porquet}, D., {Decourchelle}, A., \& {Warwick}, R.~S. 2003, \aap, 401, 197

\bibitem[{Rees} \& {Gunn}(1974)]{1974MNRAS.167....1R}
{Rees}, M.~J., \& {Gunn}, J.~E. 1974, \mnras, 167, 1

\bibitem[{Reynolds} {et~al.}(2018)]{2018ApJ...856..133R}
{Reynolds}, S.~P., {Borkowski}, K.~J., \& {Gwynne}, P.~H. 2018, \apj, 856, 133

\bibitem[{Reynolds} {et~al.}(2017)]{2017SSRv..207..175R}
{Reynolds}, S.~P., {Pavlov}, G.~G., {Kargaltsev}, O., {et~al.} 2017, \ssr, 207,
  175

\bibitem[{Roberts} {et~al.}(1999)]{1999ApJ...515..712R}
{Roberts}, M. S.~E., {Romani}, R.~W., {Johnston}, S., \& {Green}, A.~J. 1999,
  \apj, 515, 712

\bibitem[{Salter} {et~al.}(1989)]{1989ApJ...338..171S}
{Salter}, C.~J., {Reynolds}, S.~P., {Hogg}, D.~E., {Payne}, J.~M., \& {Rhodes},
  P.~J. 1989, \apj, 338, 171

\bibitem[{Sironi} \& {Spitkovsky}(2011)]{2011ApJ...741...39S}
{Sironi}, L., \& {Spitkovsky}, A. 2011, \apj, 741, 39

\bibitem[{Sironi} \& {Spitkovsky}(2012)]{2012CS&D....5a4014S}
{Sironi}, L., \& {Spitkovsky}, A. 2012, Computational Science and Discovery, 5,
  014014

\bibitem[{Sironi} \& {Spitkovsky}(2014)]{2014ApJ...783L..21S}
{Sironi}, L., \& {Spitkovsky}, A. 2014, \apjl, 783, L21

\bibitem[{Slane}(2017)]{2017hsn..book.2159S}
{Slane}, P. 2017, in Handbook of Supernovae, ed. A.~W. {Alsabti} \&
  P.~{Murdin}, 2159

\bibitem[{Tanaka} \& {Takahara}(2010)]{2010ApJ...715.1248T}
{Tanaka}, S.~J., \& {Takahara}, F. 2010, \apj, 715, 1248

\bibitem[{Tanaka} \& {Takahara}(2011)]{2011ApJ...741...40T}
{Tanaka}, S.~J., \& {Takahara}, F. 2011, apj, 741, 40

\bibitem[{Temim} {et~al.}(2006)]{2006AJ....132.1610T}
{Temim}, T., {Gehrz}, R.~D., {Woodward}, C.~E., {et~al.} 2006, \aj, 132, 1610

\bibitem[{Tibet AS{\ensuremath{\gamma}} Collaboration}
  {et~al.}(2021)]{2021NatAs...5..460T}
{Tibet AS{\ensuremath{\gamma}} Collaboration}, {Amenomori}, M., {Bao}, Y.~W.,
  {et~al.} 2021, Nature Astronomy, 5, 460

\bibitem[{Torii} {et~al.}(2000)]{2000PASJ...52..875T}
{Torii}, K., {Slane}, P.~O., {Kinugasa}, K., {Hashimotodani}, K., \& {Tsunemi},
  H. 2000, \pasj, 52, 875

\bibitem[{Torres} {et~al.}(2014)]{2014JHEAp...1...31T}
{Torres}, D.~F., {Cillis}, A., {Mart{\'\i}n}, J., \& {de O{\~n}a Wilhelmi}, E.
  2014, Journal of High Energy Astrophysics, 1, 31

\bibitem[{Ubertini} {et~al.}(2005)]{2005ApJ...629L.109U}
{Ubertini}, P., {Bassani}, L., {Malizia}, A., {et~al.} 2005, \apjl, 629, L109

\bibitem[{Van Etten} \& {Romani}(2010)]{2010ApJ...711.1168V}
{Van Etten}, A., \& {Romani}, R.~W. 2010, \apj, 711, 1168

\bibitem[{Veron-Cetty} \& {Woltjer}(1993)]{1993A&A...270..370V}
{Veron-Cetty}, M.~P., \& {Woltjer}, L. 1993, \aap, 270, 370

\bibitem[{Vorster} {et~al.}(2013)]{2013ApJ...773..139V}
{Vorster}, M.~J., {Tibolla}, O., {Ferreira}, S.~E.~S., \& {Kaufmann}, S. 2013,
  \apj, 773, 139

\bibitem[{Wach} {et~al.}(2023)]{2023arXiv230816717W}
{Wach}, T., {Mitchell}, A.~M.~W., {Joshi}, V., \& {Funk}, S. 2023, arXiv
  e-prints, arXiv:2308.16717

\bibitem[{Xin} {et~al.}(2018)]{2018ApJ...867...55X}
{Xin}, Y.-L., {Liao}, N.-H., {Guo}, X.-L., {et~al.} 2018, \apj, 867, 55

\bibitem[{Xin} {et~al.}(2019)]{2019ApJ...885..162X}
{Xin}, Y., {Zeng}, H., {Liu}, S., {Fan}, Y., \& {Wei}, D. 2019, \apj, 885, 162

\bibitem[{Zabalza}(2015)]{naima}
{Zabalza}, V. 2015, Proc.~of International Cosmic Ray Conference 2015, 922

\bibitem[{Zeng} {et~al.}(2019)]{2019ApJ...874...50Z}
{Zeng}, H., {Xin}, Y., \& {Liu}, S. 2019, Astrophys. J., 874, 50

\bibitem[{Zhang} {et~al.}(2008)]{2008ApJ...676.1210Z}
{Zhang}, L., {Chen}, S.~B., \& {Fang}, J. 2008, \apj, 676, 1210

\bibitem[{Zhou} {et~al.}(2023)]{2023RAA....23j5001Z}
{Zhou}, L.-C., {Xia}, Q., {Tian}, S.-T., {Gong}, Y.-l., \& {Fang}, J. 2023,
  Research in Astronomy and Astrophysics, 23, 105001

\bibitem[Zhu {et~al.}(2023)]{Zhu_2023}
Zhu, B.-T., Lu, F.-W., \& Zhang, L. 2023, The Astrophysical Journal, 943, 89

\bibitem[{Zhu} {et~al.}(2021)]{2021A&A...655A..41Z}
{Zhu}, B.-T., {Lu}, F.-W., {Zhou}, B., \& {Zhang}, L. 2021, \aap, 655, A41

\bibitem[{Zhu} {et~al.}(2018)]{2018A&A...609A.110Z}
{Zhu}, B.-T., {Zhang}, L., \& {Fang}, J. 2018, \aap, 609, A110

\bibitem[{Zirakashvili} \& {Aharonian}(2007)]{2007A&A...465..695Z}
{Zirakashvili}, V.~N., \& {Aharonian}, F. 2007, \aap, 465, 695

\end{thebibliography}

\end{document}